\documentclass[article,12pt]{elsarticle}
\usepackage{amssymb}
\usepackage{amsmath}
\usepackage{latexsym}
\usepackage{amsthm}
\usepackage{eucal}
\usepackage{eufrak} 
\usepackage{dsfont}
\usepackage{textcomp}
\usepackage{comment}
\usepackage{indentfirst}
\usepackage{amsthm}	
\usepackage[T1]{fontenc} 
\usepackage{eso-pic}
\usepackage{appendix}
\usepackage{float}
\usepackage{bbold}
\usepackage{graphicx,psfrag,subfig}
\usepackage{enumerate}
\usepackage{caption}
\usepackage{cancel}
\usepackage{ulem}
\usepackage{amsthm}
\usepackage{mathabx}

\DeclareMathOperator*{\Max}{Max}
\newcommand{\grafe}[1]{\left\{ #1 \right\}}
\newcommand{\tonde}[1]{\left( #1 \right)}
\newcommand{\quadre}[1]{\left[ #1 \right]}
\newcommand{\modul}[1]{\left| {#1} \right|}
\newcommand{\ope}{{I}}
\newcommand{\be}{\begin{equation}}
\newcommand{\ee}{\end{equation}}
\newcommand{\bea}{\begin{eqnarray}}
\newcommand{\eea}{\end{eqnarray}}

\newcommand{\nn}{\nonumber}

\newcommand{\cU}{\mathcal{U}}

\journal{Nuclear Physics B}

\begin{document}

\begin{frontmatter}

\title{Integrals of motion in the Many-Body Localized phase}
\author{V. Ros}
\address{SISSA, via Bonomea 265, 34136 Trieste, Italy.}
\address{INFN, Sezione di Trieste, Strada Costiera 11, 34151 Trieste, Italy. }
\author{M. M\"uller}
\address{Abdus Salam ICTP, Strada Costiera 11, 34151 Trieste, Italy}
\author{A. Scardicchio}
\address{Physics Department, Princeton University, Princeton, NJ 08544, USA}
\address{Physics Department, Columbia University, New York, NY 10027, USA}
\address{ITS, Graduate College of the City University of New York, New York, NY 10016 USA}
\address{{\tiny on leave from:} Abdus Salam ICTP, Strada Costiera 11, 34151 Trieste, Italy}
\address{INFN, Sezione di Trieste, Strada Costiera 11, 34151 Trieste, Italy. }

\begin{abstract}
We construct a complete set of quasi-local integrals of motion for the many-body localized phase of interacting fermions in a disordered potential. The integrals of motion can be chosen to have binary spectrum $\{0,1\}$,  thus constituting exact quasiparticle occupation number operators for the Fermi insulator. We map the problem onto a non-Hermitian hopping problem on a lattice in operator space. We show how the integrals of motion can be built, under certain approximations, as a convergent series in the interaction strength. An estimate of its radius of convergence is given, which also provides an estimate for the many-body localization-delocalization transition.  Finally, we discuss how the properties of the operator expansion  for the integrals of motion imply the presence or absence of a finite temperature transition.  

\end{abstract}


\end{frontmatter}

\section{Introduction}
\label{sec:introduction}
The thermodynamic description of macroscopic bodies, as shown by Boltzmann in his work on the foundations of statistical mechanics, is based on the assumption that the underlying microscopic dynamics are ergodic. More precisely, one assumes that the environment of any small subsystem of the macroscopic body acts as a thermal bath, with which the subsystem can exchange particles and energy, and which leads to the eventual thermalization of the subsystems, independently of its initial state.  In order for  thermalization to occur gradients in particle and energy density must be able to even out, which requires non-vanishing transport over arbitrarily large scales. 

However, in the absence of interaction, Anderson \cite{anderson1958absence} has shown that a sufficiently strong quenched disorder can localize quantum particles. This prevents the transport of energy and particles and therefore entails the non-ergodicity of the system. Already in Anderson's first paper, and later in the context of electron-electron interactions~\cite{AndersonFleishman82}, it was surmised that this localization might persist in the presence of interactions, despite the widespread belief that any finite interactions would restore transport, ergodicity, and thus standard thermodynamic behavior, in such systems. Later, numerical investigations of the Hubbard model~\cite{BerkovitsShklovskii} hinted indeed at the possibility of such "many-body localization" (MBL), and more recently the seminal study of disordered electrons with weak short range interactions to 
all orders of perturbation theory provided important analytical insight into this phenomenon~\cite{Basko:2006hh}, predicting that in an isolated system, decoupled from any external bath, a finite interaction is required to induce delocalization and enable transport. 
Below this delocalization threshold, truly inelastic decay processes are impossible, as the system ceases to be a heat-bath for itself, and any d.c. transport is strictly absent.
In this way many-body localized systems are crucially different from other situations where full ergodicity in phase space breaks down, such as in systems with spontaneously broken symmetries, one-dimensional integrable systems, or spin glasses. In all these examples, the thermal conductivity remains finite. \footnote{In integrable systems, transport of some quantities is often even more efficient than in non-integrable systems, being ballistic as opposed to diffusive. 
}  
In contrast, a necessary\footnote{The condition is not sufficient, since even in the absence of diffusion thermalization might occur via sub-diffusive processes. This was found empirically in one-dimensional systems close enough to the localization transition~\cite{reichman2014Absence, AgarwalAnomalousDiffusion}.  Moreover, localization can occur also in time-dependent systems (e.g., periodically driven systems) that may have no conserved local densities and thus no meaningful d.c.\ transport~\cite{dAlessio2013per, abanin2014per}. For these systems, MBL is defined more generally as a phase where any local observable does not thermalize almost certainly.
} condition for many-body localization is the vanishing of the d.c.\ transport coefficients at non-zero temperature.

Since the seminal work by Basko, Aleiner and Altshuler~\cite{Basko:2006ux,Basko:2006hh,Basko:2007hs}, the paradigm of many-body localization has attracted a lot of interest, and the phenomenology of  MBL phases and the localization transition have been explored, see for example \cite{oganesyan2007localization, vznidarivc2008many, pal2010MBL, bardarson2012unbounded, vosk2013dynamical, de2013ergodicity, Kjall:2014fj}). Many-body localization opens the interesting possibilities of protection of topological order at finite temperature or of phase transitions below the equilibrium lower critical dimension~\cite{aleiner2010finite,huse2013localization, chandran2013many, pekker2013hilbert,bahri2013localization,Yao:2013ly}). It was even proposed that MBL could survive in the absence of quenched disorder~\cite{KaganMaksimov,schiulaz2013ideal,schiulaz2014Dynamics, Huveneers, Cirac2014quasi}. (A different type of non-ergodic behavior, exhibiting, however, ballistic transport, has been conjectured in disorder free 1d systems that are close enough to integrability.~\cite{Prosen1998, Prosen1999}).

Unambiguous experimental evidence of an MBL transition or phase is however still lacking at the time of writing, despite of promising developments~\cite{errico2014observation, Ovadyahu_MWabsorption}.

An MBL phase can be seen as the prototype of a quantum glass phase, where the dynamics are slowed down indefinitely and where memory of the initial condition is retained in local observables for arbitrarily long times. This latter phenomenon has certain similarities with integrable systems  \cite{huse2013phenomenology,serbyn2013local}, in which an extensive number of conserved quantities (integrals of motion) constrain the system to evolve in a much smaller submanifold than the one determined by the conservation of energy and momenta only. The long-time relaxation then only leads to a restricted (generalized) Gibbs ensemble.
 
Given this similarity, it was conjectured that, like in the non-interacting limit, an extensive set of (quasi-)local integrals of motion should exist in the MBL phase~\cite{huse2013phenomenology,imbrie2014many,serbyn2013local}. By definition, those do not evolve with time, as they commute with the Hamiltonian. They thus constrain the dynamics to remain very close to the initial condition in which the system was prepared. The existence of such local integrals of motion was recently proven for a particular spin Hamiltonian in \cite{imbrie2014many}, under reasonable assumptions bounding potential level attraction. The notion of locality used above refers to the set of degrees of freedom, which the conserved operator affects. Conserved quantities in integrable systems are not local in this sense, as they are sums over all space of certain local terms.\footnote{Quantum mechanics provides quite trivially a large set of  mutually commuting, conserved operators in any system, namely the projectors on exact many-body eigenstates. However, those are highly non-local and have minimal rank 1. Such trivial "integrals of motion" are of no interest in the present context.} The non-locality (in our sense) of those integrals still allows for finite transport in integrable systems.
A further important difference between MBL systems and integrable ones is the fact that MBL is robust with respect to any sufficiently small perturbation of the Hamiltonian, while integrability in 1d systems is broken by generic perturbations.

The aim of this paper is to show that quasi-local integrals of motion exist for weakly interacting disordered electrons, under the same set of assumptions that were made in the original work by Basko et al.~\cite{Basko:2006hh} (henceforth referred to as BAA). We find such integrals by solving equations for conserved operators within perturbation theory. 
Our approach reduces the problem to the solution of a single-particle-like hopping problem in operator space, for which we present a solution in the strongly localized regime, and determine the radius of convergence of the construction. This furnishes an estimate of the delocalization very similar to that obtained by Basko et al.~\cite{Basko:2006hh}.
We hope that our technique will help to obtain analytic results on many-body localization in the future.
\subsection{Outline and summary of this work}
\label{sec:summary}

Here, we present a short outline of this work, summarizing the main steps, and the problems we address. 

We are seeking integrals of motion for disordered electrons with weak short range interactions, as defined in Eq.~(\ref{ham1}, \ref{deflambda}).  In Sec.~\ref{sec:model} we {coarse-grain} the model, reducing it to an array of coupled quantum dots of size of the order of the single-particle localization length.

The non-interacting model has trivial integrals of motion, namely the occupation numbers of the single-particle eigenstates. We then look for their generalization in the presence of interactions, ``dressing'' these integrals of motion (Sec.~\ref{sec:transport}). This leads us to a set of linear equations (Sec.~\ref{sec:construction}, Eq.~(\ref{fullequation})) in the space of number conserving operators, which we expand in the basis (\ref{ansatz}) of products of single particle creation and annihilation operators. 
For any strength $\lambda$ of the interaction, these equations define a unique set of conserved operators. The main question to analyze is whether they act locally, or whether they significantly affect a spatially unbounded set of degrees of freedom. We address the question of locality within the so-called forward approximation, introduced in Sec.~\ref{sec:fowardapprox}, where we only determine the leading term in perturbation theory for the expansion coefficients. Since the interaction terms act locally, for the conserved quantities to be non-local increasingly high orders of perturbation theory must contribute to the expansion, i.e., the perturbative expansion diverges (Sec.~\ref{sec:probresonances}).

We represent diagrammatically the particle-hole creation processes, which dominate the forward approximation, in Sec.~\ref{sec:diagrep}. In order to study the statistics of the diagrams at high orders, we need to solve three main technical problems. One is the estimate of their number, due to the freedom in choosing the interaction vertices and their order. We solve this (Sec.~\ref{sec:diagrep}) by introducing an integral representation that sums correlated diagrams sharing the same interaction vertices. This reduces the factorially many (in the order $N$ of the perturbation theory)  terms to an only sub-exponential number of terms, which are products of $N$ denominators. The second problem  concerns their statistical distribution. In the many-body problem the denominators are correlated even within the forward approximation, at variance with one-particle problems. Therefore determining the statistics of large deviations, which dominate the probability of creating excitations at large distance, is a challenge. We solve it using a transfer-matrix technique (Sec.~\ref{sec:largedev}). Finally, we have to count the number of processes leading to a given configuration in operator space, which is a combinatorial problem in the space of diagrams (Sec.~\ref{sec:combinatorics}). The last two ingredients allow us to determine the decay rate of the largest of these terms, which dominates the expansion. Requiring a positive spatial decay rate determines the range of convergence of the operator expansion in the forward approximation.  

After solving these technical problems, we obtain the final result in Sec.~\ref{sec:estimateradius}, namely the existence of quasi-local integrals of motion for disordered electrons for sufficiently small interaction $\lambda<\lambda_c$.  We find $\lambda_c$ in the forward approximation: in the same spirit as Anderson's ``upper bound'' approximation, this is expected to yield a lower bound for the actual phase boundary for  many-body localized phase of the lattice system at infinite temperature. In a final section, we discuss possible scenarios for a localization transition or crossover at finite temperature (Sec.~\ref{sec:finitetemp}).  

\section{Model Hamiltonian and coarse-graining}
\label{sec:model}

We consider a Hamiltonian describing weakly interacting, spinless electrons in a disordered background. At variance with the work by BAA,  we consider a model on a lattice $\Lambda$,
\bea\label{ham1}
H= \sum_{i\in \Lambda} c^\dagger_i \left[ -\frac{1}{2m}\Delta^{(\Lambda)}+{V}_{\rm dis}(i)\right]c_i + \frac{1}{2} \sum_{i,j \in \Lambda} c^\dagger_ic^\dagger_j \,U(i-j)\, c_j c_i,
\eea
where $\Delta^{(\Lambda)}$ is the lattice Laplacian, ${V}_{\rm dis}$ is a random disordered potential and ${U}$ is a short range interaction.

We choose to work with a lattice model, because in a finite volume its Hilbert space is finite, and both spectrum and energy per particle are bounded. This will allow us to take a meaningful limit of infinite temperature,  and to make statements about many-body localization in that limit.

It is convenient to write the interaction in the form 
\bea
\label{deflambda}
{U}(i-j) =\frac{\lambda}{\nu a^d} u(i-j)
\eea 
where $\nu$ is the density of states, and $u(i-j)$ is a dimensionless, normalized, short-ranged interaction kernel. The dimensionless parameter $\lambda$  measures the interaction strength. 

We consider a disorder potential such that the single particle part of the Hamiltonian possesses only fully localized wave-functions $\phi^\alpha$, $\alpha=1, ..., |\Lambda|$, with typical localization length $\xi$. Moreover, we are interested in the disorder regime relatively close to single particle delocalization, where $\xi$ is significantly bigger than the lattice spacing $a$. Let us denote by $\delta_\xi = 1/\nu \xi^d$  the average level spacing in a localization volume, and by $W$  the band width of the single particle problem.
The condition $\xi\gg a$ ensures that a large number
\begin{equation}\label{nloc}
N_{\text{loc}}=\frac{W}{\delta_\xi}
\end{equation}
of single particle wave-functions overlap significantly in space. This will provide a large parameter for our analysis.

It is convenient to switch to the basis of single particle wave-functions $\phi^\alpha$, in which the Hamiltonian assumes the form
\bea\label{hamilton}
 H&=&\sum_{\alpha} \epsilon_{\alpha} n_{\alpha} +  \sum_{\alpha <\beta, \gamma <\delta} {U}_{\alpha \beta, \gamma \delta}\, c^{\dag}_{\alpha} c^{\dag}_{\beta} c_{\gamma} c_{\delta}\nn \\
&=&\sum_{\alpha} \epsilon_{\alpha} n_{\alpha} + \frac{\lambda}{\nu a^d} \sum_{\alpha <\beta, \gamma <\delta} {u}_{\alpha \beta, \gamma \delta}\, c^{\dag}_{\alpha} c^{\dag}_{\beta} c_{\gamma} c_{\delta}\\
\label{hamilton1}
&\equiv& H_0 + U,
\eea
where $n_\alpha= c^{\dag}_{\alpha}c_\alpha$, and the Greek indices label single particle eigenstates obtained in the absence of interaction. We also choose a certain ordering relation $"<"$ among the indices $\beta,\gamma$.

Our choice of the basis $\phi_\alpha$ is different from that of BAA, who worked with Hartree-Fock (HF) orbitals. Our choice allows us to work in full generality in the operator space, while HF orbitals depend on the non-interacting occupation numbers, i.e., the many body state around which one analyzes stability with respect to interactions. In Sec.~ \ref{nodiagonals} we will argue, however,  that in the approximation in which we are working,  we can neglect the interaction vertices $U_{\alpha\beta,\gamma\delta}$ with two or more coinciding indices, even without resorting to HF, which resums most of those terms. Thus, the two different choices of basis sets lead essentially to the same combinatoric analysis of diagrams.

 To simplify the above model further, we assume the single particle energies $\epsilon_\alpha$ to be random and uncorrelated. The interaction term $U$ is antisymmetrized: $U_{\alpha \beta, \gamma \delta}=U_{\beta \alpha,  \delta \gamma}=-U_{\beta \alpha, \gamma \delta}$.
We further simplify it by taking its matrix elements $U_{\alpha \beta, \gamma \delta}$ to be local in space, i.e., they are assumed to be non-zero only if the corresponding single particle states have localization center within one localization volume. Hereby we define the localization center of a single particle state as 
\bea \label{radius}
\vec{r}_\alpha = \int d^dr \phi_\alpha^2(r) \vec{r}.
\eea
Moreover, it is known that the matrix elements decrease rather rapidly (as a power law) when the energy difference between involved levels exceeds the level spacing in the localization volume $\delta_\xi$. This motivates the use of a simplified interaction in which we take $u_{\alpha \beta, \gamma \delta} $ to be non-zero only if 
\begin{equation}\label{restriction}
 |\epsilon_\alpha-\epsilon_\delta|, |\epsilon_\beta-\epsilon_\gamma| \lesssim \delta_\xi \hspace{.3 cm}\text{     or  } \hspace{.3 cm} |\epsilon_\alpha-\epsilon_\gamma|, |\epsilon_\beta-\epsilon_\delta| \lesssim \delta_\xi.
\end{equation}
 In these cases we assume 
\bea \label{mel}
u_{\alpha \beta, \gamma \delta}=\eta_{\alpha \beta, \gamma \delta} \, \nu a^d \delta_\xi =\eta_{\alpha \beta, \gamma \delta} \left(\frac{a}{\xi}\right)^d, 
\eea
where $\eta_{\alpha \beta, \gamma \delta}$ is a random variable, box-distributed in $[-1,1]$.

\subsection{Coarse-graining}

Let us now coarse-grain the model: we assume that the interaction $U_{\alpha \beta,\gamma \delta}$   connects  wave-functions either on the same localization volume or on neighboring localization volumes. For either vertices we assume the same amplitude $\lambda$, as long as the restrictions (\ref{restriction}) on the energy levels are respected.

This differs from the coarse-graining by  BAA, who divided the sample into $d$-dimensional regions of linear size $\xi$, restricted the single particle levels $\alpha$ to those regions, but included a (small) hopping term  between  localization volumes (elastic processes). In contrast the interaction term, responsible for inelastic processes, was restricted to scattering within a given localization cell.

\section{Integrals of motion and absence of transport}
\label{sec:transport}
 In the absence of interactions ($\lambda=0$), the occupation numbers $n_\alpha$ of single particle levels are mutually commuting, conserved quantities. These operators are \emph{quasi-local} in real space, as follows immediately from their expansion in the basis of lattice operators:
\begin{equation}\label{expsingle}
 n_\alpha= \sum_{i,j} \phi_{\alpha}^{*}(i) \phi_{\alpha}(j) c^{\dag}_i c_j,
\end{equation}
where $\phi_\alpha$ is the corresponding localized single particle eigenfunction. By {\em quasi-locality} of the $n_\alpha$ we mean that an operator $ c^{\dag}_i c_j$ contributes in the expansion with a weight which decays exponentially in the distance between the localization center $\vec r_\alpha$ of $\phi_\alpha$ and the sites it acts on (its {\em support} - here the sites $i,j$). 

By truncating the sum ($\ref{expsingle}$) to terms with support only within a neighborhood of $m \xi$ of $\vec r_\alpha$ one obtains an operator, whose commutator with the Hamiltonian vanishes up to exponentially small terms. As $m\to \infty$ the operator rapidly converges (in the operator norm) to the conserved $n_\alpha$. In the non-interacting case this follows directly from the spatial localization of the single particle wave-functions. Our goal is to find an analogue of these operators in the interacting case.

That such a generalization should exist was proven by Imbrie ~\cite{imbrie2014many} under certain hypotheses on the spectrum in a 1d spin chain, for which he constructed a quasi-local unitary rotation $\cU$  which essentially diagonalizes the Hamiltonian $H$. More precisely, it brings it to  the canonical form
  \bea\label{Imbrie}
 \cU^{\dag}H \cU = -\sum_i h_i \sigma_i^z - \sum_{i<j} J_{i,j} \sigma_i^z \sigma_j^z+ \sum_{i<j<k}  J_{i,j,k} \sigma_i^z \sigma_j^z\sigma_k^z +...,
  \eea
where the $k$-spin interactions $J_{i_1,...,i_k}$ decay exponentially with the diameter of their index set. Applying the inverse unitary on the conserved spins $\sigma_i^z$ provides one with 
  integrals of motion of the original Hamiltonian $H$, $I_i = \cU \sigma_i^z \cU^{\dag}$.
  At the same time, Huse and Oganesyan~\cite{huse2013phenomenology}, and independently, Serbyn et al.~\cite{serbyn2013local} argued for the existence of such local integrals of motion in general MBL systems.
  
 Note that the set of conserved and mutually commuting quantities is by no means unique. For example, any set of independent polynomials of $\sigma_i^z$'s is conserved as well. A nice property of the set of $\sigma_i^z$, however, is the binarity of their spectrum, $\{-1,1\}$, or the property that $(\sigma_i^z)^2=1$. Knowing the eigenvalues of $N$ independent integrals of motion like this allows one to unambiguously label the $2^N$ eigenstates of the Hilbert space of an $N$-spin system~\cite{huse2013phenomenology}. 
  
An alternative construction of conserved (but non-binary) quantities is discussed in ~\cite{chandran2014} for a random spin chain, where infinite time averages of local operators are considered (such as $n_i(t) = e^{iHt} n_i e^{-iHt}$ in our case).  By definition of the time average, it  commutes with the Hamiltonian. In an MBL phase one expects the average to remain non-zero, whereas it vanishes due to diffusion in an ergodic delocalized phase.

In this paper, we make a different choice, which nevertheless defines a unique set of binary integrals of motion. Our construction consists in two steps. We will first prove the existence of local integrals of motion in perturbation theory in $\lambda$, not requiring the binarity of their spectrum. This is the most difficult task and will take the largest part of the paper. 

\subsection{Construction of exact quasiparticles of the Fermi insulator}
Since our procedure will leave us some freedom in the choice of integrals, in \ref{app:binarity} we will show how this freedom can be used to fix the spectrum to be binary, order by order in the interaction $\lambda$. 
Notice that the latter amounts to the construction of {\em exact} quasiparticle occupation numbers of  the interacting Fermi insulator.
In contrast to Fermi liquids where such exact quasiparticle operators cannot be constructed, neither in real nor in momentum space, it becomes possible in the MBL phase. Rewritten in terms of these occupation numbers $\tilde{n}_\alpha$, the Hamiltonian 
\begin{equation}
 H=\sum_{\alpha}\epsilon_\alpha \tilde{n}_\alpha+\frac{1}{2}\sum_{\alpha\neq \beta}J_{\alpha,\beta}\tilde{n}_\alpha \tilde{n}_\beta+...\,,
\end{equation}
can then be seen as an exact quasiparticle energy functional, which determines the energy $ E_\alpha^{(\text{qp})}$ of any quasiparticle as a function of the occupations of all others, as:
\begin{equation}
 E_\alpha^{(\text{qp})}\tonde{\grafe{\tilde{n}_\beta}}\equiv \frac{\partial H}{\partial \tilde{n}_\alpha }= \epsilon_\alpha+\sum_{\beta(\neq \alpha)}J_{\alpha,\beta}\tilde{n}_\beta+...\
\end{equation}

\subsection{Complete set of local integrals implies absence of transport}

Before outlining the construction of the integrals of motion, let us first show how their existence implies the absence of any d.c.\ transport, and hence many-body localization. 

In order to show the absence of d.c. transport, consider the  Kubo formula for the conductivity $\sigma$  associated with the local current density $J_r$, associated with a conserved quantity, such as charge or energy. Let
\begin{equation}
 J_r(\omega)= \sum_{r'} \sigma(r, r'; \omega) E(\omega) 
\end{equation}
be the current at frequency $\omega$ and position $r$ arising in linear response to a spatially homogeneous field $E$, and denote by 
\begin{equation}\label{avecurrent}
 J(\omega)=\frac{1}{V} \sum_{r} J_r(\omega) \equiv \sigma(\omega) E(\omega)
\end{equation}
the spatially averaged current density, $V$ being the volume of the system. At finite inverse temperature $\beta$, the dissipative part of the conductivity is given by:
\begin{equation}
{\rm Re} [\sigma(\omega)]=-\frac{1}{V} \sum_{r} \frac{{\rm Im} [\Pi(\omega,r)]}{\omega},
\end{equation}
where $\Pi(\omega,r)$ is the Fourier transform of the retarded correlation function of the current operator, with Lehmann representation:
\begin{equation}
 \Pi(\omega,r)= {\frac{1}{\mathcal{Z}} \sum_{m, m'}\sum_{r'} e^{-\beta E_{m'}} \tonde{1-e^{-\beta(E_{m}-E_{m'})}} \frac{\langle {m'} |{J_{r'+r}}| {m} \rangle \langle {m} |{J_{r'}} |{m'} \rangle}{ \omega +E_{m'}-E_{m}+i \eta}  }.
\end{equation}
Here $\mathcal{Z}$ is the partition function, and the limit $\eta \to 0$ is to be taken after the thermodynamic limit. In the d.c. limit one finds
\begin{equation}\label{KuboDC}
 {\rm Re}[\sigma(\omega\to 0)]=  \frac{ \pi \beta}{V}  \sum_{r' r} \sum_{m, m'}{\frac{e^{-\beta E_{m'}}}{\mathcal{Z}}  {\langle {m'} |{J_{r'+r}}| {m} \rangle \langle {m} |{J_{r'}} |{m'} \rangle}\ \delta_\eta\tonde{E_{m'}-E_{m}}},
\end{equation}
where we have used
\begin{equation}
 \lim_{\omega \to 0} \frac{1-e^{-\beta \omega}}{\omega}=\beta,
\end{equation}
and $\delta_\eta(x) = \pi^{-1}\eta/(x^2 +\eta^2)$ is a regularized $\delta$-function.

Let us now show first that for a {\em complete} set of {\em strictly local conserved quantities} the conductivity vanishes with probability one in the thermodynamic limit.
By strict locality operators we mean that they only act on degrees of freedom that belong to a compact spatial region with finite diameter $\zeta$. We call a set of conserved quantities complete if for any two distinct eigenstates $m\neq m'$ at least one of those integrals of motion takes a different eigenvalue.

For two eigenstates $m, m'$ let $\tilde I$ be such a distinguishing integral, with corresponding eigenvalues $\tilde I |m\rangle= \tilde I_m |m\rangle$ and $\tilde I |m'\rangle= \tilde I_{m'} |m'\rangle$, with $\tilde I_{m'}\neq \tilde I_{m}$. For a strictly local current operator and $r$ sufficiently much bigger than $\zeta$, it follows immediately that one of the two current matrix elements
\begin{equation}\label{withcomm}
\begin{split}
 &\langle m' |{J_{r'}} |m \rangle = \frac{\langle m' |\quadre{J_{r'} ,\tilde I}| m \rangle}{ \tonde{\tilde I_m- \tilde I_{m'}} },\\
  &\langle m' |{J_{r'+r}} |m \rangle = \frac{\langle m' |\quadre{J_{r'+r} ,\tilde I}| m \rangle}{ \tonde{\tilde I_m- \tilde I_{m'}} },
 \end{split}
\end{equation}
  vanishes, since one of the two commutators vanishes. Thus, in Eq.~(\ref{KuboDC}) the sum over $r$ can be restricted to $r\lesssim \zeta$. Furthermore, for any fixed eigenstate $m$ the sum over eigenstates $m'$ is restricted to a finite set, since $J_{r'}|m \rangle$ can differ only in a finite number ($\leq \exp(c \zeta^d)$, with $c=O(1)$) of integrals of motion from $| m\rangle$. Thus, in the thermodynamic limit, where we have to send $\eta\to 0$, the contribution to the $\delta$-function vanishes with probability one, and thus 
  ${\rm Re}[\sigma(\omega=0)] =0$.  Note that the potentially singular term from $m=m'$ does not contribute because $\langle m| J_r| m\rangle=0$ by time reversal invariance.
        
        This discussion is of course over-simplified since the actual integrals of motion are only quasi-local, in the sense that there are corrections to a strict locality, which decay exponentially with the diameter of their support on a typical scale $\zeta$. However, the derivation above reflects the essential mechanism by which a complete set of integrals of motion suppresses transport. Consider  the matrix elements $\langle m'| J_{r'}|m \rangle$ for eigenstates that differ significantly only in integrals of motion whose support is centered up to a distance $x \zeta$ from $r'$. These matrix elements are then not exactly zero, but exponentially small in $x$. There are also exponentially many states $m, m'$ which satisfy these criteria, and thus some energy differences $E_m-E_m'$ in ($\ref{KuboDC}$) become exponentially small.
One might worry that these exponentially small denominators can contribute to the $\delta$-function in the thermodynamic limit, leading to a non-zero conductivity. However, the very construction of the local integrals of motion outlined in the following, and the convergence of that procedure, strongly suggest that with probability tending to one as $\eta\to 0$ the exponential smallness of the energy denominators is dominated by the decay of the matrix elements in ($\ref{KuboDC}$), {in the sense that at small $\eta$ the contributions to the $\delta$-function come with weights that are almost surely much smaller than $\eta$}. If this were not the case, resonant energy denominators would systematically appear in the construction of the conserved quantities and prevent their locality. Therefore, the consistency and convergence of the following construction implies the suppression of d.c. transport in systems admitting a complete set of quasi-local conserved operators.

\subsection{Recipe for the construction of integrals of motion}  
Let us now come back to the actual construction of quasi-local conserved operators.
In order to find a generalization of the single particle occupation numbers to the interacting case one should construct an extensive set of $|\Lambda|$ functionally independent operators~\footnote{ Functional independence means that no $I_\alpha$ can be expressed as a function of all the other $I_\beta$.} $\grafe{{I}_\alpha}$, which are quasi-local and satisfy 
\begin{equation}
\quadre{{I}_\alpha, H}=0.
\end{equation}
Since the spectrum of the many-body system is almost surely non-degenerate, it follows that such conserved quantities also satisfy $\quadre{{I}_\alpha, {I}_{\beta}}=0$. Their mutual commutativity implies that they form a commutative algebra. As we discussed above, the choice of a basis spanning this algebra is not at all unique.
It is worth mentioning that if the operators $I_\alpha$ commute with $H$ and span the algebra of operators then we can write
\begin{equation}
\label{Ham_IOM}
H=\sum_{\alpha}\epsilon_\alpha I_\alpha+\sum_{\alpha,\beta}J_{\alpha,\beta}I_\alpha I_\beta+...\,.
\end{equation}
as we claimed above. The couplings $J$'s have similar exponential decay as those in (\ref{Imbrie}).

Here we present a specific construction of conserved operators, which fixes the arbitrariness in their definition in a unique way.  Our construction starts from the idea that at weak interactions the $I_\alpha$ should be expected to be a perturbed version of the $n_\alpha$. Thus, we look for a perturbative series in $\lambda$,
 \bea \label{ansatzpert}
 I_\alpha= n_\alpha + \Delta I_\alpha = n_\alpha+\sum_{n\geq 1} \lambda^n \Delta I_\alpha^{(n)}.
 \eea
 For the further discussion it is useful to introduce some natural operator subspaces. $I_\alpha$ can be sought as an element of the space $C$ of particle-conserving operators on the Hilbert space, and without loss of generality we may require it to be Hermitian. Since we will require $[H,I_\alpha] = [H_0,I_\alpha]+[U,I_\alpha]=0$, with $H_0$ and $U$ as in (\ref{hamilton1}), we consider the kernel $K$ of the linear map $f(X)= [H_0,X]$ defined for  $X\in C$, as well as its image, $O=f(C)$. The latter is the orthogonal complement of $K$ with respect to the inner product of operators, $\langle A,B\rangle = {\rm Tr}[A^\dagger B]$, $C= K\oplus  
 O$. $K$ is spanned by all possible products of $n_\alpha$'s, while
 $O$ is spanned by the normally ordered operators 
\begin{equation}
\label{Obasis}
 \mathcal{O}_{\mathcal{I,J}}= \prod_{\beta \in \mathcal{I}} c^\dag_\beta \prod_{\gamma \in \mathcal{J}} c_\gamma, \quad \mathcal{I \neq J},
\end{equation}
where the same ordering "<" as previously is chosen for the indices $\beta,\gamma$.  

At the $n$'th stage of perturbation theory one has to solve the equation 
\bea
\label{PTeq}
[U,\Delta I_\alpha^{(n-1)}]+ [H_0, \Delta I_\alpha^{(n)}]=0. 
\eea
In order for this equation to have a solution one has to make sure that $[U,\Delta I_\alpha^{(n-1)}] \in O$\footnote{Note that it is not obvious from the outset that this simple perturbative scheme should work and produce a local operator. Indeed we construct perturbation theory for an extensive set of operators which are all null eigenvectors of $[H_0,.]$. In principle one should thus use degenerate perturbation theory for all these operators simultaneously, which could turn out to require a non-local change of basis. The further steps below show, however, that this is not the case.}. If this is the case, $\Delta I_\alpha^{(n)}$ is determined up to an element of $K$.  
In \ref{app:binarity} we show how to use this freedom to impose binarity of the spectrum of $I_\alpha$, ${\rm spec}(I_\alpha)=\{0,1\}$, i.e., $I_\alpha^2=I_\alpha$. The latter allows these operators to be interpreted as generalized quasiparticle number operators of the interacting Fermi insulator.

Below we describe the construction of conserved $I_\alpha$ based on a simpler choice, however. In particular, we claim that if our Hamiltonian is time-reversal invariant, and thus has real matrix elements in the basis of  single particle eigenstates, there is a \emph{unique} solution of (\ref{PTeq}) with $\Delta I_\alpha\in O$. This choice implies that the only term in the expansion of $I_\alpha$ that commutes with $H_0$ will be the very first one, $n_\alpha$. 
To prove this at the perturbative level, we have to show that one always finds $[U,\Delta I_\alpha^{(n-1)}] \in O$, or equivalently, that, 
$x(\Psi_0) :=\langle \Psi_0| [U,\Delta I_\alpha^{(n-1)}]|\Psi_0\rangle =0$ for every eigenstate $\Psi_0$ of $H_0$. 
One can easily check that at each stage of perturbation theory $\Delta I_\alpha$ has real coefficients in the occupation number basis (\ref{Obasis}).  Thus $x(\Psi_0)$ is real. On the other hand, from the anti-Hermiticity of $[U,\Delta I_\alpha^{(n-1)}]$ it follows that $x(\Psi_0)$ is purely imaginary, and thus vanishes indeed. 

From the above it follows that we can express the solution of Eq.~(\ref{PTeq}) formally as
\begin{equation}\label{inversee}
 \begin{split}
  \Delta I_\alpha^{(n)}&=i \lim_{\eta \to 0} \int_{0}^{\infty} dt e^{-\eta t} e^{i H_0 t} [U, \Delta\ope_\alpha^{(n-1)}] e^{-i H_0 t},
  \end{split}
 \end{equation}
 which determines the successive terms in perturbation theory recursively.

As we show in \ref{app:binarity}, the recipe to construct a binary operator consists in modifying order by order the terms in the perturbative expansion
 \begin{equation}\label{shift}
  \Delta I_\alpha ^{(n)}  \longrightarrow  \Delta B_\alpha ^{(n)}=\Delta I^{(n)}_\alpha + \Delta K^{(n)}_\alpha,
 \end{equation}
by adding to each $\Delta I_\alpha ^{(n)}$ a diagonal operator  $\Delta K^{(n)}_\alpha \in K$, which is determined  by the previous orders in perturbation theory as:
 \begin{equation}\label{ExpDiag}
  \Delta K^{(n)}_\alpha=\tonde{1-2 n_\alpha} \quadre{\sum_{m=1}^{n-1} \Delta{B}_\alpha ^{(m)} \Delta{B}_\alpha ^{(n-m)} + \grafe{n_\alpha-\frac{1}{2}, \Delta I^{(n)}_\alpha}}.
 \end{equation}
  It is plausible that the convergence for binary operators is essentially the same as for the operators constructed below.

Based on the above perturbative argument, we make the following ansatz for the conserved quantities:
\begin{equation}\label{ansatz}
 I_\alpha= n_\alpha + \sum_{N \geq 1} \sum_{
  \begin{subarray}{l}
\hspace{.4 cm}\mathcal{I \neq J}\\
|\mathcal{I}|=N=|\mathcal{J}|  \end{subarray}} \mathcal{A}^{(\alpha)}_{\mathcal{I,J}} \tonde{\mathcal{O}_{\mathcal{I,J}} + \mathcal{O}^\dag_\mathcal{I,J}},
\end{equation}
where the sets $\mathcal{I,J}$ run over all sets of indices 
 $\grafe{\beta_1< \cdots < \beta_N}$ of single particle states. Linear constraints on the coefficients $\mathcal{A}^{(\alpha)}_{\mathcal{I,J}}$ are found by imposing the conservation condition $\quadre{\ope_\alpha, H}=0$. The coefficients result as $\lambda$-dependent functions of the random energies ${\epsilon_\alpha}$ and of the random matrix elements $U_{\alpha \beta, \gamma \delta}$, which  vanish in the limit $\lambda=0$. Since the resulting operators $I_\alpha$ are functionally independent for $\lambda=0$, we expect the same to hold for any finite $\lambda$ before the delocalization transition. Indeed it is hard to see how a polynomial of $I_{\beta \neq \alpha}$'s could contain only a single diagonal term $n_\alpha$.

It is important to note that the expansion (\ref{ansatz}) should not be seen as an expansion in $\lambda$, but rather as an expansion in the support on which the operators $\mathcal{O}_{\mathcal{I,J}}$ act. A formal expansion in $\lambda$ must always be re-summed locally when rare, but very small denominators are encountered, implying that the naive perturbative series (\ref{PTeq},\ref{inversee}) has vanishing  radius of convergence in $\lambda$~\cite{anderson1958absence}. In \ref{app:resum} we discuss a simple example where such a re-summation is necessary.

We point out that in any finite system the above ansatz, even though motivated by a perturbative consideration, uniquely determines a conserved operator even if perturbation theory does not converge, despite of re-summations. In that case $I_\alpha$ is defined as the finite (possibly exponentially large) sum (\ref{ansatz}) whose coefficients satisfy the linear system of equations~(\ref{fullequation}) below. In a delocalized regime that operator will have support on the whole system.

\subsection{Convergence criterion}

We argue that for sufficiently small $\lambda$ the expansion ($\ref{ansatz}$) converges in the operator norm. The convergence holds in probability, that is, for any $\epsilon>0$: 
\begin{equation}\label{condconvergenceR}
\lim_{R\to\infty}\mathbb{P} \tonde{
\sum_{\begin{subarray}{c}
\mathcal{I \neq J}\\
|\mathcal{I}|=
|\mathcal{J}|\\ r({\cal I, J}) >R \end{subarray}} \modul{\mathcal{A}^{(\alpha)}_{\mathcal{I,J}}}< \epsilon}=1,
\end{equation}
where $r({\cal I, J}) = {\rm max}_{\beta \in {\cal I}\cup {\cal J}} |\vec r_\alpha-\vec r_\beta|$ is the maximal distance between the localization center of the state $\alpha$ and any of the states $\beta$ that are acted upon by the operator $\mathcal{O}_{\mathcal{I},\mathcal{J}}$. $\mathbb{P}$ is the probability measure over the disorder realizations.
This ensures that the series defining the operator $I_\alpha$ converges almost surely, since $||\mathcal{O}_{\mathcal{I},\mathcal{J}}||=1$ for all $\mathcal{I,J}$. 

The resulting operator $I_\alpha$ is quasi-local in the sense defined above. As will become clear below, cf. Sec.~\ref{sec:probresonances}, one can associate a length scale to the support of these operators like for the non-interacting case: truncating the expansion at that length scale yields operators that are conserved up to exponentially small corrections. This scale is essentially the localization length pertaining to the interacting problem.

The many-body delocalization transition is expected to happen at a sharply defined critical  value $\lambda=\lambda_c$ of the interaction strength, at which thermalization and ergodicity are restored. It is natural to expect that this coincides with the delocalization of physically defined conserved quantities, such as the time average of local operators. There is also a sharply defined interaction strength $\lambda=\lambda_c'$  at which our integrals $I_\alpha$ become non-local with probability one. Logically we cannot exclude that $\lambda_c'$ is slightly smaller than $\lambda_c$ (since it might be possible to find a prescription for conserved quantities that leads to more local operators than ours); however, we believe that within the approximations we are making, see Sec.~\ref{sec:fowardapprox}, $\lambda_c$ and $\lambda_c'$ cannot be distinguished. 
We  therefore use the notation $\lambda_c$ indistinctly for both critical values.

To discuss the convergence (\ref{condconvergenceR}), we map the problem of constructing conserved quantities into an equivalent problem of a particle hopping on a disordered lattice whose sites are labeled by the Fock indices ($\mathcal{I,J}$). In particular, the exponential decay of the coefficients of $\ope_\alpha$ corresponds to the localization of the particle on that lattice, in analogy with the non-interacting case ($\ref{expsingle}$). In turn, the delocalization of the particle corresponds to the divergence of the operator expansion ($\ref{ansatz}$).

\section{Explicit construction of the integrals of motion}
\label{sec:construction}

In this section we present the equations defining $\mathcal{A}_{\mathcal{I},\mathcal{J}}$ in (\ref{ansatz}) and discuss how to solve them. To illustrate the procedure, we first solve exactly a non-interacting case and then proceed with the interacting problem.

\subsection{Non-interacting single-particle example}
Consider a non-interacting one-dimensional disordered Hamiltonian:
\begin{equation}\label{handerson}
 H_{\rm And}=\sum_{i} \epsilon_i n_i -t \sum_{i} \tonde{c^\dag_i c_{i+1}+ c^\dag_{i+1} c_{i}},
\end{equation}
where $\epsilon_i$ are random energies and the hopping $t$ is treated perturbatively. In this case, the ansatz
\begin{equation}\label{andans}
 I_{k}= n_k + \sum_{i<j} \mathcal{A}^{(k)}_{ij} \tonde{c^\dag_i c_j + c^\dag_{j} c_i},
\end{equation}
is consistent. Imposing $[H,  I_{k}]=0$, we obtain a set of linear equations for the coefficients $\mathcal{A}^{(k)}_{ij}$, one equation for each index $k$. If for identical indices we define:
\begin{equation}\label{anddiag}
 \mathcal{A}^{(k)}_{ii} \equiv \delta_{k,i}
\end{equation}
then the equations for $\mathcal{A}^{(k)}_{ij}$ with $i \neq j$ can be compactly written as:
\begin{equation}\label{anderson}
 \begin{split}
(\epsilon_{i}-\epsilon_{j})\mathcal{A}^{(k)}_{ij} - t \tonde{\mathcal{A}^{(k)}_{i-1 j}+\mathcal{A}^{(k)}_{i+1 j}-\mathcal{A}^{(k)}_{ij-1} -\mathcal{A}^{(k)}_{ij+1}} =0.
 \end{split}
\end{equation}

In view of these equations, one may re-interpret $\mathcal{A}^{(k)}_{ij}$ as the wave-function amplitudes of a particle on a square lattice with sites $(i,j)$, and correlated on-site disorder $\mathcal{E}_{i,j}=\epsilon_{i}-\epsilon_j$, subject to the constraint ($\ref{anddiag}$). An explicit expression for them can be given in terms of the eigenfunctions $\phi_{\alpha}$ of the Anderson problem ($\ref{handerson}$) as:
\begin{equation}\label{sol}
 \mathcal{A}^{(k)}_{ij}=\sum_{\alpha} \omega_{\alpha}^k \phi_{\alpha}(i) \phi_{\alpha}(j),
\end{equation}
where the $\omega_{\alpha}^k$ have to be determined from the constraint
\begin{equation}\label{omegacond}
  \sum_{\alpha} \omega_\alpha^k [\phi_{\alpha}(i)]^2= \delta_{k,i}.
\end{equation}

The exponential decay of the amplitudes ($\ref{sol}$) in the distance between the sites $i,j$ follows from the localization in space of the eigenstates $\phi_{\alpha}$. It implies the convergence of the expansion ($\ref{andans}$). Therefore, the operators $I_k$ are quasi-local conserved operators, similarly to the particle number operators $n_\alpha$ in ($\ref{expsingle}$). 

However, note that these two sets of operators differ, in particular ($\ref{andans}$) does not contain any diagonal terms ($i=j \neq k$). Using (\ref{sol}), (\ref{omegacond}) one can also explicitly check that the operators ($\ref{andans}$) do not coincide with the time average of the operators $n_k(t)$.

\subsection{Interacting case}

We now return to the interacting case. Since the operators $I_\alpha$ will contain strings of $c^\dag$'s and $c$'s of arbitrary length, we need a way to deal with large index sets. We introduce the following notation: for any index set $\mathcal{X}=(x_1 \cdots x_N)$, we define  diagonal coefficients as zero, except if $\mathcal{X} = \{\alpha\}$:
\begin{equation}\label{diagamp}
\mathcal{{A}}^{(\alpha)}_{\mathcal{X}, \mathcal{X}} \equiv \delta_{\mathcal{X},\{\alpha\}} . 
\end{equation}
 Moreover, for any $l, m$ (with $l<m$) and any single particle labels $\gamma,  \delta$ (with $\gamma<\delta$), define the index sets:
\begin{equation}
 \begin{split}
  \mathcal{X}_{l }&\equiv (x_1 \cdots  \cancel{x_l}  \cdots x_N),\\
   {\mathcal{X}}_{l m}^{ \gamma}&\equiv (\gamma \hspace{.1 cm} x_1 \cdots  \cancel{x_l}  \cdots  \cancel{x_m} \cdots x_N  ),\\
 {\mathcal{X}}_{l m}^{ \gamma \delta}&\equiv (\gamma \hspace{.1 cm} \delta \hspace{.1 cm} x_1 \cdots  \cancel{x_l}  \cdots  \cancel{x_m} \cdots x_N ).
 \end{split}
\end{equation}
In general, the set $\mathcal{{X}}_{\cdots}^{\cdots}$ is obtained from $\mathcal{X}$ by eliminating the indices in the subscript and appending the ones in the superscript on the left. Note that the resulting sets are thus not ordered. Let $\sigma \quadre{\cdot}$ denote the sign of the permutation which orders the set, and define:
\begin{equation}
 \begin{split}
 s\quadre{ \mathcal{X}_{l }}&\equiv l,\\
   s \quadre{{\mathcal{X}}_{l m}^{ \gamma}}&\equiv l+m + \sigma \quadre{{\mathcal{X}}_{l m}^{ \gamma}} ,\\
s \quadre{ \mathcal{X}_{l m}^{ \gamma \delta}}&\equiv l+m +\sigma \quadre{{\mathcal{X}}_{l m}^{ \gamma \delta}} .
 \end{split}
\end{equation}
Finally, for index sets with $|\mathcal{Y}|=|\mathcal{Z}|$, define the modified amplitudes:
\begin{equation}
 \mathcal{\tilde{A}}^{(\alpha)}_{\mathcal{Y}, \mathcal{Z}} \equiv (-1)^{s \quadre{\mathcal{Y}}+s \quadre{\mathcal{Z}}} \mathcal{{A}}^{(\alpha)}_{\mathcal{Y}, \mathcal{Z}}\,.
\end{equation}

With this notation, the condition $\quadre{H, I_\alpha}=0$ is equivalent to the following set of linear equations for $\mathcal{A}^{(\alpha)}_{\mathcal{I,J}}$:

\begin{equation}\label{fullequation}
 \begin{split}
 0=&\left(\sum_{n=1}^{N}  \frac{\epsilon_{\alpha_n}- \epsilon_{\beta_n}}{\delta_\xi} \right)\mathcal{A}^{(\alpha)}_{\mathcal{I}, \mathcal{J}} +\\
    +&\lambda \sum_{ \begin{subarray}{l}
l,m=1\\
\hspace{.1 cm}l<m  \end{subarray}}^{N} \quadre{\sum_{\gamma < \delta} \tonde{ \eta_{\alpha_l \alpha_m, \gamma \delta} \mathcal{\tilde{A}}^{(\alpha)}_{\mathcal{{I}}_{ l m}^{ \gamma \delta}, \mathcal{J}}- \eta_{ \gamma \delta, \beta_l \beta_m} \mathcal{\tilde{A}}^{(\alpha)}_{\mathcal{I}, \mathcal{{J}}_{ lm}^{\gamma \delta} }}}+\\
+& \lambda  \sum_{ \begin{subarray}{l}
l,m=1\\
\hspace{.1 cm}l<m  \end{subarray}}^{N} \sum_{n=1}^N (-1)^{N+1}  \quadre{\sum_{\gamma}   \tonde{ \eta_{\alpha_l \alpha_m, \gamma \beta_n } \mathcal{\tilde{A}}^{(\alpha)}_{\mathcal{{I}}_{lm} ^{\gamma}, \mathcal{{J}}_{n}} -
\eta_{\gamma \alpha_n , \beta_l \beta_m}  \mathcal{\tilde{A}}^{(\alpha)}_{\mathcal{{I}}_{n}, \mathcal{{J}}_{ l m}^{ \gamma }}}},
 \end{split}
\end{equation}
where $(\mathcal{I,J})= (\alpha_1 \cdots \alpha_N, \beta_1 \cdots \beta_N)$ and $\mathcal{I} \neq \mathcal{J}$. The diagonal coefficients appearing on the right-hand side are defined in (\ref{diagamp}).

\subsubsection{Topology of the operator lattice}

\begin{figure}[!ht]
  \captionsetup[subfigure]{labelformat=empty}
  \centering
  \subfloat[\hspace{2.5 cm}(a)]{\raisebox{.13 cm}{\includegraphics[width=.43\textwidth]{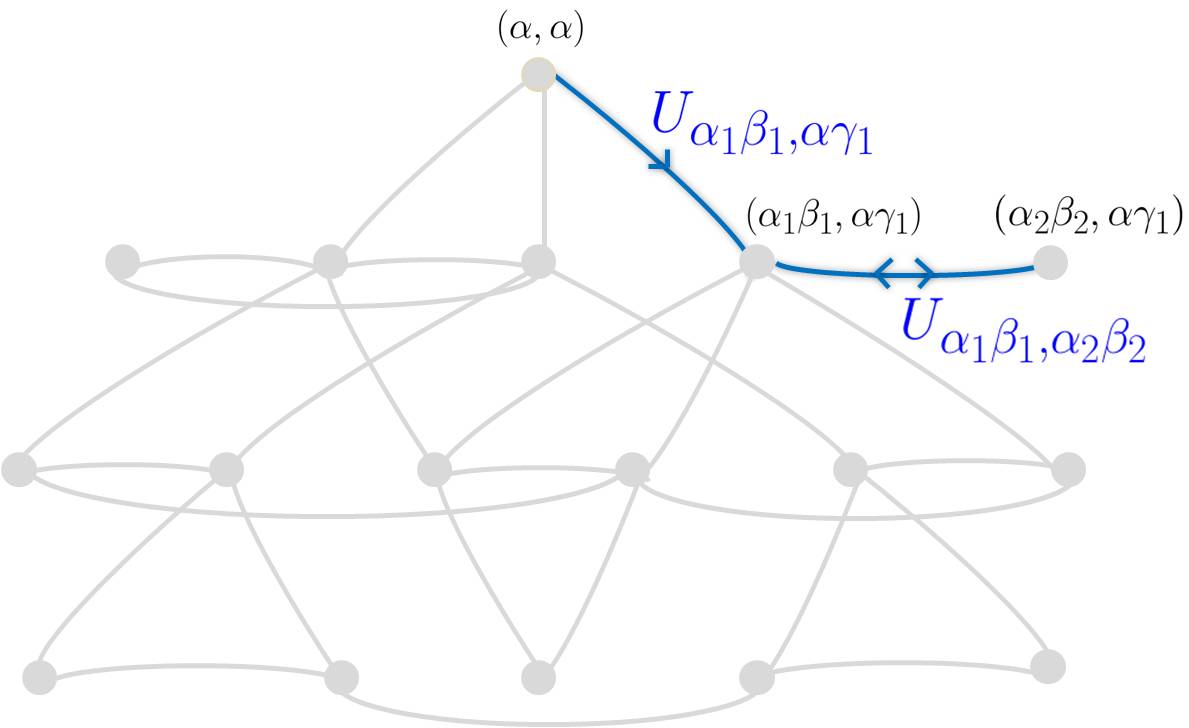}}} \quad
   \subfloat[]{\includegraphics[width=.08\textwidth]{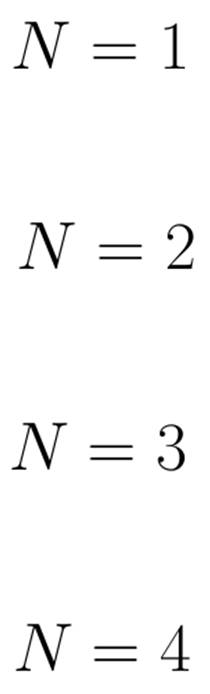}}\quad
  \subfloat[\hspace{2.5 cm}(b)]{\raisebox{.1 cm}{\includegraphics[width=.4\textwidth]{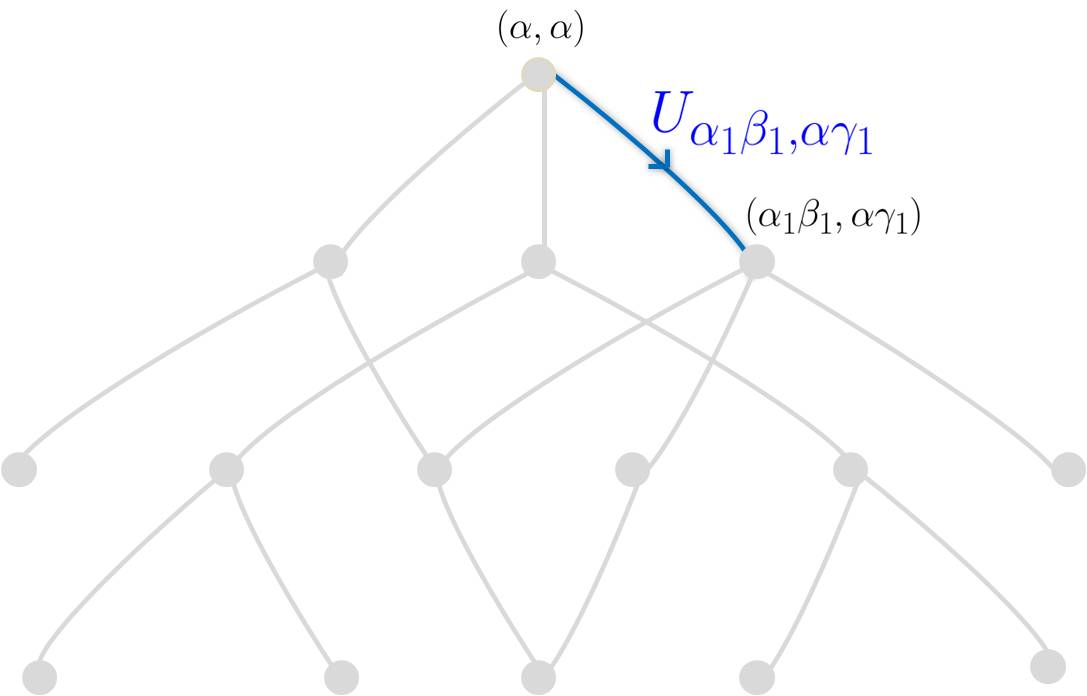}}}
  \caption{Structure of the operator lattice before (a) and after (b) making the forward approximation. Vertices correspond to Fock indices ($\mathcal{I,J}$); links are drawn between index pairs, which are connected by the interaction $U$, that is, if the pairs appear simultaneously in at least one of the Eqs.~(\ref{fullequation}). }
  \label{fig:lattice}
\end{figure}

Similarly as in the previous single-particle example, Eq.~($\ref{fullequation}$) can be thought of as a  hopping problem for a single particle on a lattice with sites given by the Fock indices ($\mathcal{I,J}$) and local, correlated disorder  $\mathcal{E}_{\mathcal{I,J}}=\sum_{n=1}^{N}  ({\epsilon_{\alpha_n}- \epsilon_{\beta_n}})$. The hopping is provided by the interaction $U$, see Fig.~$\ref{fig:lattice}$a. The non-interacting limit  corresponds to the wave-function $\mathcal{A}^{(\alpha)}$ being completely localized on the site $({\cal I, J}) = (\alpha, \alpha)$. 

The lattice topology, as determined by the interactions, is rather complicated. However, Eqs.~($\ref{fullequation}$) have a clear hierarchical structure: the equation for index sets $\cal I,J$ of length $N$ are coupled only to amplitudes with index sets of equal or shorter length. Therefore, the sites can be organized into generations, according to the length of their index sets. Hopping is possible only within the same generation (second term in equation $(\ref{fullequation})$) or between consecutive ones (third term in equation $(\ref{fullequation})$). In the latter case, the hopping is unidirectional, and thus {the hopping problem is non-Hermitian}.

The connectivity of the lattice is determined by the restrictions in energy, Eq.~($\ref{restriction}$), and space (particles need to be in the same or in an adjacent localization volume) of the matrix elements $U_{\alpha \beta, \gamma \delta}$.  Hoppings from a site  $(\mathcal{I,J})$ in generation $N$ to a site $(\mathcal{I',J'})$ in generation $N+1$ requires a particle (or hole) in a state $\alpha$ to scatter to the closest energy level $\gamma$ above or below $\alpha$, while another particle-hole pair of adjacent levels $(\beta,\delta)$ is created. The particle $\beta$ can be chosen in $N_{\text{loc}}$ ways with $N_{\text{loc}}$ given in ($\ref{nloc}$), and there are two choices for $\gamma$ and $\delta$, respectively. Therefore, the number of Fock states ($\mathcal{I',J'}$) accessible from ($\mathcal{I,J}$) via the decay of a given quasiparticle $\alpha$ is: 
 \begin{equation}\label{connectivity}
 \begin{split}
  \mathcal{K}=4 \frac{W}{\delta_\xi}=4 N_{\text{loc}}.
  \end{split}
 \end{equation}
In contrast, hoppings from  $(\mathcal{I,J})$ to a site of the same generation correspond to processes where each member of a pair of particles (or holes) scatter to one of the two closest energy levels: there are $4$ possible final states to which a given pair can decay.

At this point we emphasize that we are not restricting ourselves to a specific many-body state or energy sector. Thus no assumption about the occupation of the levels or about the position of the Fermi level $E_F$ is made. This gives the largest possible connectivity $\mathcal{K}$. It will be reduced to an effective connectivity once we consider the restriction of the integrals $I_a$ to subspaces of a definite energy by means of a projector over many-body states, $\tilde I_a=PI_aP$, where
\begin{equation}
P=\sum_{E_a\in [E-\delta E/2,E+\delta E/2]}|E_a\rangle\langle E_a|.
\end{equation}
This projection will alter the connectivity $\mathcal{K}$, so as to reflect the higher probability for some processes to be Fermi-blocked, since the involved levels might already be occupied. This yields an effective connectivity $\mathcal{K}_{\rm eff}$, whose typical value depends both on the average energy density of the states $E_a$ and the average filling fraction of the band. It is not difficult to see that if we use typical values for occupation numbers as given by the Fermi distribution (without assuming the underlying states to be thermal), repeating the above considerations at finite temperature $T\ll E_F$ we obtain $\mathcal{K}_{\rm eff}\sim T/\delta_\xi$, in analogy to the analysis in \cite{Basko:2006hh}.

\section{The forward approximation}
\label{sec:fowardapprox}

\subsection{Simplifications due to large connectivity, $\xi\gg a$}
The requirement of convergence of the operator expansion, Eq. ($\ref{condconvergenceR}$), can be interpreted as a localization condition for the hopping problem on the disordered lattice of Fock indices. In order to investigate under which conditions localization occurs,  we introduce the main approximation of this work: we neglect the second term of the equation ($\ref{fullequation}$), that accounts for the hopping between sites in the same generation.

 This approximation is motivated by the following consideration, assuming that the number of single particle levels per localization volume, and thus $\cal K$, is large: for operator sites with a density of Fock indices per localization volume much smaller than the maximally possible $\sim \cal K / \xi^d$,  
  the connectivity within the same generation is much smaller than the connectivity $\mathcal{K}$ among sites in different generations ($\ref{connectivity}$).   
  Note, however, that transitions from a given state $(\mathcal{I,J})$ due to the second term of Eq.~(\ref{fullequation}) can involve any pair of particles or holes in the same localization volume. Therefore, for operators with a high density of indices per localization volume those transitions are as numerous as the third class of terms in Eq.~(\ref{fullequation}). Our approximation of dropping the second term is therefore not fully controlled at sufficiently high orders in perturbation theory where operators with a high density of indices per localization volume appear.  We postpone further discussions of the subtleties related to this approximation to Sec.~\ref{sec:finitetemp}.

Once the second term in ($\ref{fullequation}$) is dropped, the equations reduce to recursive equations for increasing generations, with the initial condition $\mathcal{A}_{\alpha_1,\beta_1}^{(\alpha)}=\delta_{\alpha_1,\beta_1}\delta_{\alpha_1, \alpha}$.  However, only some of the amplitudes $\mathcal{A}^{(\alpha)}_{\mathcal{I,J}}$ in  ($\ref{ansatz}$) are determined through the recursion, while we approximate all other amplitudes to be zero: in generation $N$, the non-zero amplitudes correspond to sites ($\mathcal{I,J}$) that can be reached from ($\alpha, \alpha$) via \emph{directed} paths of length $N-1$. Retaining only these sites simplifies the structure of the lattice of Fock indices very substantially, see Fig.~$\ref{fig:lattice}$b.

The amplitudes on these sites ($\mathcal{I,J}$) can be written as the sum over all directed paths that connect them to the root ($\alpha, \alpha$) in Fig.~$\ref{fig:lattice}$b:
\begin{equation}\label{ampath}
 \mathcal{A}^{(\alpha)}_{\mathcal{I}, \mathcal{J}}=\sum_{
 \begin{subarray}{c}
\text{directed paths:}\\
\mathcal{(\alpha,\alpha)}\to \mathcal{(I,J)}
 \end{subarray}} \omega_{\text{path}}.
\end{equation}
The path weights $\omega_{\text{path}}$ are of  the form: 
\begin{equation}\label{pathweight}
{\omega_{\text{path}}} \equiv (-1)^{\sigma_{\text{path}}} \prod_{i=1}^{N-1} {\frac{ \lambda  {\eta_{\alpha_i \beta_i, \gamma_i \delta_i}}\delta_\xi}{\sum_{k=1}^i \mathcal{E}_{\alpha_i \beta_i, \gamma_i \delta_i}}}. 
\end{equation}
in close analogy to forward approximations in single particle problems~\cite{anderson1958absence, nguyen1985tunnel, nguyen1985aharonov,Kardarbook, muller2013magnetoresistance}.

The factor $(-1)^{\sigma_{\text{path}}}$ takes into account the global fermionic sign associated with the path, arising from the sign factors in Eq.~(\ref{fullequation}). However, we will see below that these signs are immaterial at the level of our approximation.

Note that the resulting expression for $\mathcal{A}^{(\alpha)}_{\mathcal{I}, \mathcal{J}}$ is of order $\lambda^{N-1}$, that is, the lowest possible order in $\lambda$ for amplitudes of operators involving $2 N$ particle-hole indices. Indeed, at least $N-1$ interactions are needed to create the corresponding excitations.

\subsection{Probability of resonances on the operator lattice}
\label{sec:probresonances}
Let us discuss the configuration in real space of the indices $(\mathcal{I,J})$ with $\modul{\mathcal{I}}=N$, which are retained within the forward approximation, cf. Fig.~$\ref{fig:lattice}$b. Since the amplitudes  $\mathcal{A}^{(\alpha)}_{\mathcal{I,J}}$ are of order $\lambda^{N-1}$ and the interaction is local, the indices satisfy $r(\mathcal{I,J}) \leq N \xi$: amplitudes involving single particle states sufficiently far away from the localization center $\alpha$ must belong to sufficiently high generations. Within the approximations made, the convergence criterion ($\ref{condconvergenceR}$) can then be restated in terms of the generation number $N$ as:

\begin{equation}\label{pres0}
\lim_{N^* \to \infty} \mathbb{P}\tonde{ \sum_{N > N^*} \sum_{ \begin{subarray}{l}
\hspace{.4 cm}\mathcal{I \neq J}\\
|\mathcal{I}|=N=|\mathcal{J}|  \end{subarray}} \modul{A^{(\alpha)}_{\mathcal{I}, \mathcal{J}}} < \epsilon}=1
\end{equation}
for arbitrary $\epsilon>0$.

A sufficient condition for Eq.~$(\ref{pres0})$ to hold is that for some $z<1$ and for $N^*$ sufficiently big:
\begin{equation}\label{pres}
 \mathbb{P}\tonde{ \forall N>N^*, \sum_{ \begin{subarray}{l}
\hspace{.4 cm}\mathcal{I \neq J}\\
|\mathcal{I}|=N=|\mathcal{J}|  \end{subarray}} \modul{A^{(\alpha)}_{\mathcal{I}, \mathcal{J}}} < z^{N-1}}=1-\zeta(N^*)
\end{equation}
with
\begin{equation}\label{pres2}
 \lim_{N^* \to \infty} \zeta(N^*)=0.
\end{equation}

The left hand side of Eq.~($\ref{pres}$) can be interpreted as the probability that no resonance\footnote{A resonance is said to occur at a site ($\mathcal{I,J}$) if $\mathcal{A}_\mathcal{I,J}$ is comparable with the amplitude at the origin $(\alpha, \alpha)$, i.e., if it is of order $O(1)$.} occurs at large distance from the unperturbed localization center ($\alpha, \alpha$). Whenever it holds, it implies the quasi-locality of the operators $I_\alpha$  within the forward approximation: indeed, Eq.~(\ref{pres}) implies that the first appearance of operators $c_\beta, c^\dag_\beta$'s in $I_\alpha$, with $|\vec{r}_\beta-\vec{r}_\alpha|\approx N\xi$ and $N \gg 1$ is with high probability exponentially small in $N$.

 In the following we will show that Eq.~($\ref{pres}$) holds in a regime of small couplings $\lambda$; the critical value $\lambda_c$ at which ($\ref{pres}$) ceases to hold gives an estimate for the radius of convergence of the operator series, and thus for the boundary of the many-body localized phase.
 
 \subsection{Similarities and differences with localization problems on trees}

The similarity to a one-particle problem allows us to revisit  analogies and differences between many-body localization and single particle problems on lattices which have some features of a Cayley tree \cite{altshuler1997quasiparticle} (see also \cite{de2013support,de2014anderson} and references therein). Indeed, in the simplified lattice of Fig.~$\ref{fig:lattice}$b, the number of sites at distance $N$ from the localization center $(\alpha, \alpha)$ grows as $\mathcal{K}^N$ with $\mathcal{K}$ given in ($\ref{connectivity}$). This exponential growth is analogous to the growth on trees and other hierarchical lattices, see e.g. \cite{laumann2014many}.
However, we caution the reader that, despite superficial similarities, the calculation we will perform does not reduce to studying an equivalent single particle problem on a Cayley tree as in \cite{abou1973selfconsistent}. Indeed, in the latter problem there is a unique path leading from the root to a given site and thus there are no loops. In contrast, in the operator lattice, there are typically exponentially many diagrams (or effective paths) leading to a given site, and thus plenty of loops, similarly as in finite dimensional lattices.
Nevertheless, it is usually the case that among those many paths only very few dominate the sum over all paths - an observation we will heavily rely on in the sequel.   

Our present problem also differs from the study of the decay of excitations in a zero-dimensional quantum dot, as considered in \cite{altshuler1997quasiparticle}. There, no genuine delocalization can take place due to the finite available phase space. Instead, 
it is essential that our operator expansion leave the localization volume of the initial state $\alpha$, for delocalization to be possible beyond a critical interaction strength $\lambda_c$.

 \subsection{Connection with many-body diagrammatic perturbation theory}
\label{sec:diagrep} 
\begin{figure}[ht!]
  \captionsetup[subfigure]{labelformat=empty}
  \centering
  \subfloat[]{%
                
  \raisebox{.3 cm}{ \includegraphics[width=0.5\textwidth]{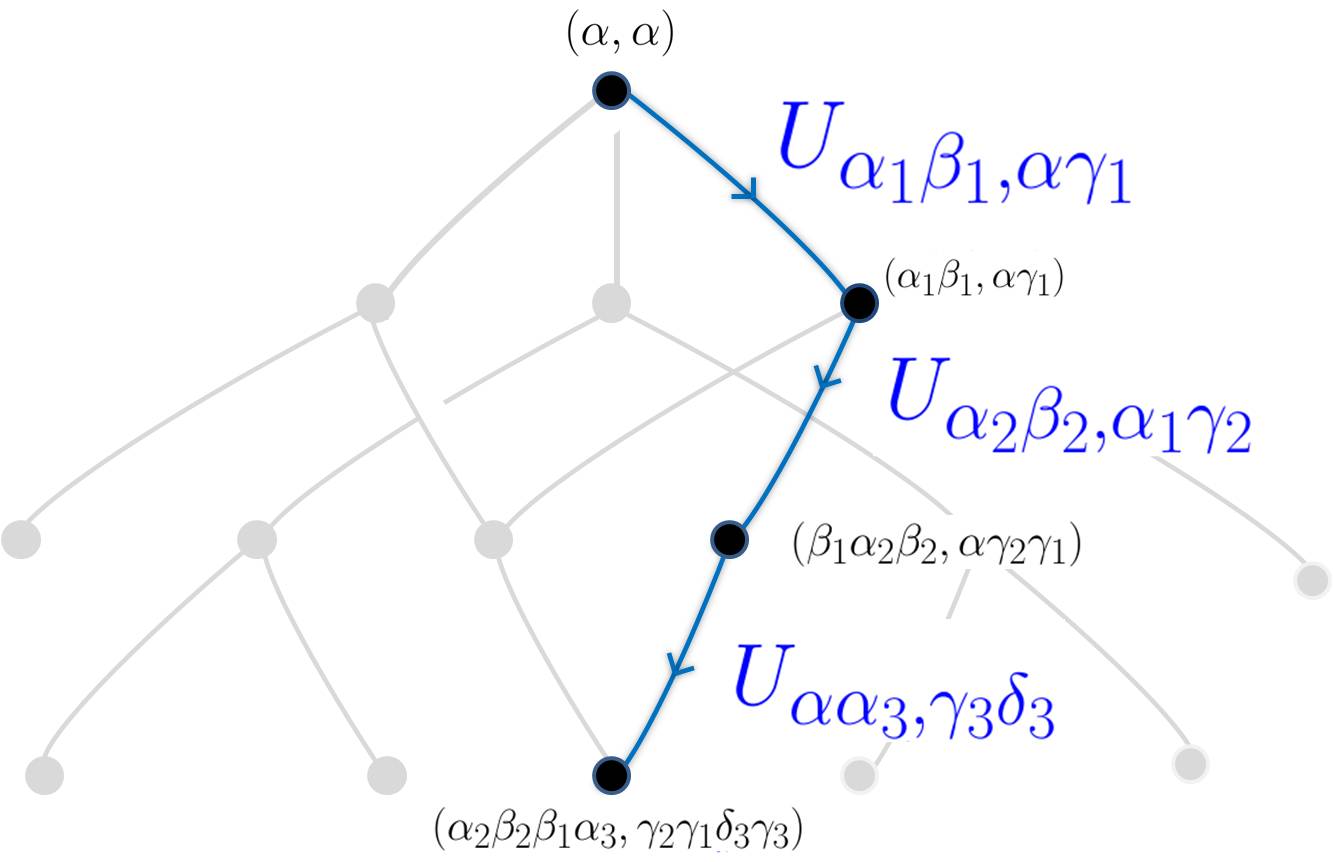}
        }}%
        \qquad
        \subfloat[]{%
\includegraphics[width=0.47\textwidth]{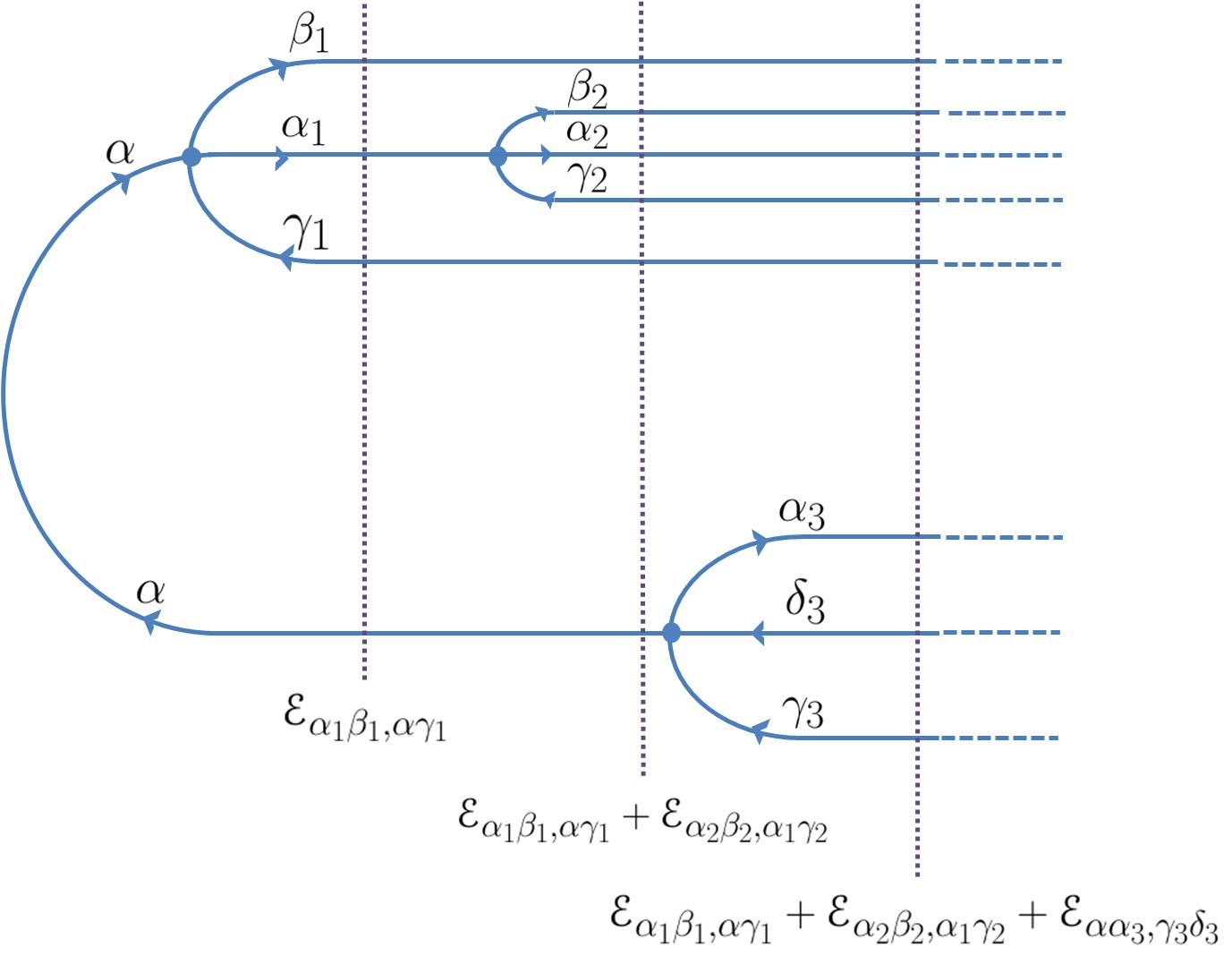}
        } 
    \caption{%
      Directed path in the operator lattice and associated ordered scattering graph. The sites ($\mathcal{I,J}$) along the path correspond to the intermediate states of the graph, indicated by dashed lines.  Hoppings on the lattice correspond to vertices  $U_{\alpha_1 \alpha_2, \beta_1 \beta_2}$ in the graph.  The energy $\mathcal{E}_{\mathcal{I,J}}$ of an intermediate state is the sum of the energy differences $\mathcal{E}_{\alpha_1 \alpha_2, \beta_1 \beta_2}= \epsilon_{\alpha_1}+\epsilon_{\alpha_2}-\epsilon_{\beta_1} -\epsilon_{\beta_2}$ associated with all preceding scatterings. The three  excitations emanating from a vertex are associated to the outgoing legs as follows: the excitation with energy level adjacent to the incoming one is associated with the central leg. The upper and lower leg correspond to the particle and the hole, respectively, of the additionally created pair. The condition ($\ref{restriction}$) requires them to have an energy difference of the order of $\delta_\xi$.
           }%
   \label{fig:IntroDiagrams}
\end{figure}

Insight into the meaning of the forward approximation at the level of the many-body system is given by a diagrammatic representation of the paths, as shown in Fig.~$\ref{fig:IntroDiagrams}$.

To any path of length $N$ in the operator lattice we uniquely associate an ordered graph with $N$ vertices. These graphs have two main branches representing the decay of the operators $c_\alpha$ and $c^\dag_\alpha$ of the initial operator $n_\alpha$. Directed paths of length $N$ on the lattice translate into graphs having the geometry of a tree, with a root and $N$ nodes corresponding to the creation of particle-hole pairs. 
The intermediate states of the graph correspond to the sites $(\mathcal{I,J})$ along the path in the operator lattice, their energy being $\mathcal{E}_{\mathcal{I,J}}$. Note that the order of the sites along the path fixes the order of the interaction vertices in the graph.

\begin{figure}[ht!]
     \begin{center}
\captionsetup[subfigure]{labelformat=empty}
  \centering
  \subfloat[]{%
          
           \includegraphics[width=0.6\textwidth]{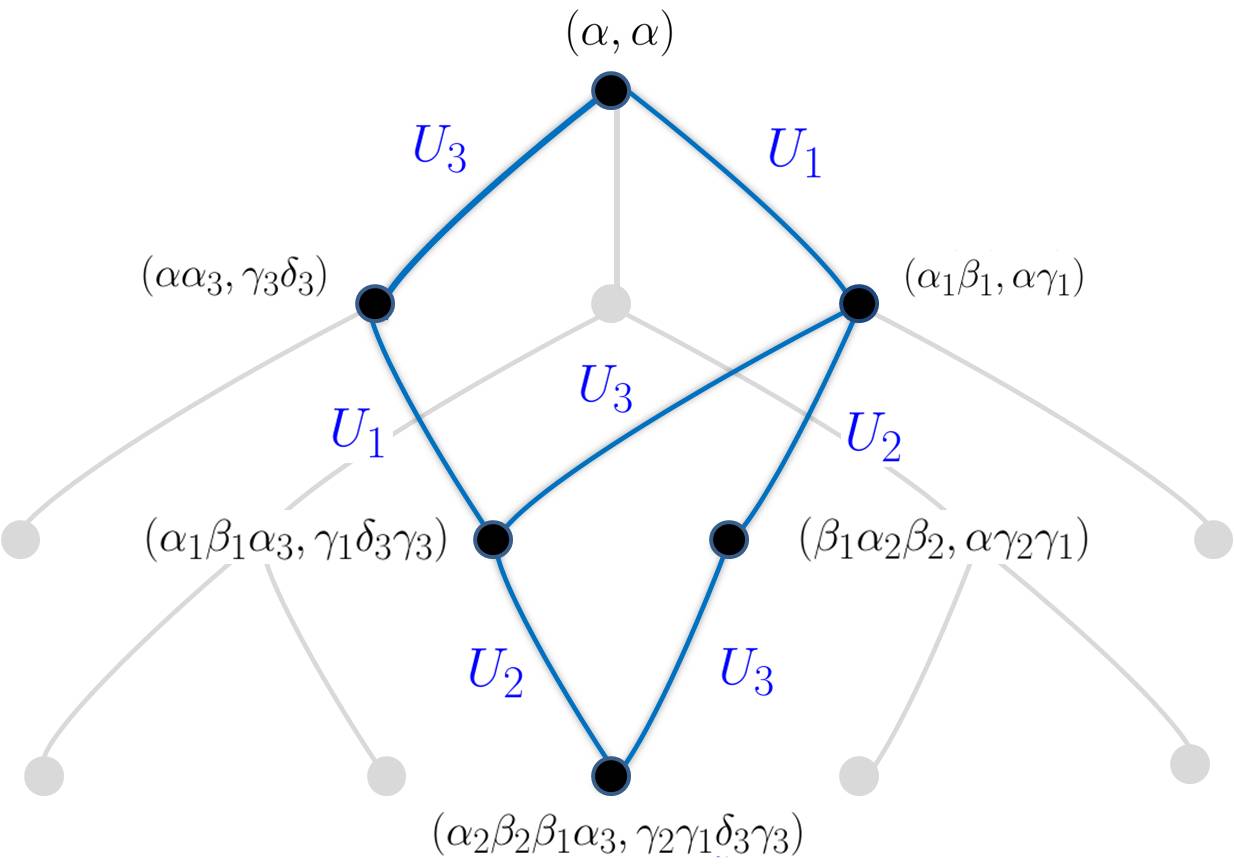}
        } \\
            \subfloat[ \hspace{2 cm} (a)]{%
           \includegraphics[width=0.3\textwidth]{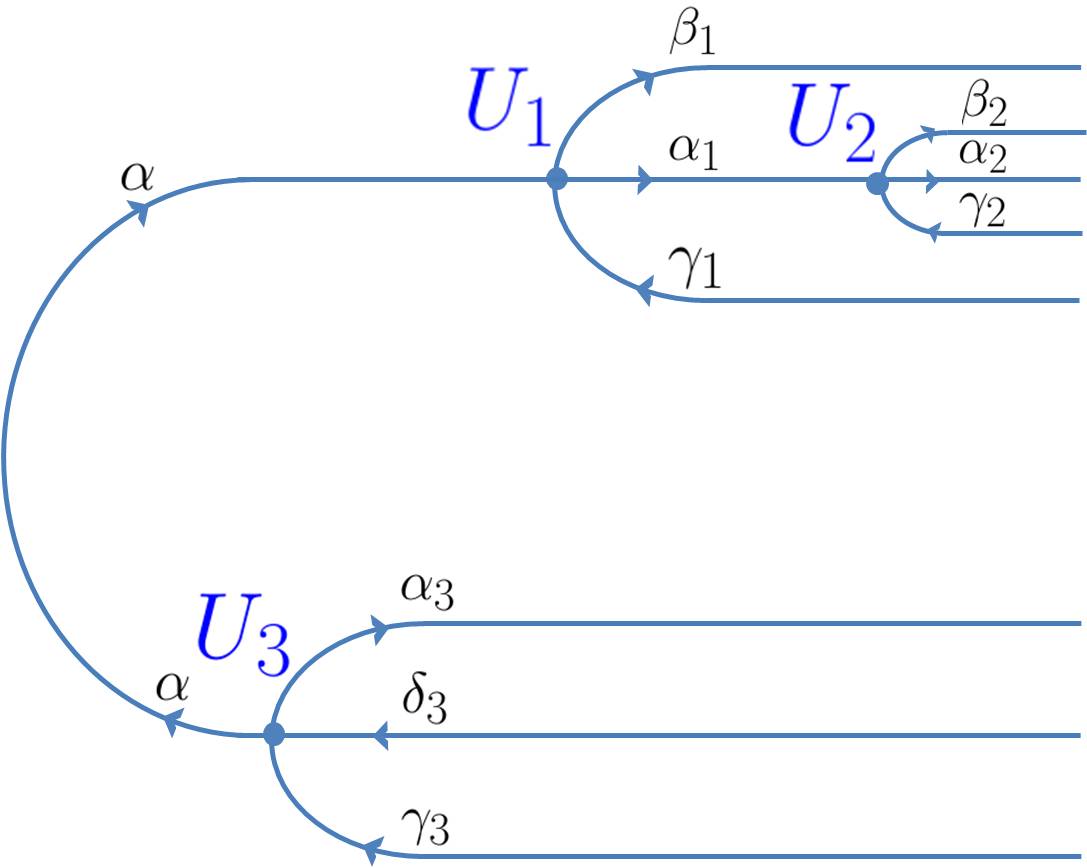}
        } 
        \subfloat[\hspace{2 cm} (b)]{%
            \includegraphics[width=0.3\textwidth]{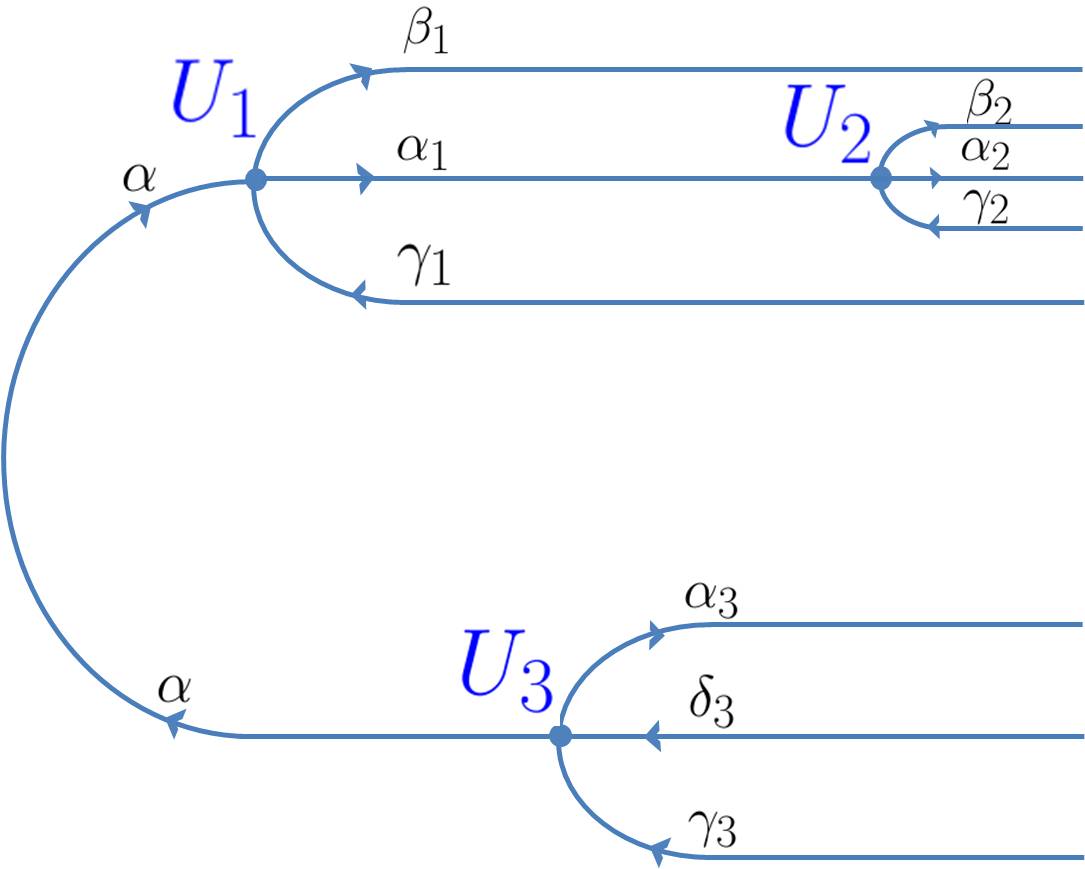}
        }%
        \subfloat[\hspace{2 cm} (c)]{%
            \includegraphics[width=0.3\textwidth]{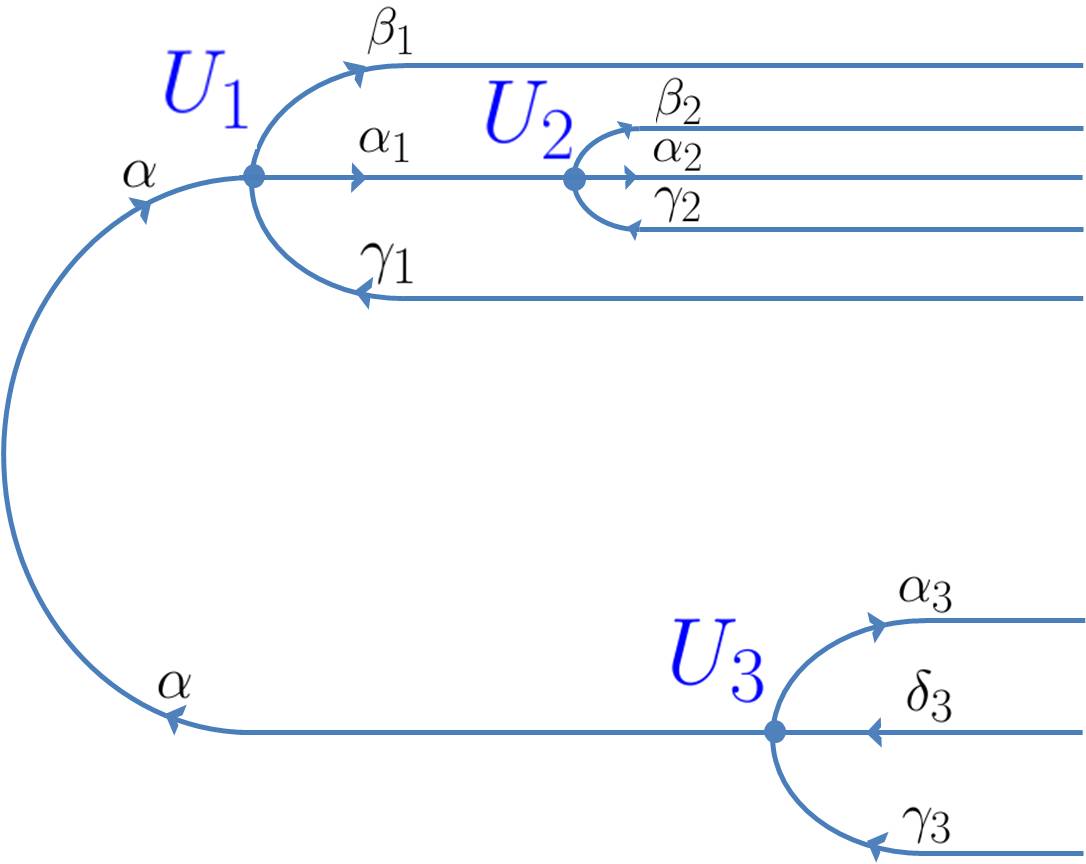}
        }%
    \end{center}
    \caption{%
        Loops in the many-body lattice corresponding to different processes with the same final state, and the corresponding ordered graphs. The graphs differ only in the order in which the interactions $U_1$, $U_2$, $U_3$ act. The weights of such paths are strongly correlated: they are all proportional to the same product of matrix elements, $U_1 U_2 U_3$, and have highly correlated denominators. The sum over all these ordered graphs constitutes a {\em diagram}.
     } %
     \label{fig:Paths2}
  \end{figure}

Such graphs can be grouped into {\em diagrams}: members of the same diagram only differ in the ordering of vertices, while sharing the same geometry and labeling of the legs; they are obviously highly correlated among each other. An example is shown in Fig.~$\ref{fig:Paths2}$, where all three paths connect the state ($\mathcal{I,J})=(\alpha_2 \beta_2 \beta_1 \alpha_3, \gamma_2 \gamma_1 \delta_3 \gamma_3$) to the root ($\alpha, \alpha$), and involve the same interaction matrix elements.  

Such correlated paths exist for all diagrams with {\em branchings} (i.e., vertices where more than one of the outgoing excitations undergo further scattering). The order of the subsequent interactions on different branches can be permuted. This corresponds to different paths on the lattice and different ordered graphs, respectively. 

Obviously we should sum over all possible vertex order permutations of branched diagrams with fixed geometry and labeling of legs.

\subsubsection{Singly branched diagrams}
Consider the sum of the energy denominators\footnote{The global sign of amplitudes of tree-like diagrams without loops does not depend on the order in which the interactions act. This is because the associated four-fermion interaction terms mutually commute, which implies that the signs arising from eventually bringing the operators into the normal order are the same for all vertex orders.} of the three path weights in the example of Fig.~$\ref{fig:Paths2}$. It is immediate to check that the following holds:
\begin{equation}\label{totalfac}
\begin{split}
 \Sigma&\equiv \frac{1}{\mathcal{E}_1(\mathcal{E}_1+\mathcal{E}_2)(\mathcal{E}_1+\mathcal{E}_2+\mathcal{E}_3)}+ \frac{1}{\mathcal{E}_1(\mathcal{E}_1+\mathcal{E}_3)(\mathcal{E}_1+\mathcal{E}_2+\mathcal{E}_3)}\\
 &+\frac{1}{\mathcal{E}_3(\mathcal{E}_3+\mathcal{E}_1)(\mathcal{E}_3+\mathcal{E}_1+\mathcal{E}_2)} = \frac{1}{\mathcal{E}_3} \frac{1}{\mathcal{E}_1(\mathcal{E}_1+\mathcal{E}_2)},
\end{split}
\end{equation}
where $\mathcal{E}_i$ is the energy difference between out- and in-going states at the vertex $i$. Thus, the sum over the three paths weights in Fig.~$\ref{fig:Paths2}$ can be written as a single term $\tilde{\omega}_\Gamma$:
\begin{equation}\label{effpath1}
 \frac{\tilde{\omega}_\Gamma}{(\lambda \delta_\xi)^3} \equiv \frac{\eta_3}{\mathcal{E}_3} \frac{\eta_1 \eta_2}{\mathcal{E}_1 (\mathcal{E}_1+\mathcal{E}_2)},
\end{equation}
where $\eta_i$ is the random variable associated the vertex $i$. More precisely, $\tilde{\omega}_\Gamma$ is the product of two weights of the form ($\ref{pathweight}$), describing the independent decay of the particle $c^\dag_\alpha$ and the hole $c_\alpha$, respectively. It can  easily be checked by induction that this factorization generalizes to an arbitrary number of interactions in such singly branched diagrams: for any of them, a weight of the form (\ref{effpath1}) is obtained by summing over all the path weights. We refer to $\tilde{\omega}_\Gamma$ as the weight of the \emph{effective path} associated to the diagram, and denote the latter by $\Gamma$.   

 \begin{figure}[ht!]
     \centering
  \centering
   
   \subfloat[\hspace{3.5 cm}]{%
           \label{Single}
           \includegraphics[width=0.55\textwidth]{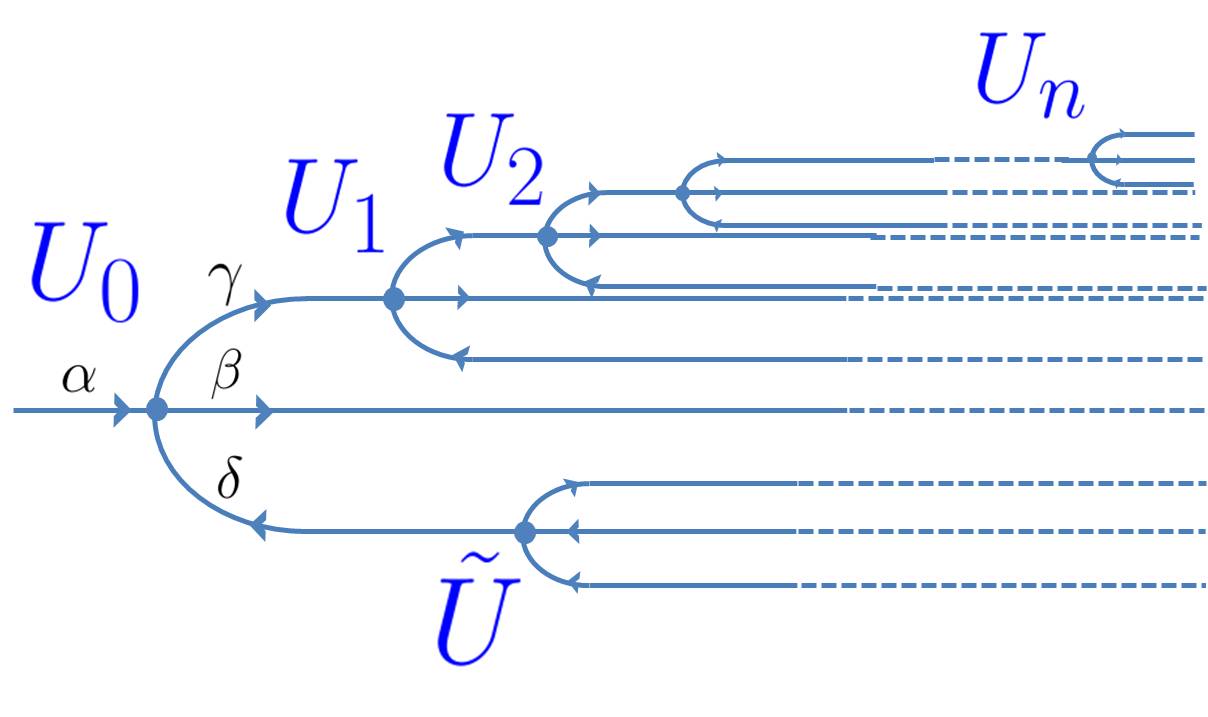}}
   \hspace{.8 cm}
   \subfloat[\hspace{2 cm}]{%
   \includegraphics[width=.3\textwidth, angle=0]{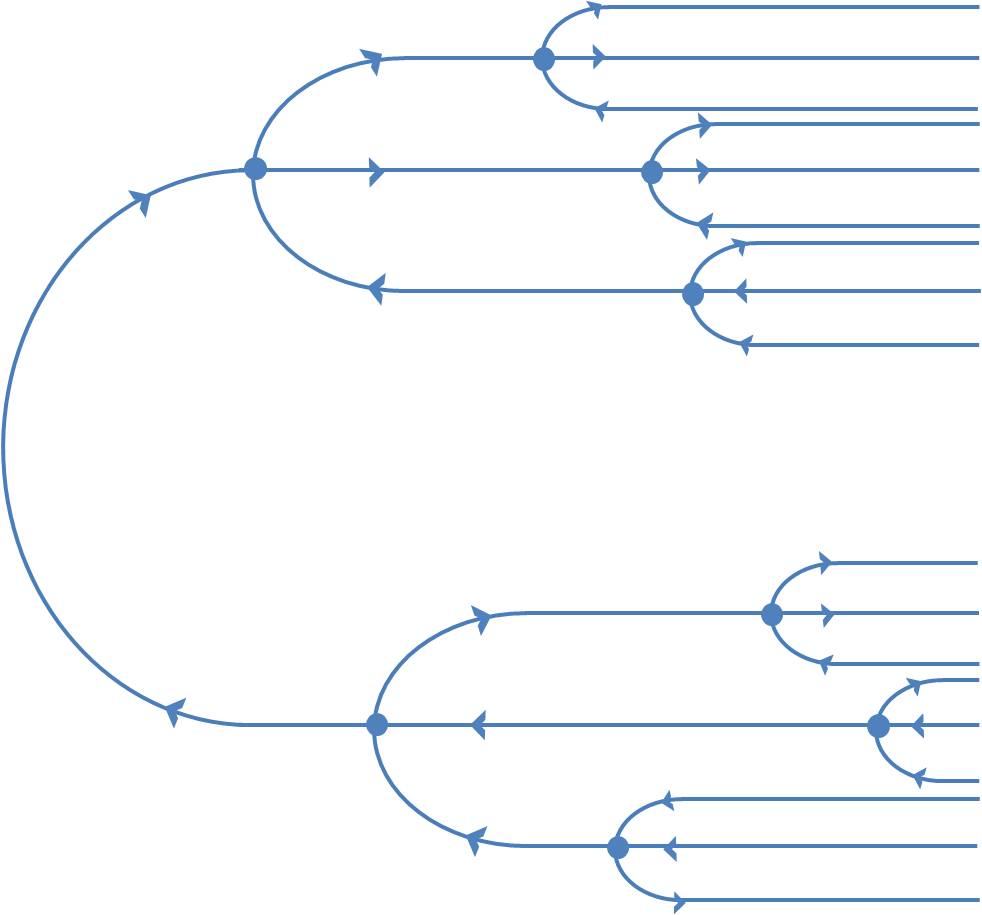}
\label{MaxBranch}
                        
        } %

  \caption{(a) Branched decay of a single particle.  (b) Geometry of a diagram with the maximal possible number of branchings for a fixed number of interactions.}
     \label{fig:MaxBranch}
  \end{figure}

\subsubsection{Multiply branched diagrams}
Let us now discuss further branchings in the sub-diagrams describing the independent decays of the particle $c_\alpha^\dag$ and the hole $c_\alpha$. Consider a multi-branched decay of the single particle $c_\alpha^\dag$, as shown in Fig.~$\ref{Single}$. There the particles $\gamma$ and $\delta$, which are produced in the first scattering, decay further through $n$ vertices $U_{i=1,...,n}$, and the vertex ${\tilde{U}}$, respectively. The possible orderings of this diagram correspond to $n+1$ correlated paths, which differ by the relative position of the vertex ${\tilde{U}}$ with respect to the $U_i$. Their sum,
\begin{equation}\label{totalsum}
\begin{split}
\Sigma'=& \frac{1}{\mathcal{E}_0(\mathcal{E}_0+\mathcal{\tilde{E}})(\mathcal{E}_0+\mathcal{\tilde{E}}+\mathcal{E}_1)\cdots (\mathcal{E}_0+\mathcal{\tilde{E}}+\cdots +\mathcal{E}_n) }+\\
& \frac{1}{\mathcal{E}_0(\mathcal{E}_0+\mathcal{E}_1)(\mathcal{E}_0+\mathcal{E}_1+\mathcal{\tilde{E}})\cdots (\mathcal{E}_0+\mathcal{E}_1\cdots +\mathcal{E}_n) }+\cdots +\\
&  \frac{1}{\mathcal{E}_0(\mathcal{E}_0+\mathcal{E}_1)(\mathcal{E}_0+\mathcal{E}_1+\mathcal{E}_2)\cdots (\mathcal{E}_0+\mathcal{E}_1\cdots +\mathcal{\tilde{E}}) },
\end{split}
\end{equation}
does not simply factorize, but it can nevertheless be written in compact form through an integral representation,
\begin{equation}\label{integral}
\Sigma'=\lim_{\epsilon \to 0} \int \frac{d \omega_1 d \omega_2 \delta\tonde{\omega_1 +\omega_2 -\mathcal{E}_0}}{\omega_1^- (\omega_1^- + \mathcal{\tilde{E}}) \cdot \omega_2^- (\omega_2^-+\mathcal{E}_1) \cdots (\omega_2^- +\mathcal{E}_1+\cdots+\mathcal{E}_n)},
\end{equation}
where $\omega_i^-=\omega_i -i\epsilon$.   
Indeed, the sum $\Sigma'$ (multiplied by the matrix elements of the correspondent vertices) must be equal to the retarded Green function associated to the independent, parallel decay of the particle $\gamma$  and the hole $\delta$, computed in the forward scattering approximation and at energy $\mathcal{E}_0$. For loop-free graphs like the one of Fig.~\ref{Single}, 
the decay processes of the particle $\gamma$ and the hole $\delta$ are independent. In the time domain, the Green function of their joint decay is the product of the individual Green functions, which leads to the convolution (\ref{integral}) in frequency space.

The above formula is rather natural when relating with standard many-body perturbation theory. Indeed, after the summation over orderings of vertices, the diagrams of a fixed geometry are in direct correspondence with the diagrams obtained by BAA in the perturbative expansion of the Keldysh self energy in the imaginary self consistent Born approximation.  The latter neglects the renormalization of the real part of the self energy and retains only processes where at each vertex an additional particle-hole pair is created. In our formalism, this corresponds to the directed paths jumping from generation to generation, see also the discussion in \ref{app:resum}. 
Not surprisingly, the statistical analysis of this class of diagrams will give an estimate of the radius of convergence for the operator expansion ($\ref{ansatz}$) which is similar to the criterion for the breakdown of stability of the localized phase found by BAA, or to its extension to infinite temperature~\cite{reichman2014dynamics}. Our further analysis is also very similar to the calculation in Ref.~\cite{gornyi2005interacting}, but differs in some points, which will be indicated.

The expression ($\ref{integral}$) for a branched diagram is a random variable, whose probability distribution is hard to analyze. However, the analytic structure of the integrand can be exploited to rewrite $\Sigma'$ as a sum over a much smaller number of terms than the number of orderings in Eq.~(\ref{totalsum}). After performing the integral over $\omega_2$ in Eq.~($\ref{integral}$), we find a number of poles in the complex plane of $\omega_1$. Using the residue theorem, we can write ($\ref{integral}$) as the sum over residues of the poles in the half plane, which contains less poles.
In the particular example considered, closing the contour on the upper half plane yields the algebraic identity:
\begin{equation}\label{smaller}
\begin{split}
 \Sigma'=&\frac{1}{\mathcal{\tilde{E}}}\frac{1}{\mathcal{E}_0(\mathcal{E}_0+\mathcal{E}_1)(\mathcal{E}_0+\mathcal{E}_1+\mathcal{E}_2)\cdots (\mathcal{E}_0+\mathcal{E}_1+\cdots +\mathcal{E}_n) }-\\
 &\frac{1}{\mathcal{\tilde{E}}}\frac{1}{(\mathcal{E}_0+\mathcal{\tilde{E}})(\mathcal{E}_0+\mathcal{\tilde{E}}+\mathcal{E}_1)(\mathcal{E}_0+\mathcal{\tilde{E}}+\mathcal{E}_1+\mathcal{E}_2)\cdots (\mathcal{E}_0+\mathcal{\tilde{E}}+\mathcal{E}_1+\cdots +\mathcal{E}_n) }.
 \end{split}
\end{equation}
 The two terms in ($\ref{smaller}$) have a similar structure as the denominators in the original path weight ($\ref{pathweight}$). For the considered sub-diagram, the sum over all the $n+1$ orderings of vertices could thus be reduced to the sum of only two "effective path" weights.

\subsubsection{General branched diagrams}
A convolution formula analogous  to Eq.~(\ref{integral}) can be written for any branched diagram: 
to each branching one associates an integral of the form ($\ref{integral}$) with one auxiliary frequency per decaying branch, as well as an energy conserving $\delta$-function for the vertex (see \ref{app:effpaths} for an example).
Then one eliminates the $\delta$-functions by integrating over the frequency variable, that occurs most often in the denominators.  Using the residue theorem, the remaining integrals can be carried out, and the sum over all orderings of a diagram with fixed geometry can be expressed as a much smaller sum of weights of effective paths, as in the example above.  
The number of such terms is given by the product of the number of residues obtained for each auxiliary frequency. 

The number of effective paths associated to a general diagram depends on its structure; to obtain an upper bound on this number, consider the diagram with the maximal number of branchings at fixed order $N$, see Fig.~$\ref{MaxBranch}$. In \ref{app:effpaths}, we show that in this case the number of effective paths scales  as $\text{exp} \quadre{\log 3 \,(\log N)^2 + \ O (\log N \log (\log N))}$. 
This upper bound implies that the number of effective paths associated to an arbitrary diagram is always \emph{sub-exponential} in $N$.

\section{Summing diagrams}
\label{sec:correlation}

In this section we show that in the localized region, at a any given order of the expansion, a few terms dominate the operator sum. The term with the largest coefficient in turn is dominated by the maximal diagram contributing to it.

\subsection{Summing over diagrams and their effective paths}
Let $\mathcal{D}_{\mathcal{I,J}}$ denote the set of all diagrams with final state $\cal I,J$, each diagram being characterized by its geometry and the labeling of its segments. For any diagram $d \in \mathcal{D}_{\mathcal{I,J}}$, let $\mathcal{P}(d)$ be the set of effective path weights $\tilde{\omega}_\Gamma$ associated to it, following the procedure described in the previous section. The corresponding amplitude on the operator lattice can then be written as
\begin{equation}\label{effectivepaths}
\mathcal{A}^{(\alpha)}_{\mathcal{I,J}} =\sum_{d \in \mathcal{D}_{\mathcal{I,J}}} \tonde{\sum_{
 \Gamma \in \mathcal{P}(d)} \tilde{\omega}_{\Gamma}}\equiv \sum_{d \in \mathcal{D}_{\mathcal{I,J}}} S(d).
\end{equation}

As we shall prove in the following section, the $ \tilde{\omega}_{\Gamma}$ are random variables with fat-tailed distributions. The effective paths associated to a diagram $d \in \mathcal{D}_{\mathcal{I,J}}$ all involve the same set of energies in their denominators and are thus correlated. Nevertheless, we argue that the tail of the distribution of their sum, $S(d)$, is still very similar to the tail distribution of a single effective path, since in the case of a large deviation, $S(d)$ is very likely to be dominated by the effective path with the biggest weight.
Indeed, consider a rare set of energies ${\cal E}_i$, which produces an atypically large value of $S(d)$. There is typically one single effective path for which all denominators become simultaneously small, while the combination of energies in the denominators of other effective paths are very likely to be suboptimal for a fraction of the denominators. Therefore, with high probability, $S(d)$ will approximately be equal to the maximum over all effective paths weights: $
 S(d) \approx \max_{\Gamma \in \mathcal{P}(d)} \tilde{\omega}_\Gamma.$

The set of energies ${\cal E}_i$ that optimize distinct effective paths are typically different, and thus 
these rare events can be approximated as being independent from each other.
 Hence, the tail of the distribution of $S(d)$ is enhanced with respect to the tail of a single path weight by a factor $\modul{\mathcal{P}(d)}$. We shall see, however, that due to the sub-exponential scaling of the number of effective paths, this enhancement is immaterial for the estimate of the radius of convergence of the operator series.

Inspecting the explicit examples of Eq.~(\ref{smaller}) or Eq.~(\ref{appendix:effpath}), one can see that there exist energy realizations for which cancellations occur between effective paths with significant weight. 
This happens when the single path weights are individually big, but $\tilde{\mathcal{E}}$ is much smaller than all the other energy variables $\mathcal{E}_i$, which leads to a cancellation between  effective paths.  However, such configurations require an atypically small $\tilde{\mathcal{E}}$ and do not occur with  significant probability. Therefore the suppression of the tail distribution due to such effects is hardly relevant.

Correlations between effective path weights of different diagrams are even weaker than those above, since they share at most a fraction of all ${\cal E}_i$. Therefore we may approximate rare deviations of $S(d)$ and $S(d')$  as independent if $d \neq d'$. Given that the $S(d)$ are themselves fat-tailed random variables, the sum over diagrams is dominated by the largest term. Therefore, the full operator amplitude $\mathcal{A}^{(\alpha)}_{\mathcal{I,J}}$ is likely to be dominated by one single effective path:
\begin{equation}\label{effectivepaths}
\mathcal{A}^{(\alpha)}_{\mathcal{I,J}} \approx \max_{d \in \mathcal{D}_{\mathcal{I,J}}} \tonde{\max_{\Gamma \in \mathcal{P}(d)} \tilde{\omega}_{\Gamma}} \approx \max_{\begin{subarray}{c}
\Gamma: \mathcal{(\alpha,\alpha)}\to \mathcal{(I,J)}
 \end{subarray} }\tilde{\omega}_\Gamma
\end{equation}
where on the right hand side the maximum is taken over all effective paths from $(\alpha, \alpha)$ to $(\mathcal{I,J})$. As a consequence, for the tail of the probability distribution we obtain the approximation
\bea \label{Peffpath}
P(\mathcal{A}^{(\alpha)}_{\mathcal{I,J}}=a) \approx |{\cal D}_{\mathcal{I,J}}| \overline{\mathcal{P}(d)} P(\tilde{\omega}_{\Gamma}=a),
\eea
where $\overline{\mathcal{P}(d)}$ is an average number of effective paths contributing to a diagram.

\subsection{Summing over amplitudes: probability of resonances}
Similarly to the effective path weights of different diagrams, also the amplitudes $\mathcal{A}^{(\alpha)}_{\mathcal{I,J}}$ associated to different sites $\mathcal{I,J}$ are weakly correlated, and we treat them as independent random variables. Let us now consider the probability in ($\ref{pres}$):
\begin{equation}\label{sum1}
\mathbb{P} \tonde{ \forall N>N^*,  
\sum_{
  \begin{subarray}{l}
\hspace{.6 cm}\mathcal{I \neq J}\\
|\mathcal{I}|=N+1=|\mathcal{J}|  \end{subarray}} 
\modul{\mathcal{A}^{(\alpha)}_{\mathcal{I,J}}}<z^N}\approx  \prod_{N>N^*}\mathbb{P} \tonde{ \sum_{
  \begin{subarray}{l}
\hspace{.6 cm}\mathcal{I \neq J}\\
|\mathcal{I}|=N+1=|\mathcal{J}|  \end{subarray}} 
\modul{\mathcal{A}^{(\alpha)}_{\mathcal{I,J}}}<z^N}.
 \end{equation}
 Here we approximated the probability to satisfy the condition at each generation to be independent from the previous generations.
As follows from (\ref{Peffpath}) and from the fact that the effective paths $\tilde{\omega}_\Gamma$ have fat tails, the amplitudes $\mathcal{A}^{(\alpha)}_{\mathcal{I,J}}$ have themselves a fat-tailed distribution. Their sum is therefore dominated by the maximal amplitude, and each factor on the right hand side ($\ref{sum1}$) can be computed as:
\begin{equation}\label{chain}
\begin{split}
\mathbb{P} \tonde{ \Max\limits_{
  \begin{subarray}{l}
\hspace{.6 cm}\mathcal{I \neq J}\\
|\mathcal{I}|=N+1=|\mathcal{J}|  \end{subarray}} 
\modul{\mathcal{A}^{(\alpha)}_{\mathcal{I,J}}}<z^N
 } &=\prod_{
  \begin{subarray}{l}
\hspace{.6 cm}\mathcal{I \neq J}\\
|\mathcal{I}|=N+1=|\mathcal{J}|  \end{subarray}} \tonde{1-\mathbb{P} \tonde{
\modul{\mathcal{A}^{(\alpha)}_{\mathcal{I,J}}}>z^N
 }}\\
&\approx  \text{exp} \tonde{-\sum_{
  \begin{subarray}{l}
\hspace{.6 cm}\mathcal{I \neq J}\\
|\mathcal{I}|=N+1=|\mathcal{J}|  \end{subarray}} \mathbb{P} \tonde{
\modul{\mathcal{A}^{(\alpha)}_{\mathcal{I,J}}}>z^N}}.
  \end{split}
\end{equation}
Using $(\ref{Peffpath})$, the exponent in ($\ref{chain}$) is re-written as:
\begin{equation}\label{short}
 \sum_{
  \begin{subarray}{l}
\hspace{.6 cm}\mathcal{I \neq J}\\
|\mathcal{I}|=N+1=|\mathcal{J}|  \end{subarray}} \mathbb{P} \tonde{
\modul{\mathcal{A}^{(\alpha)}_{\mathcal{I,J}}}>z^N}=\sum_{
  \begin{subarray}{l}
\hspace{.6 cm}\mathcal{I \neq J}\\
|\mathcal{I}|=N+1=|\mathcal{J}|  \end{subarray}} \modul{\mathcal{D}_{\mathcal{I,J}}} \overline{\mathcal{P}(d)} \mathbb{P} \tonde{
\modul{\tilde{\omega}_{\Gamma}}>z^N}.
\end{equation}

The probability in ($\ref{short}$) is a large deviation probability: indeed, the weights  $\tilde{\omega}_\Gamma$ of effective paths are of order $ O \tonde{\lambda ^{N}}$: in order for $\tilde{\omega}_\Gamma$ to be bigger than $z^N$ (with $z$ arbitrarily close to $1$), this decay factor must be compensated by an atypical smallness of the energy denominators. We devote the following section to the computation of the probability of these large deviation events. The calculation will reveal that, for $\lambda$ sufficiently small, the probability decays exponentially with $N$. This decay competes with the exponential growth of the total number of effective paths of length $N$:
\begin{equation}\label{totdia}
\mathcal{N}_N \equiv \sum_{
  \begin{subarray}{l}
\hspace{.6 cm}\mathcal{I \neq J}\\
|\mathcal{I}|=N+1=|\mathcal{J}|  \end{subarray}} \modul{\mathcal{D}_{\mathcal{I,J}}}\overline{\mathcal{P}(d)},
\end{equation}
which we estimate in Sec.~\ref{sec:combinatorics} below. The competition between these two terms leads to a transition at a given critical value of $\lambda$, which we determine in  Sec.~\ref{sec:estimateradius}.

\section{Large deviations of paths with correlated denominators}
\label{sec:largedev}

In the previous section we argued that the large deviations of operator amplitudes are essentially determined by the large deviations of effective path weights. The weight of any effective path is the products of two terms, describing the decay of $c_\alpha^\dag$ and $c_\alpha$, respectively. In each of those terms, (cf. ($\ref{smaller}$) e.g.), the functional dependence on the ${\cal E}_i$ is similar to that in the original path weights ($\ref{pathweight}$). We will first discuss the latter and then show that general effective paths behave essentially identically.
 
 Because of the energy restrictions ($\ref{restriction}$) the energy differences $\mathcal{E}_{\alpha \beta, \gamma \delta}/\delta_\xi$ are random variables of order $O(1)$. For simplicity, we take them  as independent Gaussian random variables with zero mean and unit variance.
The denominators in ($\ref{pathweight}$) are partial sums of such energies, and we may write:
\begin{equation}\label{pathweight2}
|\omega_{\text{path}}|
=\prod_{i=1}^{N-1} \frac{\lambda |\eta_{\alpha_i \beta_i, \gamma_i \delta_i}|}{|s_i|},
\end{equation}
where $s_i=( \mathcal{E}_1 + \cdots+ \mathcal{E}_i)/\delta_\xi$, with $\mathcal{E}_i\equiv \mathcal{E}_{\alpha_i \beta_i, \gamma_i \delta_i}$.

In path weights of the form (\ref{pathweight2}) we are mostly interested in characterizing the distribution of the product of denominators. The numerator behaves as $\sim (\lambda \eta_{\rm typ})^{N-1}$, with $\eta_{\rm typ}=\exp[\langle \log|\eta| \rangle]=1/e$, and we neglect the Gaussian fluctuations of its logarithm. 

The fact that the denominators in ($\ref{pathweight2}$) are correlated distinguishes the many-body problem from single particle localization. These correlations are a feature that any perturbative treatment of MBL has to deal with, and it is thus important to develop a method to calculate the large deviations in this case.

The distribution function $P_N(y)$ of the logarithm of the product of denominators, 
\begin{equation}\label{vari}
  Y_N \equiv -\sum_{i=1}^N \log |s_i|,                                                                               
\end{equation}
can be obtained from its generating function,  
\begin{equation}
 G_N(k) \equiv \mathbb{E} \quadre{e^{-k Y_N}},
\end{equation}
by inverse Laplace transform,
\begin{equation}\label{inverse}
P_N(y)= \frac{1}{2 \pi i}\int_{\mathcal{B}} e^{y k} G_N(k) dk,
\end{equation}
where $\mathcal{B}$ is the Bromwich path in the complex $k$-plane.

In the present case, the relevant $y$ scales linearly with $N$, and thus we define $\tilde{y} = y/N$, and
\begin{equation}
P_N(N\tilde y)=\frac{1}{2 \pi i}\int_{\mathcal{B}} e^{N\phi_N} dk,
\end{equation}
where the function
\begin{equation}\label{cramer}
\phi_N(\tilde{y},k)=\tilde{y} k + \frac{ \log G_N(k)}{N}\stackrel{N\to \infty}{\to} \phi(\tilde{y},k)
\end{equation}
has a well-defined limit, $\phi(\tilde{y},k)$, for large $N$. In that limit, the integral over $k$ can be done by a saddle point approximation. The contour has to be deformed to pass parallel to the imaginary axis through $k^*=k^*(\tilde{y})$, which satisfies:
\begin{equation}\label{saddle}
 \tilde{y}= -\frac{d}{dk} \quadre{\lim_{N\to \infty}\frac{\log G_N(k)}{N}}_{k= k^*(\tilde{y})}.
\end{equation}
Large deviations correspond to $\tilde y=O(1)$. In the case of parametrically small interaction strength $\lambda$ (which is relevant in the case of large connectivity ${\cal K}$) we will see that we can restrict our attention to $\tilde y\gg 1$, see Sec.~\ref{sec:estimateradius}. For large values of $\tilde y$, we will see that the saddle point tends to $k^*\to -1$. 

The computation of the generating function $G_N$ is given in \ref{app:largedev}. Here it suffices to say that the recursive structure of the denominators $s_i$ lends itself naturally to a transfer matrix expression for $G_N$, which grows as the $N$th power of the largest eigenvalue.  

The final result for the exponent at the saddle point is
\begin{equation}
\begin{split}
 \phi(\tilde{y}, k^*(\tilde y))=& -\tilde{y} +\log\left(\frac{2e\tilde{y}}{\sqrt{2\pi}}\right) +\frac{\gamma}{2\tilde y} +\ O\tonde{\frac{1}{\tilde{y}^2}},
 \end{split}
\end{equation}
for $\tilde{y} \gg 1$. From this we obtain the large deviation probability:
\begin{equation}\label{zz2}
\begin{split}
P_N\left(N\tilde{y}=\log\left[\prod_{i=1}^N\frac{1}{|s_i|}\right]\right)= C(\tilde{y},N) \tonde{{\frac{2e}{\sqrt{2 \pi}}}}^N \tilde{y}^N e^{-N\mathcal{F}(\tilde{y})},
 \end{split}
\end{equation}
where $C$ contains only negligible logarithmic corrections to the exponent, and

\begin{equation}\label{effe}
 \mathcal{F}(\tilde{y})={\tilde{y}  -\frac{\gamma}{2 \tilde{y}}+\ O\tonde{\frac{1}{\tilde{y}^2}}}.
\end{equation}

\subsection{Comparison between correlated and uncorrelated denominators}
It is interesting to compare the large deviation distribution ($\ref{zz2}$) with the tails of the distribution of the random variable:
\begin{equation}\label{indi}
 Y'_N \equiv -\sum_{i=1}^N \log |X_i| 
\end{equation}
where $X_i$ are i.i.d. Gaussian random variables with zero mean and unit variance. As derived in \ref{app:largedev}, at leading order in $N$, up to
a correction ${\cal F} \to {\cal F} - {\log 2} / (2\tilde{y}) + O(1/\tilde{y}^2)$,  both have the same form (\ref{zz2}).

Physically, this result can be understood as follows. By restricting to $\tilde y\gg 1$, we are concentrating on very 
rare realizations of $Y_N$. Those are insensitive to the details in the structure of the denominators. Indeed, atypically big values of objects like $\tonde{\prod_{i=1}^N s_i}^{-1}$ arise from restraining the random walk  $(s_1, \cdots, s_N)$ to the vicinity of the origin. This boils down to computing the probability that $s_i$ is small \emph{conditioned} on the fact that $s_{i-1}$ was small. To leading order in the typical smallness of such denominators, one obtains the same result as by minimizing $N$ denominators independently. The leading correction with respect to the case of  i.i.d. denominators consists in a small suppression of the tail, since it is slightly less probable to encounter small denominators, when they are correlated.

The above reasoning can be extended to more general weights $\tilde{\omega}_\Gamma$, associated with effective paths. Indeed, the corresponding denominators are still products of single energies or partial sums (see Eq.~($\ref{smaller}$) or Eq.~(\ref{appendix:effpath})). In the limit of very large deviations ($\tilde y\gg 1$) they all share the same tail distribution ($\ref{zz2}$), the only relevant parameter being the total number $N$ of denominators. Therefore, approximating the numerator in $\tilde{\omega}_\Gamma$ with its typical value  $ (\lambda \eta_{\rm typ})^{N}$ and using (\ref{zz2}), we finally obtain:
\begin{equation}\label{touse}
 \mathbb{P} \tonde{\frac{\log \modul{\tilde{\omega}_\Gamma}}{N} = {\tilde{x} + \log \lambda  \eta_{\rm typ}}} \approx C(\tilde{x},N) \tonde{{\frac{2e}{\sqrt{2 \pi}}}}^N \tilde{x}^N e^{-N\mathcal{F}(\tilde{x})},
\end{equation}
with $\mathcal{F}$ given in (\ref{effe}).

\section{Counting diagrams}
\label{sec:combinatorics}

\subsection{Justification of neglecting interaction vertices with equal indices}
\label{nodiagonals}
We recall  that we have neglected interaction terms $U_{\alpha\beta,\gamma\delta}$ where two or more indices are identical. This will significantly simplify the combinatorics of counting  diagrams.
Let us now give a justification a posteriori for this approximation, by showing that such terms would make contributions which are down by factors of $1/{\cal K}$.  
Consider the various scattering processes with one pair of equal indices among the four legs of a vertex, whereby we restrict to one ingoing and three out-going particles.
Consider first the scattering $\alpha\to\beta$ with the simultaneous creation of a pair $(\gamma,\alpha)$. The constraints $|\epsilon_\alpha-\epsilon_\gamma|<\delta_\xi,|\epsilon_\alpha-\epsilon_\beta|<\delta_\xi$ imply that all levels have to lie within $\delta_\xi$ from each other. The phase space for such events is smaller by a factor of $1/\mathcal{K}$ with respect to generic scattering processes where $\gamma$ is unrestricted. 

The second case is more subtle. It consists in a scattering $\alpha\to \beta$ from a particle $\gamma$, which remains in place. If this is to be a resonant contribution one needs the energy increment $\Delta{\cal E}$ of the vertex to be $|\Delta{\cal E}|= |\epsilon_\alpha-\epsilon_\beta|\lesssim\delta_\xi/\mathcal{K}$. In a scattering where $\gamma$ switches to a neighboring state $\delta$, with $|\epsilon_\gamma-\epsilon_\delta|\sim \delta_\xi$, one can optimize $\alpha, \beta$ among the ${\cal K}$ different choices, such as to make  $\Delta{\cal E}$ of order $\delta_\xi/{\cal K}$. However, if $\gamma$ remains in place, the optimum over the $\cal K$ choices for $\alpha, \beta$ will yield a parametrically bigger  $\Delta{\cal E} = \epsilon_\alpha-\epsilon_\beta$, because of the repulsion between the neighboring levels $\alpha,\beta$. Therefore such processes are systematically much less resonant than processes involving four distinct levels.~\footnote{BAA dropped such terms for a different reason, working directly with Hartree-Fock orbitals, which implicitly depend on the interaction strength and the initial state to be studied. The latter introduces  
  a slight dependence of the localization length $\xi$, and hence of the level spacing $\delta_\xi$, on $\lambda$. This in turn might induce a small shift of $\lambda_c$.
 However, since their subsequent analysis boils down to dropping the same terms as we have argued above, this shift is expected to be a $1/\mathcal{K}$ correction.}

\subsection{Combinatorics of  diagrams}
We now estimate the total number of diagrams $\mathcal{N}_N$ at a given order $N$, cf. Eq.~(\ref{totdia}). For simplicity, we restrict here to the case of spatial dimension $d=1$.

Consider any amplitude $\mathcal{A}^{(\alpha)}_{\mathcal{I,J}}$ with index set $(\mathcal{I,J})=(\alpha_1 \cdots \alpha_N, \beta_1 \cdots \beta_N)$. The localization centers $r_{\alpha_i}, r_{\beta_i}$, cf. Eq.~(\ref{radius}), of the single particle indices are distributed over a certain number of localization volumina of length $\xi$ around $r_\alpha$, with a given number of single particle indices per localization length. Due to the energy restrictions imposed on the interactions, particles and holes belonging to the same localization volume are organized in pairs: members of a pair are produced in the same scattering process, and have an energy difference of order $\delta_\xi$.

Due to the fact that the interaction is local, only particle-hole pairs in nearby localization volumina can be involved in the same interaction vertex: this imposes some constraints on the geometry of the diagrams representing the scattering processes with $(\mathcal{I,J})$ as final state. For example, states $(\mathcal{I,J})$ having only one particle-hole pair per localization length must be associated to diagrams with no branchings in the decays of $c_\alpha^\dag$ and $c_\alpha$, since the particle-hole pairs must be created in a fixed order dictated by their spatial sequence, and thus no permutation is possible. In contrast, final states with several pairs per localization length can be reached by a variety of diagrams.

In the following, we construct the subset of diagrams corresponding to scattering processes with a "necklace structure", in which the particle-hole pairs are created in a sequence of $n$ groups of $m_{i=1,...,n}$ pairs, each group belonging to a single localization volume. This furnishes a lower bound on the number of all diagrams. Note that $m_i$ is bounded by the maximal number of particle-hole pairs per localization volume ($N_{\text{loc}}=\mathcal{K}/4$), and $\sum_{i=1}^n m_i=N$. Due to locality, pairs belonging to the $i$th and $(i+1)$th group  belong to neighboring localization volumina in real space; pairs belonging to different groups $i,j \neq \grafe{i-1, i+1}$ might belong to the same localization volume.

\begin{figure}
   
  \captionsetup[subfigure]{labelformat=empty}
  \centering
  \subfloat[\scriptsize{(a) 
  The $m_1$ pairs of the first group belong to the localization volume containing the localization center $r_\alpha$, with one pair ($\tilde{\alpha} \tilde{\beta}$ in the Figure) with energies close to $\epsilon_\alpha$. The remaining $m_1$ pairs ($m_1-1$ in the same volume and one in the adjacent volume) are produced in $m_1$ scatterings organized in diagrams with all possible geometries ($\mathcal{T}_{m_1}$ of them). For each geometry, a factor $m_1$ comes from the choice of the vertex (red in the Figure) that produces a pair in the subsequent localization volume.}]{
                   \raisebox{.0 cm}{ \includegraphics[width=0.46\textwidth]{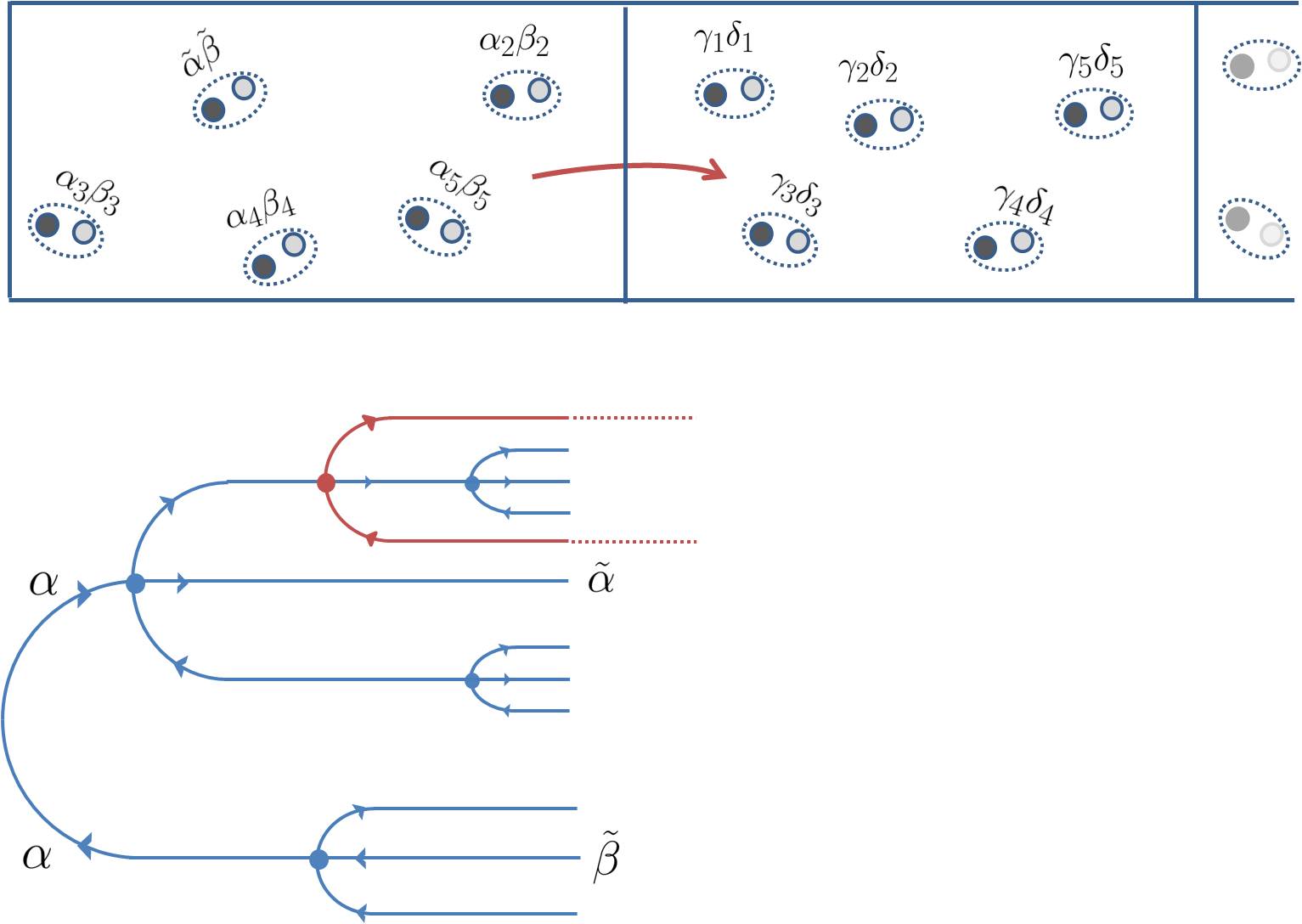}
        }}%
        \quad
        \subfloat[\scriptsize{(b) The indices of the remaining $m_1-1$ pairs in the localization volume are assigned to the legs. This fixes all labels up to the internal legs. For the first localization volume, the possible internal indices satisfying the energy restrictions are $2^{m_1-2}$, $\alpha$ being fixed.}]{%
%
\includegraphics[width=0.46\textwidth]{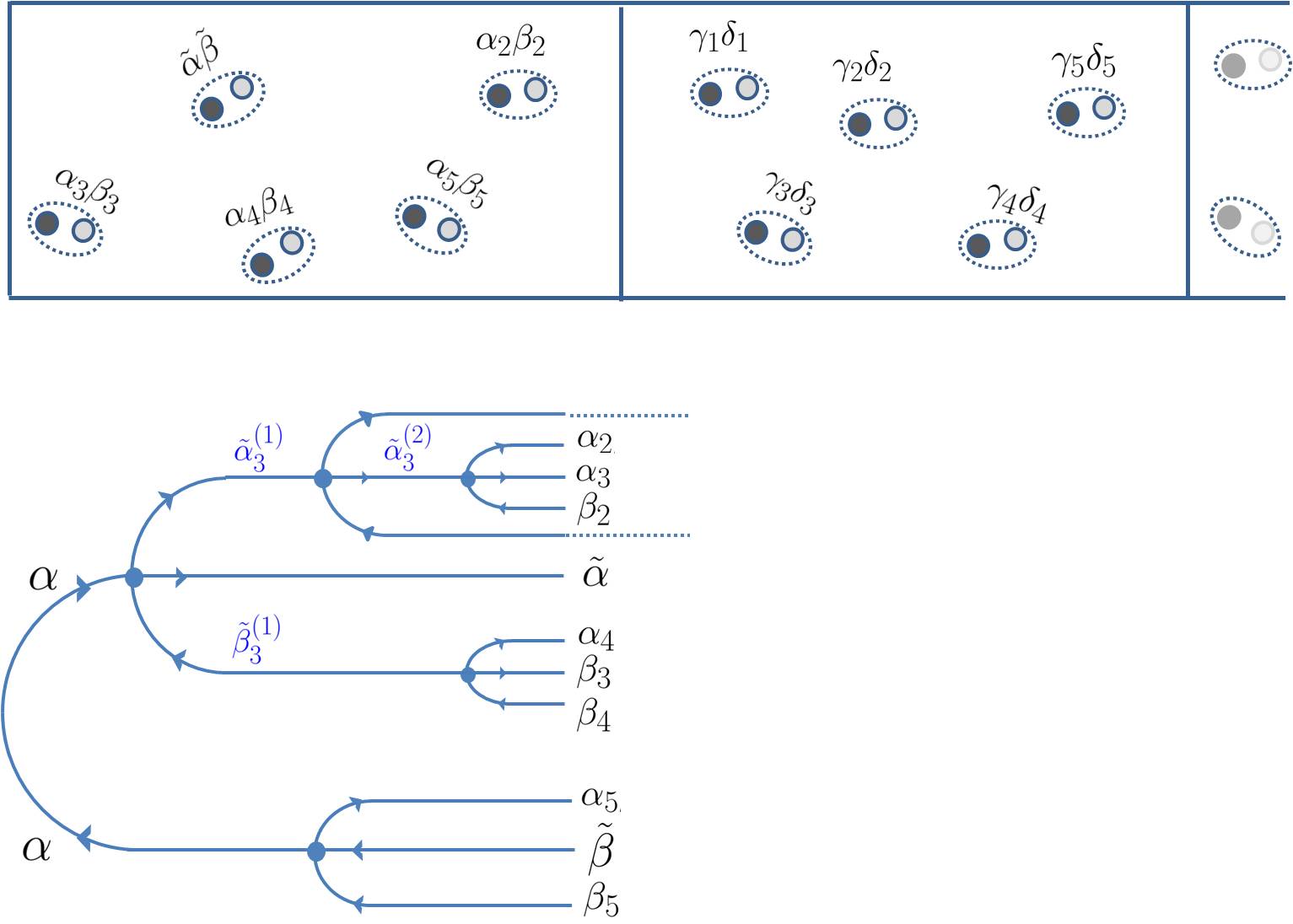}
} 
               
                \subfloat[\scriptsize{(c)  For a fixed geometry and a set of final labels, the permutation of assignments of labels to the legs gives rise to an independent diagram with the same final state, since the matrix elements of the interactions change. There are $(m_1-1)!$ such permutations (the pair $\tilde{\alpha} \tilde{\beta}$ is fixed). When legs are permuted, the corresponding internal indices must change as well in order to satisfy the energy restriction of the interactions (in the Figure, $\tilde{\alpha}^{(i)}_3 \to \tilde{\alpha}^{(i)}_5 $ and $\tilde{\beta}^{(i)}_3 \to \tilde{\beta}^{(i)}_5 $).}]{%
%
\includegraphics[width=0.46\textwidth]{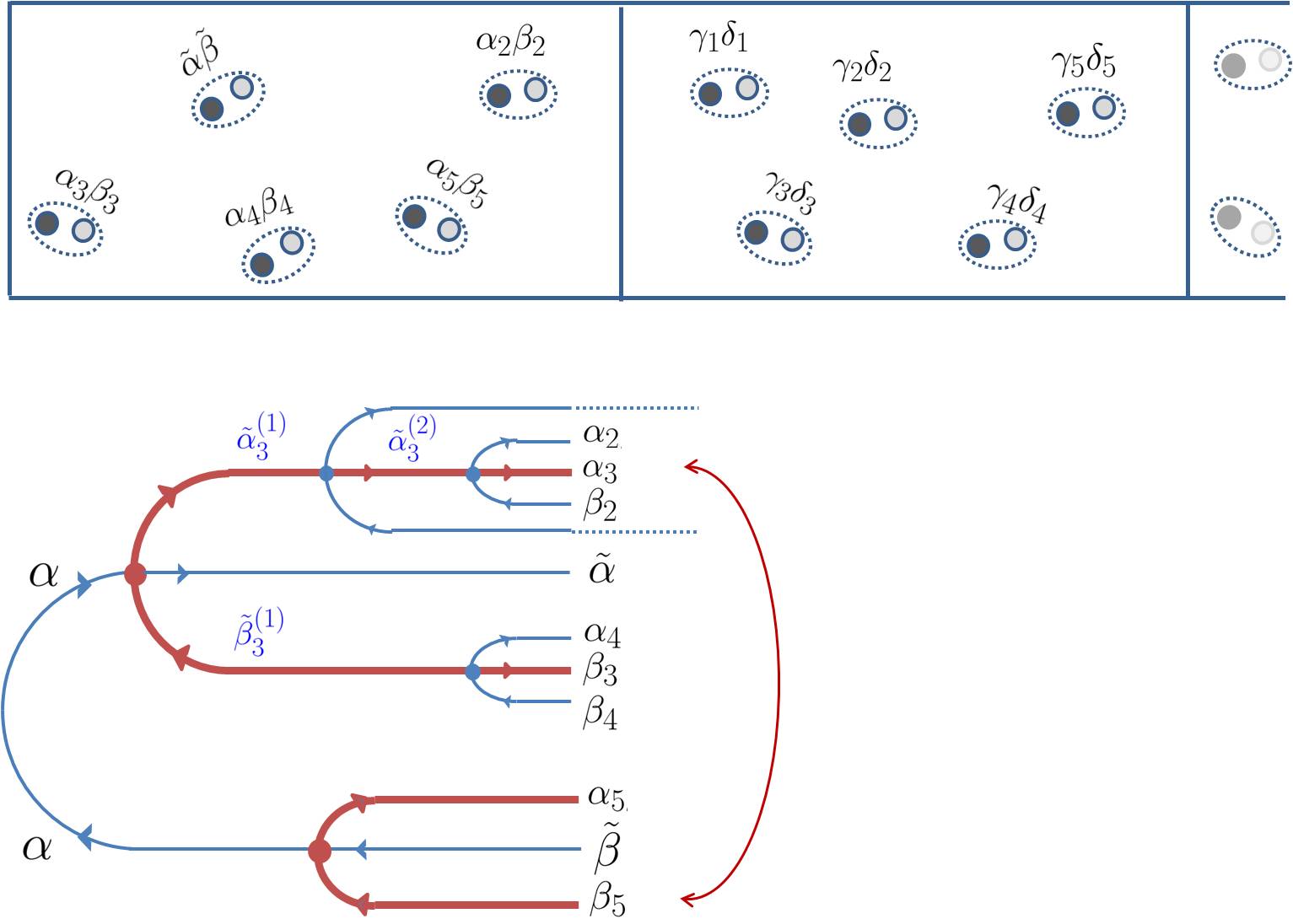}
} \quad
 \subfloat[\scriptsize{(d) Diagrams corresponding to the decay processes in the adjacent localization volume are attached to a pair of legs selected in (a). Again, $m_2$ interactions occur on two branches, in a total of $\mathcal{T}_{m_2}$ possible distinct geometries. There are $m_2$ choices to select the pair of legs to which to attach the next subdiagram. For any of the $m_2!$ labelings of the remaining external legs, there are two choices for each internal index, corresponding to whether the incoming particle scatters up or down in energy. In total there are $m_2 2^{m_2} m_2! \mathcal{T}_{m_2}$ different diagrams associated to this group of pairs. The same counting holds true for the subsequent groups.}]{%
\includegraphics[width=0.46\textwidth]{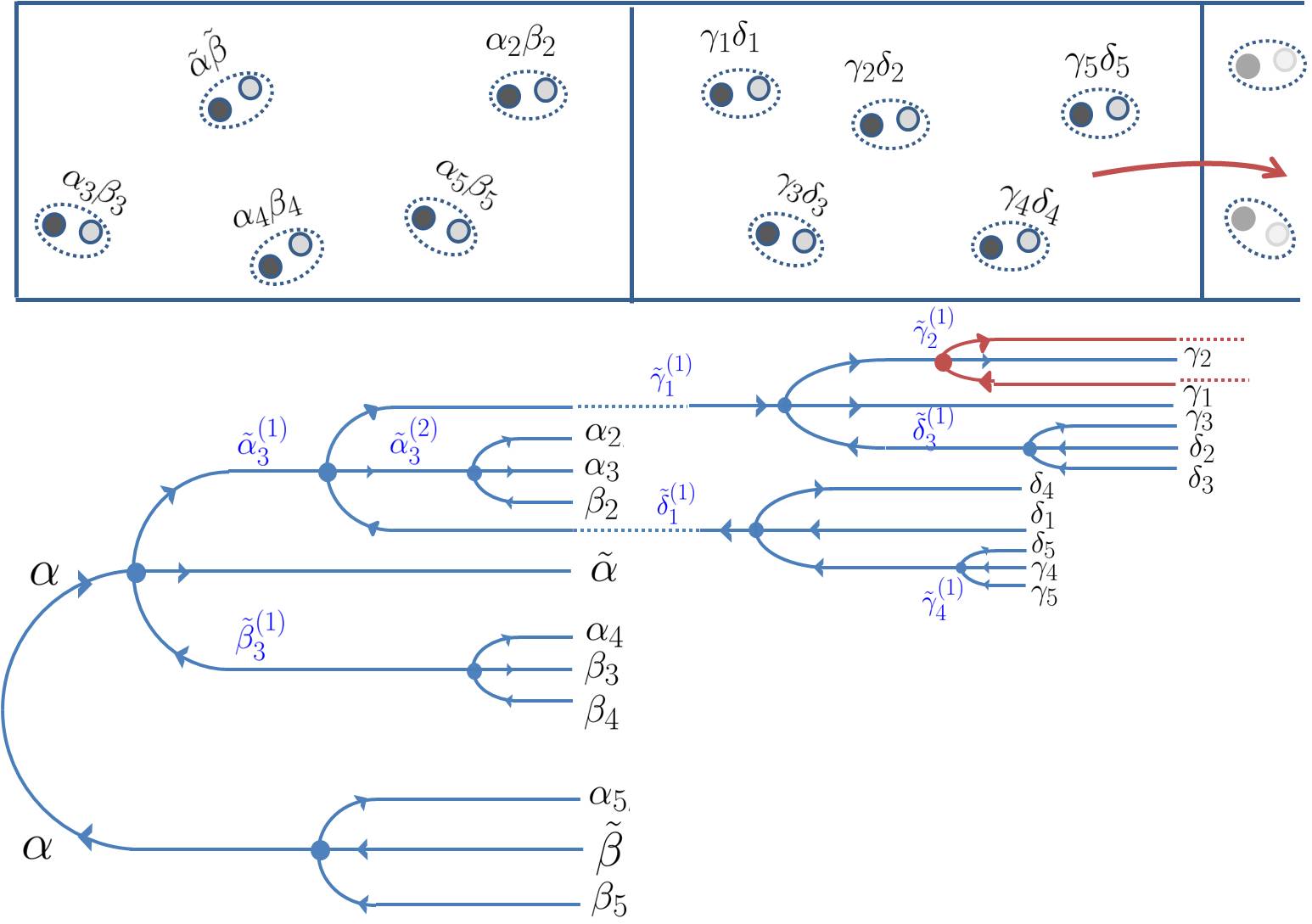}
} 
    \caption{%
    Construction of the diagrams representing the decay of groups of $m_{i}$ particle-hole pairs, where members of the same group belong to the same localization volume. The diagrams are constructed by connecting sub-diagrams describing the decay of each single group of pairs. We restrict the combinatorics to only one scattering vertex connecting the sub-diagrams of different groups.
           }%
   \label{Mirlin}
\end{figure}

 This construction is done in two steps: first, for every group $i$ we build all possible sub-diagrams with final indices corresponding to the indices of the $m_i$ pairs, as illustrated in Fig.~$\ref{Mirlin}$. In a second step, we connect sub-diagrams of neighboring groups by a single scattering vertex. We thus obtain a global necklace diagram, and count how many different diagrams with this structure there are.  The counting is similar to Ref.~\cite{gornyi2005interacting}, but here we include diagrams corresponding to final states with a non-uniform density of particle and hole indices per localization length, since these have a larger abundance.

A central ingredient for the combinatorics is the number of all possible geometries of diagrams with $m$ interactions in a given localization volume, see Fig.~$\ref{Mirlin}$. We denote this number by $\mathcal{T}_m$. It equals the number of trees with one root (of connectivity $2$) and $m$ nodes (of connectivity $4$). As we derive in \ref{app:combinatorics}: 
\begin{equation}\label{rec1}
\begin{split}
 \mathcal{T}_m=  \frac{3^{\frac{3}{2} + 3m}}{\pi} \frac{\Gamma \tonde{m + \frac{2}{3} }\Gamma \tonde{m + \frac{4}{3}}}{\Gamma \tonde{2m +3}}\sim \frac{3 }{4 }\sqrt{\frac{3}{\pi}} \frac{1}{m^{\frac{3}{2}}} \tonde{\frac{27}{4}}^m.
\end{split}
 \end{equation}

Following the reasonings explained in Fig.~~$\ref{Mirlin}$, we find  the number of necklace diagrams associated with fixed groups of $m_i$ pairs to be
\bea
\label{neck}
n_{\rm neck} = \prod_{i=1}^{n} \quadre{m_i 2^{m_i}  m_i! {\cal T}_{m_i} }.
\eea
The origin of the various factors is explained in detail in Fig.~\ref{Mirlin}: the  factor $m_i$ counts the number of pairs which are created subsequently to the first pair entering the volume associated to the group $i$. One of those $m_i$ pairs belongs to the adjacent localization volume and creates the subsequent cascade of pair creations there. The second factor describes the choice of two levels (the level closest in energy above or below) to which an incoming quasiparticle may scatter at a vertex. The factorial term comes from the choice of assigning the $m_i$ pairs to the final legs of a given tree diagram in the localization volume of group $i$. 

Consider first the case in which only a single group $i$ of pairs occupies a given localization volume.
The number of choices of $\grafe{m_i}$ particle-hole pairs is then given by 
\begin{equation}\label{ns}
 n_{s}(\{m_i\},\mathcal{K}) 
 \equiv  \prod_{i=1}^n  2^{m_i} \binom{N_{\text{loc}}-m_i}{m_i}= \prod_{i=1}^n \quadre{2^{m_i} \binom{{\cal K}/4-m_i}{m_i}}.
\end{equation}

Indeed, a configuration of $m_i$ pairs of (disjoint) adjacent levels, and the remaining $N_{\text{loc}}-2m_i$ untouched levels in the same localization volume form a set of $N_{\text{loc}}-m_i$ objects, out of which $m_i$ are pairs. This explains the binomial factor. For each pair, one can choose how to assign the two levels to particle and hole, respectively. This yields the factor $2^{m_i}$. 

As we will see below, the relevant $m_i$ are of order $O(1)\ll {\cal K}$. We therefore approximate:
\begin{equation}
\label{approx}
\binom{{\cal K}/4-m_i}{m_i} \approx \frac{({\cal K}/4)^{m_i}}{m_i!}.
\end{equation}

Note that the necklace structure will in general fold back and forth in real space, such that several groups will get to lie in the same volume.
Nevertheless, the above approximation remains good as long as the total number of pairs created in a given localization volume is significantly smaller than $\cal K$.

Combining Eqs.~(\ref{neck}-\ref{approx}), the total number of necklace diagrams is: 
\begin{equation}
\label{totalnumbdia}
 \mathcal{N}_N \approx  \overline{\mathcal{P}(d)}\sum_{\{m_i\}|\sum_i m_i=N}   \frac{1}{2} \prod_{i=1}^n  \quadre{2{\cal K}^{m_i}m_i  {\cal T}_{m_i}} ,
\end{equation}
where the average number of effective paths per diagram, $\overline{\mathcal{P}(d)}$, scales sub-exponentially with $N$. The factors of $2$ arise due to freedom of each group to scatter to the left or the right of the preceding group as long as there is still significant phase space in the corresponding localization volumina. The correction due to the finiteness of ${\cal K}\gg  1$ is small and was thus neglected.
 
We now determine the distribution of group sizes $\grafe{m_i}$ which dominates the sum ($\ref{totalnumbdia}$), writing
\bea \label{ni}
{\cal N}_N&=& \frac{1}{2}\sum_{\{m_i\} | \sum_i m_i = N}  \prod_i 2{\cal K}^{m_i} m_i {\cal T}_{m_i} = \frac{{\cal K}^N}{2} \sum_{\{m_i\} | \sum_i m_i = N}  \prod_i  2m_i {\cal T}_{m_i}\nn\\
&=& \frac{{\cal K}^N}{2} \sum_{\{n_m\} | \sum_m m n_m  = N} \binom{\sum_m n_m}{n_1,n_2,...,n_m}\prod_m (2m {\cal T}_m)^{n_m},
\eea
where $n_m = \sum_i \delta_{m,m_i}$ is the number of groups $i$ with $m$ pairs. For the relevant $m$'s, $n_{m}\sim N \gg 1$; therefore, at large $N$ the sum (\ref{ni}) is dominated by the saddle point over the $n_m$. Imposing the constraint $\sum_m m n_m  = N$ with a Lagrange multiplier $\mu$ yields the saddle point equations:
\bea
\mu m= -\log(n_m)+\log(\sum_m n_m) + \log(2 m {\cal T}_m),
\eea
and thus
\bea 
\frac{n_m}{\sum_{m'} n_{m'}} =  2 m {\cal T}_m e^{-\mu m}.
\eea

The Lagrange multiplier $\mu$ is fixed by the constraint:
\begin{equation}
\label{Eqforemu}
1=\sum_m 2 m {\cal T}_m e^{-\mu m} =-2\frac{d}{d\mu} [\mathcal{T}(x=e^{-\mu})],
\end{equation}
 with $\mathcal{T}(x)= \sum_{m} {\cal T}_m x^{ m}$. As discussed in \ref{app:combinatorics},  $\mathcal{T}(x)= [T(x)]^2$, where $T(x)$ is the generating function of 3-branched trees satisfying $T(x)=1 +x T^3(x)$. The solution of Eq.~(\ref{Eqforemu}) is: 
\bea
e^{-\mu}=0.0941.
\eea

\begin{figure}
   
  \captionsetup[subfigure]{}
  \centering
  \subfloat[]{\label{fig:Distm1}
                   \raisebox{.0 cm}{ \includegraphics[width=0.459\textwidth]{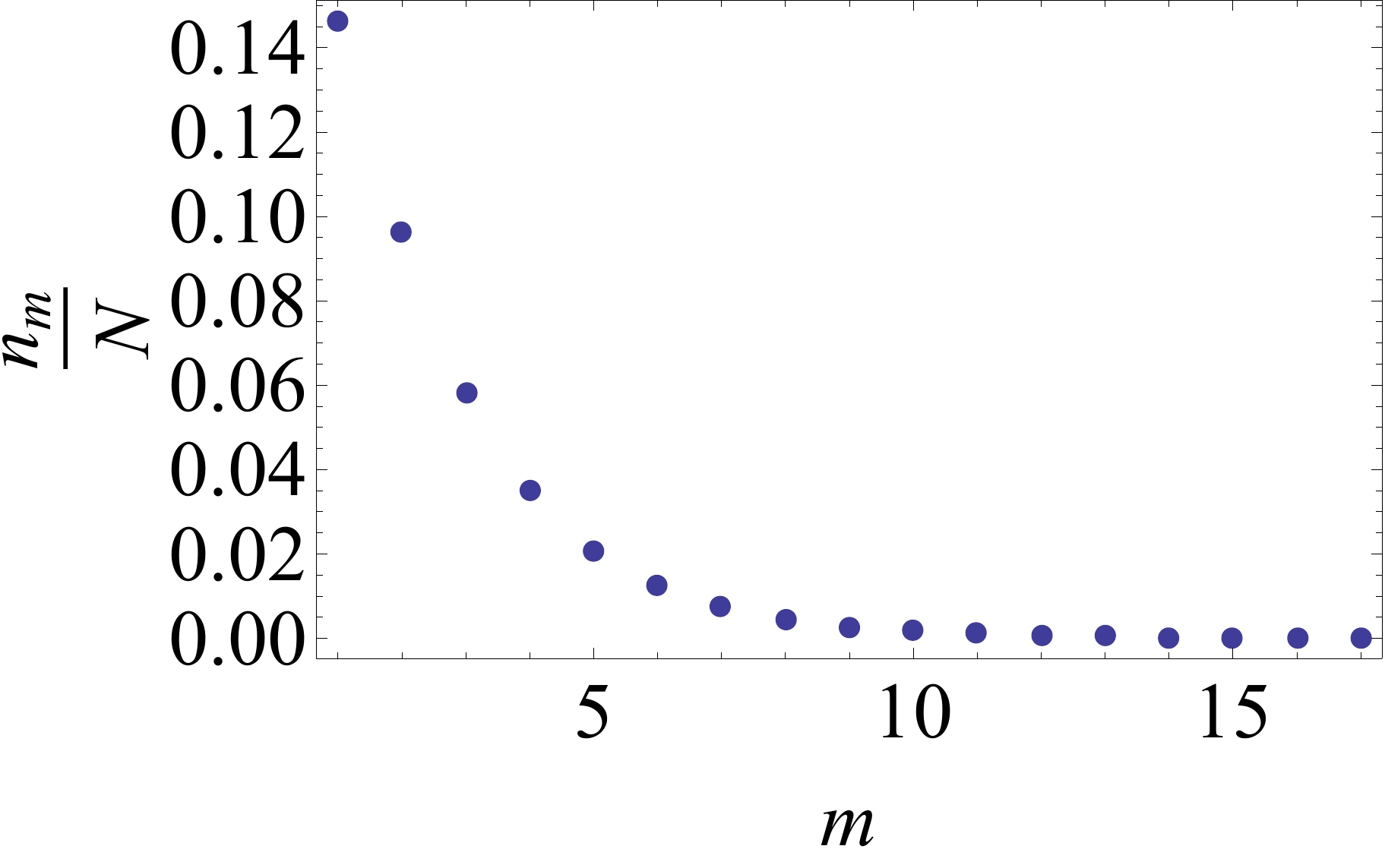}
        }}%
        \quad
        \subfloat[]{%
           \label{fig:Distm2}
\includegraphics[width=0.459\textwidth]{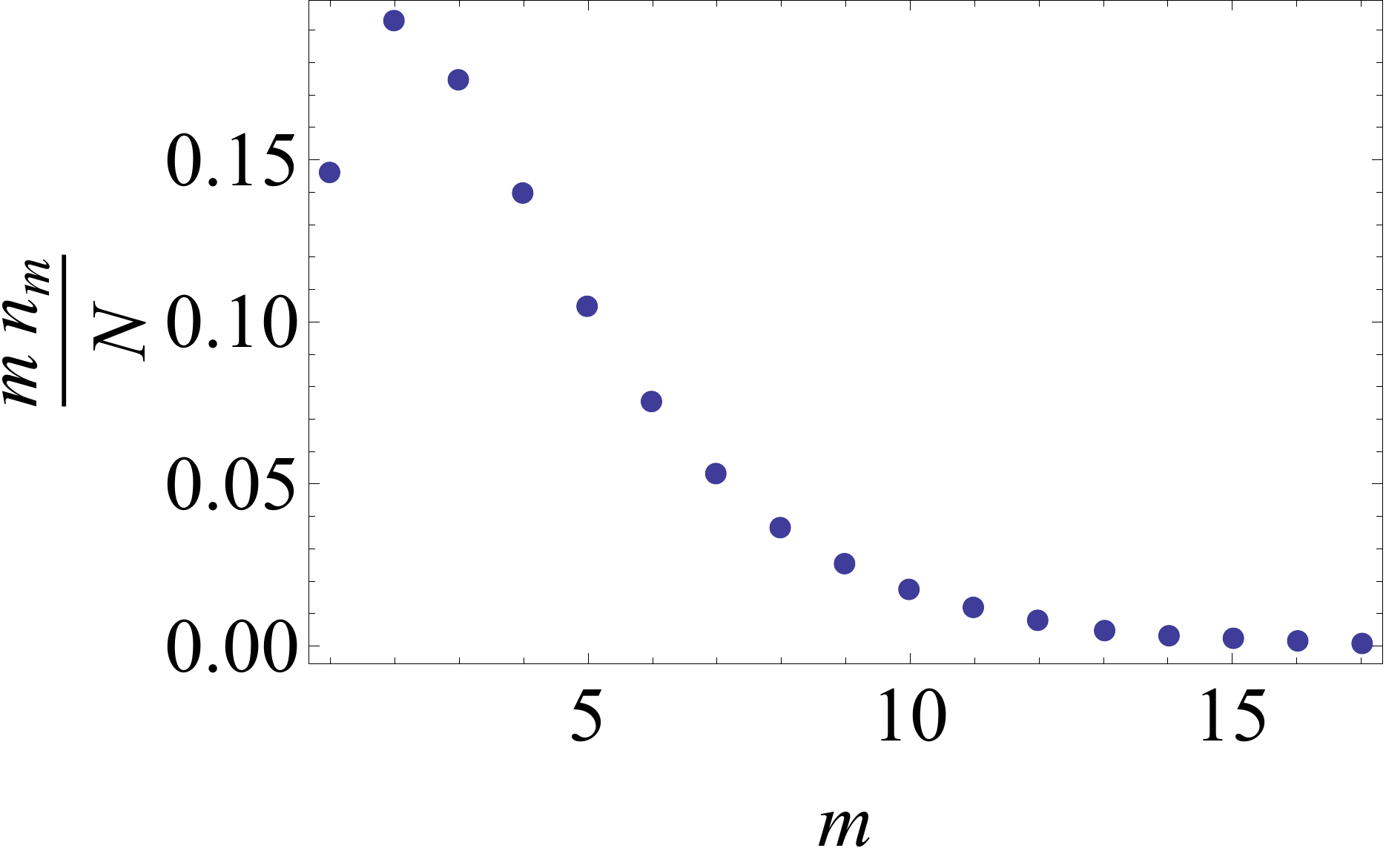}
} 

    \caption{%
        The plot (a) shows the distribution of the number $n_m/N$ of groups of $m$ particle-hole pairs in necklace diagrams 
         dominating $\mathcal{N}_N$. The plot (b) shows the probability   $m n_m/N$ that a given pair belongs to a group containing $m$ pairs.
           }%
\end{figure}

The saddle point solution can thus be written as
\bea\label{solsaddle}
\frac{n_m}{N} = A m {\cal T}_m e^{-\mu m},
\eea
where $1/A= d^2/d\mu^2[T(x=e^{-\mu})^2]= 1/0.778$, as follows from the constraint $\sum_m m n_m =N$. The resulting values for $n_m/N$ are shown in Fig.~\ref{fig:Distm1}. 
The probability that a given pair is created in a scattering process involving a total of $m$ pairs in the same localization volume is plotted in Fig.~\ref{fig:Distm2}. We see that most pairs are created together with a few more pairs within the same localization volume.

Plugging (\ref{solsaddle}) into the saddle point for ${\cal N}_N$, we find the number of diagrams to grow like (dropping pre-exponential factors)
\bea\label{Ntot}
{\cal N}_N \approx  ({\cal K}e^\mu)^N \approx (10.6 \, {\cal K})^N. 
\eea

This result is based on the approximation that we only allow for diagrams with a necklace structure, where groups of $m_i$ pairs are connected by a single scattering between subsequent localization volumina. Performing the calculation without this restriction is difficult since it is less easy to control the spatial constraints. However, we can easily obtain an upper bound by realizing that all possible diagrams consist in all geometrically possible labellings of trees of size $N$. The number of trees grows as $(27/4)^N$. For each label one has roughly $3{\cal K}$ possibilities, as the pair must lie in a localization volume adjacent to or identical with the one of the pair preceding it on the tree. This yields the simple upper bound
\bea
\label{upperbound}
{\cal N}_N < (3\cdot \frac{27}{4} {\cal K})^N \approx (20.25 \, {\cal K})^N, 
\eea
which yields a growth factor which is only about a factor of $2$ bigger than the much more conservative estimate (\ref{Ntot}).
Let us thus write
\bea
{\cal N}_N \approx (C \, {\cal K})^N,
\eea
with 
\bea
10.6 < C < 20.25.
\eea

\subsection{Effect of Fermi blocking}
\label{Fermibocking}

The above counting is still not entirely complete. Indeed, eventually the operators we have constructed should act on some many body states, and get annihilated when attempting to create particles on occupied states or holes on already empty states. In an infinite temperature state, and at a filling fraction $\nu$ each particle-hole creation operator has a chance to annihilate the state with probability $1- \nu(1-\nu)$, or, in other words, only a fraction of $[\nu(1-\nu)]^N$ of all operators will not annihilate a typical infinite temperature state. One should thus modify the number of relevant diagrams to 

\begin{equation}\label{enne}
 \mathcal{N}_N\to \mathcal{N}_N {\simeq} \tonde{C \ \nu(1-\nu) \mathcal{K}}^N.
\end{equation}
In the next section  we use this result to determine the radius of convergence of the operator series. Similar considerations apply to finite temperature as we will discuss below.  

\subsection{Structure of the dominant operator terms}
Our result differs from the similar analysis in Ref.~\cite{gornyi2005interacting}. The main difference consists in our assumption that the sum of diagrams that add up to the amplitude of a given operator $O_{\cal I,J}$  is dominated by the biggest term (provided the considered amplitude is among the largest ones at that order). In contrast, the authors of \cite{gornyi2005interacting} assumed that the exponentially many diagrams have comparable amplitudes, but random signs, and applied the central limit theorem to the sum. Moreover, we allow for fluctuations of the number of pairs generated in each localization volume instead of imposing a homogeneous spatial density. We find that in the restricted set of necklace diagrams the optimal distribution of group sizes $m_i$s is peaked at values of order $O(1)$, but still clearly larger than one. Upon folding of the necklace, the number of pairs per localization volume will become even more significantly larger than 1.
Thus we see that multiple scattering processes within a localization volume significantly  enhance the delocalization tendency. This shows that the many-body problem is genuinely different from an effective one-body problem, in which a simple excitation  would propagate nearly ballistically, by shedding one particle-hole excitation in every localization volume.

\section{Estimate of the radius of convergence}
\label{sec:estimateradius}

We now have all the ingredients to estimate the probability of resonances at generation $N$, in order to prove that for $\lambda$ sufficiently small there are no delocalizing resonances and ($\ref{pres}$) holds true. 

Consider the probability in expression (\ref{short}). Using (\ref{touse}), we estimate:
\begin{equation}
 \mathbb{P} \tonde{\modul{ \tilde{\omega}_\Gamma} >z^N}\approx \tonde{{\frac{2e}{\sqrt{2 \pi}}}}^N \int_{\log \tonde{\frac{z}{\lambda  
 \eta_{\rm typ}}}}^{\infty}  C(\tilde{x},N)  \tilde{x}^N e^{-N\mathcal{F}(\tilde{x})} d \tilde{x}.
\end{equation}
Note that the large deviation result applies since $\tilde{x} \geq \log \tonde{\frac{z}{\lambda  \eta_{\rm typ}}} \gg 1$. Approximating the integral with the value of the integrand at the extremum, setting $z=1$ and neglecting sub-exponential terms in $N$ we obtain: 
\begin{equation}\label{finalprob}
\begin{split}
  \mathbb{P} \tonde{\modul{ \tilde{\omega}_\Gamma} >1} \approx \tonde{{\frac{2e}{\sqrt{2 \pi}}}\log \tonde{\frac{1}{\lambda  
 \eta_{\rm typ}}}}^N e^{- N \quadre{ \log \tonde{\frac{1}{\lambda  
 \eta_{\rm typ}}} +  O \tonde{1/\log \tonde{\frac{1}{\lambda  
 \eta_{\rm typ}}}}}}.
 \end{split}
\end{equation}
Substitution of (\ref{finalprob}) and (\ref{enne}) into (\ref{short}) yields:
\begin{equation}
 \sum_{
  \begin{subarray}{l}
\hspace{.6 cm}\mathcal{I \neq J}\\
|\mathcal{I}|=N+1=|\mathcal{J}|  \end{subarray}} \mathbb{P} \tonde{
\modul{\mathcal{A}^{(\alpha)}_{\mathcal{I,J}}}>1} \simeq  \text{exp}\quadre{N \log \mathcal{G}(\lambda, \mathcal{K}) +{o}(N)}, 
\end{equation}
with
\begin{equation}\label{coeff}
 \mathcal{G}(\lambda, \mathcal{K})=\nu(1-\nu) \frac{ 2 e\,  C  \eta_{\rm typ}}{ \sqrt{2 \pi  }} \lambda \mathcal{K} \log \tonde{\frac{ {1}}{\lambda \eta_{\rm typ}}}.  
\end{equation}

Taking into account (\ref{sum1}) and (\ref{chain}), we finally obtain:

\begin{equation}\label{definitiva}
\mathbb{P} \tonde{ \forall N>N^*,  
\sum_{
  \begin{subarray}{l}
\hspace{.6 cm}\mathcal{I \neq J}\\
|\mathcal{I}|=N+1=|\mathcal{J}|  \end{subarray}} 
\modul{\mathcal{A}^{(\alpha)}_{\mathcal{I,J}}}<1}= \prod_{N>N^*} \text{exp}\quadre{- e^{ {N \log \mathcal{G}(\lambda, \mathcal{K}) +{o}(N)}}}.
\end{equation}

If $\mathcal{G}(\lambda, \mathcal{K})<1$, then, for $N^*$ sufficiently big, each of the factors in (\ref{definitiva}) is arbitrarily close to $1$. Therefore, their product converges to $1$ in the limit $N^* \to \infty$ (see also \cite{ThoulessReview} for a similar reasoning). This allows us to conclude that, for all values of $\lambda$ for which $\mathcal{G}(\lambda, \mathcal{K})<1$ holds,  ($\ref{pres}$) holds, too, and the series in operator space (\ref{ansatz}) converges to a quasi-local operator. In this regime, the excitation of the single particle level $\alpha$, localized at $\vec{r}_\alpha$, is very unlikely to create a distant disturbance at $\vec{r}_\beta$ with large $L=|\vec{r}_\beta-\vec{r}_\alpha|$, its probability tending to zero exponentially as $L\to \infty$: there is no diffusion at small $\lambda$.

The critical value for $\lambda$ is given by $ \mathcal{G}(\lambda_c, \mathcal{K})=1$.  For large $\mathcal{K}$, it equals to:
\begin{equation}\label{lcritical}
 \lambda_c =\frac{\sqrt{2 \pi}}{C \hspace{.1 cm} \nu(1-\nu) 2 e\, }\frac{1}{\mathcal{K} \log{\mathcal{K}}},
\end{equation}
where we used $\eta_{\rm typ}=1/e$. 

\subsection{Comparison with a single particle on the Bethe lattice}
It is interesting to note that the delocalization threshold (\ref{lcritical}) looks identical to the critical ratio between hopping  and disorder strength for a single particle problem on a Bethe lattice (see  Eq.(5.8) in \cite{abou1973selfconsistent}) with effective connectivity  $\mathcal{K}_{\text{eff}}=\nu(1-\nu) (C/ \sqrt{2 \pi}) \mathcal{K}$, which is a significantly bigger than the connectivity associated with each vertex, $\nu (1-\nu) \mathcal{K}$.
This reflects the fact that in the many-body problem the same final state can be reached with many different decay processes.
The results are nevertheless similar, because both problems are dominated by very few resonant paths, whereby the large local connectivity in the many-body problem ensures that different resonant paths are likely to be uncorrelated, even if they lead to the same final state.

\subsection{Possible implications for delocalization in higher dimensions}
According to the above calculation, in the dominating decaying processes only groups of $\mathcal{O}(1)$ particle-hole pairs are created at  the same time in a localization volume. This suggests that the necklace-type diagrams are diffusing back and forth a lot. This contrasts with the model of BAA, where the hopping strength between adjacent volumina was assumed to be parametrically smaller than $\lambda$, which favored the particle-hole creation cascade to fully explore a localization volume before moving on to the next volume. The latter led them to conjecture a critical exponent for the localization length  in higher dimensions
by relating the decay processes of single particle excitations to self-avoiding random walks.
  This scenario hardly holds in our model, as the optimal processes are not of this kind.

\section{Finite temperature}
\label{sec:finitetemp}

So far we have been discussing the convergence of the  expansion of integrals of motion in the forward approximation. If the expansion converges, we have succeeded in constructing a complete set of quasi-local conserved quantities which entail the absence of transport in whatever state the system is, in particular at any temperature, including the limit $T\to \infty$. Note again, that the latter limit is meaningful because we work on a lattice on which the energy density is bounded.

An interesting question arises when we ask about transport at finite temperature, and the possibility of a MBL transition as a function of temperature, as predicted by BAA. How would this reflect at the level of integrals of motion? If there is a finite temperature transition, one expects that the localized low $T$ phase is still governed by local conservation laws which inhibit transport, while local integrals of motion do not exist at higher temperature. Clearly the latter rules out the convergence of the conserved operators in the operator norm. Rather one has to invoke that the norm of operators $\cal O_{I,J}$, when restricted to typical low temperature states,  becomes exponentially small in $N=|\cal I|$, if the index sets $\cal I, J$ contain a finite fraction of hole excitations above $E_F+T$ or particle excitations below $E_F-T$. This effect may enhance the convergence of the series expansion. This is certainly so at the level of the forward approximation where the temperature $T$ essentially replaces the bandwidth in the analytical estimates of our expansion. This will lead to a larger domain of (weak) convergence of the operator expansion, suggesting the possibility of a delocalization transition at finite temperature.

A similar consideration shows that the transition (\ref{lcritical}) at $T=\infty$ takes place in a regime where the operator expansion is not convergent in the operator norm, but converges only weakly on typical high energy states. This is due to the Fermi blocking discussed in Sec.~\ref{Fermibocking}.

 \begin{figure}[ht!]
     \centering
  \centering
 \includegraphics[width=.8\textwidth, angle=0]{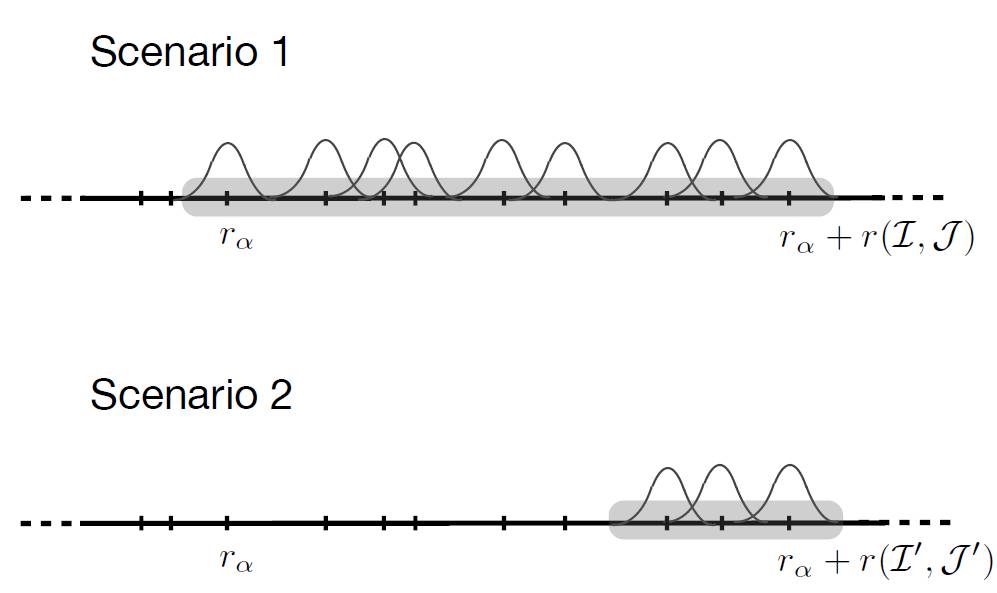}

  \caption{\footnotesize{Pictorial representation of the supports of two different operators $\mathcal{O}_{\mathcal{I,J}}$ and $\mathcal{O}_{\mathcal{I',J'}}$ contributing to the series expansion ($\ref{ansatz}$). In the pictures, the wave-functions are the single particle states contributing to ($\mathcal{I,J}$) and ($\mathcal{I', J'}$). Both operators involve degrees of freedom whose maximal distance to the localization center $r_\alpha$ is the same: $r(\mathcal{I,J})=r(\mathcal{I',J'})$; however, the length of the support $N$ of the operators (shaded in the picture) increases when $N$ grows in the first case, while it remains bounded in the second case. }}
 \label{fig:delocscenarios}
  \end{figure}

However, a different scenario  is possible as well. The operator series in Eq.~(\ref{ansatz}), or subsequences of it,  can diverge for two reasons: (i) Either the amplitude of terms with growing $N$ do not decrease sufficiently fast, and thus the diameter of the support of these terms grows indefinitely. (ii) There can be subsequences of (\ref{ansatz}) whose terms have bounded index level $N$, but supports which wander off to infinity. These two possibilities are illustrated in Fig.~\ref{fig:delocscenarios}.  

Possibility (i) is what is obtained within the forward approximation. The fraction of terms at $\lambda_c$, which survive when applied on finite $T$ states,
decreases rapidly with $N$. However, such a projector would not affect the convergence properties of a subsequence of type (ii). Upon restricting to finite $T$ states the norm of the relevant operators is typically reduced  by a factor, which remains  bounded from below. Therefore the series will continue to diverge despite the projection.

To address the question of whether or not a finite temperature transition is possible one has to consider the interaction strength $\lambda_c$ at which the infinite $T$ transition takes place, i.e., where the integrals of motion delocalize. If there is a subsequence of type (ii), which diverges at this point, the delocalization transition is a function of $\lambda$ only, but independent of $T$. In the delocalized phase ($\lambda>\lambda_c$) transport would always remain finite, even though it  may become very inefficient and strongly activated at low $T$.
If instead there is no subsequence with bounded index cardinality, which diverges at $\lambda_c$, a transition in temperature should be expected, as predicted by BAA. Such a transition was recently reported by a numerical study~\cite{Kjall:2014fj}.
 
Physically the scenario (ii) corresponds to transport and delocalization driven by rare, compact, but mobile regions with a local "temperature" above the putative $T_c$. At first sight one is tempted to rule this out because one would expect such a hot bubble to diffuse and loose its extra energy forever to the environment. However, the environment being in the supposed MBL phase cannot transport the extra energy to infinity, and thus there should be a finite  recurrence time until the hot bubble forms again. Whether such a bubble would nevertheless have to remain localized, or whether its internally delocalized state would allow it to move around is a difficult open question. Recently, it was argued that big enough bubbles could undergo resonant delocalization~\cite{Roeck2014:Scenario}. At the level of integrals of motion these two scenarii translate into the above dichotomy about critical subsequences.
     
Note that a divergence of type (ii) by a set of operators with bounded support is made less likely by the large parameter $\mathcal{K}$. We in fact invoked this large parameter to neglect these terms, similarly as BAA. However, it is difficult to exclude that there is no such divergent subsequence which contributes with a finite, but with a relative weight which is parametrically small in ${\cal K}$. In that case, numerical approaches such as \cite{Kjall:2014fj, Bauer:2013rw} would not capture this divergence.

It would be interesting to revisit the question of the finite $T$ transition also as a function of density. In the low density limit, the effective connectivity  $\mathcal{K}_{\rm eff}$ (resulting from projection onto typical states) can be reduced to $K_{\rm eff }\ll 1$, in which propagation channels of type (ii) become parametrically favorable, and may be the ones to induce delocalization - if interactions can induce a transition at all under such circumstances. 

\section{Conclusion}

In this work we have constructed explicit quasi-local integrals of motion within the weakly interacting regime, which we argued to imply the absence of any d.c. transport. We reduced the problem of constructing such operators to a non-Hermitian hopping problem in operator space, an idea that we hope to have potential for further more rigorous studies. We have also obtained an explicit recipe for constructing generalized occupation numbers of a Fermi insulator order by order in perturbation theory.

We have used the large parameter ${\cal K}$ (proportional to the number of sites in a single particle localization volume) to concentrate on processes where  one more particle-hole pair is created at every order of perturbation theory. Within this forward approximation, and based on an analysis of rare resonances at large distance, we found an analytical estimate of the radius of convergence of this perturbative construction, yielding a critical value of the reduced interaction strength
$ \lambda_c =\sqrt{2 \pi}/ \tonde{C \hspace{.1 cm} \nu(1-\nu) 2 e\, \mathcal{K} \log{\mathcal{K}}}$ with $10.6 < C < 20.25$,  at infinite $T$ and filling fraction $\nu$, similar to the prediction by BAA based on the analysis of the life time of a single particle injection.

We believe that the spatial structure of our integrals of motion provides a good picture for the "quantum avalanche" created by injection of  an extra particle.  We have found that the optimal way of its propagation is by exciting a necklace of groups of $O(1)$ particle-hole pairs per localization volume. Due to the meandering of the necklace structure, several groups of such pairs may be created in the same localization volume, an effect which is enhanced in low dimensions.

The convergence of our construction for the local integrals of motion implies the absence of transport and equilibration at any temperature and density. Taken as such, it appears to be blind to potential phase transitions upon varying those parameters. However, projecting the operator series onto typical states with thermal single particle occupations, one may discuss the {\em weak} convergence of the operator expansion. In this vein, we have discussed the question of the existence of a genuine finite temperature transition, depending on the properties of the operator series at its critical point at $T=\infty$. Further investigations of this question would be interesting.

\section{Acknowledgment}
We would like to thank Denis Basko, David Huse, Vadim Oganesyan and Dimitry Abanin, for discussions. V.\ Ros thanks the Princeton Center for Theoretical Science at Princeton University for hospitality, where part of this work was done. A.\ Scardicchio is in part supported by the NSF grant PHY-1005429.

\appendix

\section{Imposing binary spectrum}
\label{app:binarity}

In this Appendix we show how one can modify, order by order in $\lambda$, the previously obtained integrals of motion in order to fix their spectrum to be that of occupation numbers, i.e., $\{0,1\}$. This is equivalent to the condition:
\begin{equation}\label{fullCond}
I_\alpha^2=I_\alpha.
\end{equation}

This procedure leads to a modified expansion for $I_\alpha$:
\begin{equation}\label{modExp}
I_\alpha=n_\alpha+\sum_{m \geq 1} \lambda^m \Delta B_\alpha^{(m)},
\end{equation}
with $\Delta B_\alpha ^{(m)}$ given explicitly in Eqs.(\ref{shift}),(\ref{ExpDiag}).

In the following, we work by induction on $m$. We set $\Delta B^{(0)}_\alpha=n_\alpha$ and we omit the index $\alpha$ for simplicity. We define the truncation to $m$th order of $I$:
\begin{equation}
I^{\leq m} \equiv n+\sum_{i=1}^m \lambda^i \Delta B^{(i)},
\end{equation}
and assume that the property (\ref{fullCond}) holds to order $O(\lambda^{m-1})$, namely:
\begin{equation}\label{binOr}
(I^{\leq m-1})^2=I^{\leq m-1} + \ o (\lambda^{m-1}).
\end{equation}
 Note that $I^{\leq 0}$ is naturally binary, with $(I^{\leq 0})^2=I^{\leq 0}$. 

We denote with $ \Delta  \hat{I}^{(m)}$  the solution of the equation:
\begin{equation}\label{condCons}
[H_0,\Delta \hat{I}^{(m)}]+[U,\Delta B^{(m-1)}]=0
\end{equation}
in the subspace $O$, cf. Eq. (\ref{inversee}), and define
 \begin{equation}
 \hat{I}^{\leq m} \equiv I^{\leq m-1}+ \lambda^m \Delta \hat{I}^{(m)}.
 \end{equation}

The operator $\hat{I}^{\leq m} $ is not binary to order $\ O (\lambda^m)$; however, we show that it is possible to add to $ \Delta \hat{I}^{(m)}$ a suitably chosen operator $\Delta K^{(m)}$ in the kernel $K$ of the linear map $f(X)=[H_0, X]$, so that 
\begin{equation}
I^{\leq m} =  \hat{I}^{\leq m} + \lambda^m \Delta K^{(m)} \equiv {I}^{\leq m} + \lambda^m \Delta B^{(m)} 
\end{equation}
is binary to order  $\  O (\lambda^{m})$. To show this, it is sufficient to show that the difference $\tonde{\hat{I}^{\leq m}}^2- \hat{I}^{\leq m}$, truncated to order $\  O (\lambda^m)$, is an element of the subspace $K$, i.e.:
\begin{equation}\label{condBin}
[H_0,(\hat{I}^{\leq m})^2]=[H_0,\hat{I}^{\leq m}] +o(\lambda^{m}).
\end{equation}

This holds, since:
\begin{equation}\label{quadrato}
 \begin{split}
 (I^{\leq m})^2=&[(I^{\leq m-1})^2]_{m-1}+\lambda^m {\sum_{a=0}^{m} \Delta B^{(a)}\Delta B^{(m-a)}}+o(\lambda^m)\\
=&I^{\leq m-1}+\lambda^m {\sum_{a=0}^{m} \Delta B^{(a)}\Delta B^{(m-a)}}+o(\lambda^m), 
 \end{split}
\end{equation}
where $[X]_{m-1}$ denotes the restriction of the Taylor series of $X(\lambda)$  to terms up to order $\lambda^{m-1}$. Using the inductive step $m-1$ we have from (\ref{quadrato})
\begin{equation}\label{inde}
[H_0,(\hat {I}^{\leq m})^2]=[H_0,I^{\leq m-1}]+\lambda^m [H_0,{\sum_{a=0}^{m} \Delta B^{(a)}\Delta B^{(m-a)}}]+\ o (\lambda^m),
\end{equation}
where in the terms with $a=0, m$ we have replaced $\Delta \hat{I}^{(m)}$ with $\Delta B^{(m)}$, since Eq.(\ref{inde}) does not depend on the choice of $\Delta K^{(m)}$.
Given that
\begin{equation}
\begin{split}
[H_0,\Delta B^{(a)}\Delta B^{(m-a)}]&=\Delta B^{(a)}[H_0,\Delta B^{(m-a)}]+[H_0,\Delta B^{(a)}]\Delta B^{(m-a)}\nonumber\\
&=-\Delta B^{(a)}[U,\Delta B^{(m-a-1)}]-[U,\Delta B^{(a-1)}]\Delta B^{(m-a)},
\end{split}
\end{equation}
summing over $a$ we get
\begin{equation}
[H_0,{\sum_{a=0}^{m} \Delta B^{(a)}\Delta B^{(m-a)}}]=-[U,{\sum_{a=0}^{m-1} \Delta B^{(a)}\Delta B^{(m-a)}}].
\end{equation}
Using that (\ref{fullCond}) at the inductive step $m-1$ implies
\begin{equation}
\sum_{a=0}^{m-1} \Delta B^{(a)}\Delta B^{(m-a)}=\Delta B^{(m-1)},
\end{equation}
and using (\ref{condCons}), we find 
\begin{equation}
\begin{split}
[H_0,(\hat {I}^{\leq m})^2]&=[H_0,I^{\leq m-1}]+\lambda^m[H_0,\Delta \hat{I}^{(m)}]+\ o (\lambda^m)\\
&=[H_0,\hat{I}^{\leq m}]+\ o (\lambda^m),
\end{split}
\end{equation}
which proves (\ref{condBin}).

A simple computation shows that by choosing:
\begin{equation}
{I}^{\leq m} \equiv \hat{I}^{\leq m} + \lambda^m \tonde{1-2 \Delta{B} ^{(0)}}\quadre{\tonde{\hat{I}^{\leq m}}^2- \hat{I}^{\leq m}}_m
\end{equation}
the condition (\ref{binOr}) is fulfilled to order $\ O (\lambda^m)$. Eq. (\ref{ExpDiag}) follows from noticing that:
\begin{equation}
 \quadre{\tonde{\hat{I}^{\leq m}}^2- \hat{I}^{\leq m}}_m= \sum_{i=1}^{m-1} \Delta{B} ^{(i)}\Delta{B} ^{(m-i)} + \grafe{\Delta{B} ^{(0)}-\frac{1}{2}, \Delta \hat I^{(m)}}.
\end{equation}

\section{Local re-summation in the case of small denominators}
\label{app:resum}

In the following we present a simple example in which  the perturbative expansion in $\lambda$, Eq.~(\ref{ansatzpert}), diverges. 
Suppose that at order $n$ the series expansion contains the term: 
\begin{equation}
 J_n \equiv \mathcal{J}_n  O c_\alpha,
\end{equation}
 where $O= c_{i_1}^\dag \cdots c_{i_m}^\dag  c_{j_1} \cdots c_{j_{m-1}}$ is a string of operators with $i,j \neq \grafe{\alpha, \beta, \gamma,\delta}$, and that the amplitude  $\mathcal{J}_n=O(\lambda^n)$ therefore contains the energy denominator:
\begin{equation}
 \mathcal{J}_n  \propto \tonde{\sum_{k=1}^m \epsilon_{i_k}- \sum_{k=1}^{m-1} \epsilon_{j_k}-\epsilon_\alpha}^{-1} \equiv \tonde{\Delta \mathcal{E}}^{-1}.  
\end{equation}
Suppose $\Delta \mathcal{E}$ to be atypically small. One then easily finds a subsequence of the series (\ref{ansatzpert}), which contains arbitrarily high powers of the small denominator. Indeed, let us restrict the interaction to the term $U_{\alpha \beta, \gamma \delta} \tonde{c_\alpha^\dag c_\beta^\dag c_\gamma c_\delta + h.c.}$ in the interaction $U$; higher order terms in the perturbative expansion are obtained by subsequent application of (\ref{inversee}) to $J_n$; this produces:  
\begin{equation}
\label{Jeqs}
\begin{split}
 J_{n+1} &\equiv  \mathcal{J}_n \frac{U_{\alpha \beta, \gamma \delta}}{\Delta \mathcal{E} + \mathcal{E}_{\alpha \beta, \gamma \delta}}  \hspace{.2 cm}O c_\beta^\dag c_\gamma c_\delta \equiv  \mathcal{J}_{n+1}   \hspace{.2 cm}O c_\beta^\dag c_\gamma c_\delta,\\
  J_{n+2} &\equiv  -\mathcal{J}_{n+1} \frac{U_{\alpha \beta, \gamma \delta}}{\Delta \mathcal{E}} \hspace{.2 cm} O \tonde{n_\beta(1-n_\gamma)(1-n_\delta)+(1-n_\beta)n_\gamma n_\delta} c_\alpha,\\
  J_{n+3} &\equiv  \mathcal{J}_{n+1} \quadre{\frac{U_{\alpha \beta, \gamma \delta}}{\Delta \mathcal{E}} \hspace{.1 cm} \frac{U_{\alpha \beta, \gamma \delta}}{\Delta \mathcal{E} + \mathcal{E}_{\alpha \beta, \gamma \delta}}} O c_\beta^\dag c_\gamma c_\delta,
 \end{split}
\end{equation}
with $\mathcal{E}_{\alpha \beta, \gamma \delta}=\epsilon_\alpha+ \epsilon_{\beta} -\epsilon_\gamma -\epsilon_\delta$. By iteration of this procedure,  a sub-sequence of operators containing arbitrarily high powers of $\tonde{\Delta \mathcal{E}}^{-1}$ is generated, preventing the convergence of the series if the term in brackets is larger than $1$.

Divergences of this kind are of the same nature as local resonances encountered in single particle localization ~\cite{anderson1958absence}. They have to be properly re-summed for the series expansion to make sense. For example, all terms multiplying $O c^\dag_\beta c_\gamma c_\delta$  re-sum into a self-energy correction of the denominator in the first line of (\ref{Jeqs}):
\begin{equation}\label{resummedd}
 J \equiv \mathcal{J}_n  \quadre{\frac{U_{\alpha \beta, \gamma \delta}}{\tonde{\Delta \mathcal{E}+\mathcal{E}_{\alpha \beta, \gamma \delta}} - \frac{U_{\alpha \beta, \gamma \delta}^2}{\Delta \mathcal{E}}}}O c_\beta^\dag c_\gamma c_\delta \equiv \mathcal{J}  \hspace{.2 cm} O c_\beta^\dag c_\gamma c_\delta.
\end{equation}
The term in square brackets in (\ref{resummedd}) contains a very large self energy correction ${U_{\alpha \beta, \gamma \delta}^2}/{\Delta \mathcal{E}}$, which compensates the divergence in $\mathcal{J}_n$ when $\Delta \mathcal{E} \to 0$.

Self-energy corrections like this are neglected in the forward approximation. 
Their main effect is to weaken the role of small denominators: As noticed by Anderson,  small denominators essentially neutralize themselves by introducing enormous self-energies for the neighboring sites which then appear as very large denominators.~\cite{anderson1958absence}. The resummation thus increases the convergence as compared to the naive perturbative expansion in forward approximation in~\cite{anderson1958absence}. In single particle localization problems with large connectivity, the critical hopping is increased by a factor $e/2$~\cite{abou1973selfconsistent}, and a similar effect is expected here~\cite{Basko:2006hh}.

\section{Evaluating diagrams as sums over effective paths: a more involved example}
\label{app:effpaths}

\begin{figure}[ht!]
     \begin{center}
 \captionsetup[subfigure]{labelformat=empty}
  \centering
  \   \includegraphics[width=.7\textwidth, angle=0]{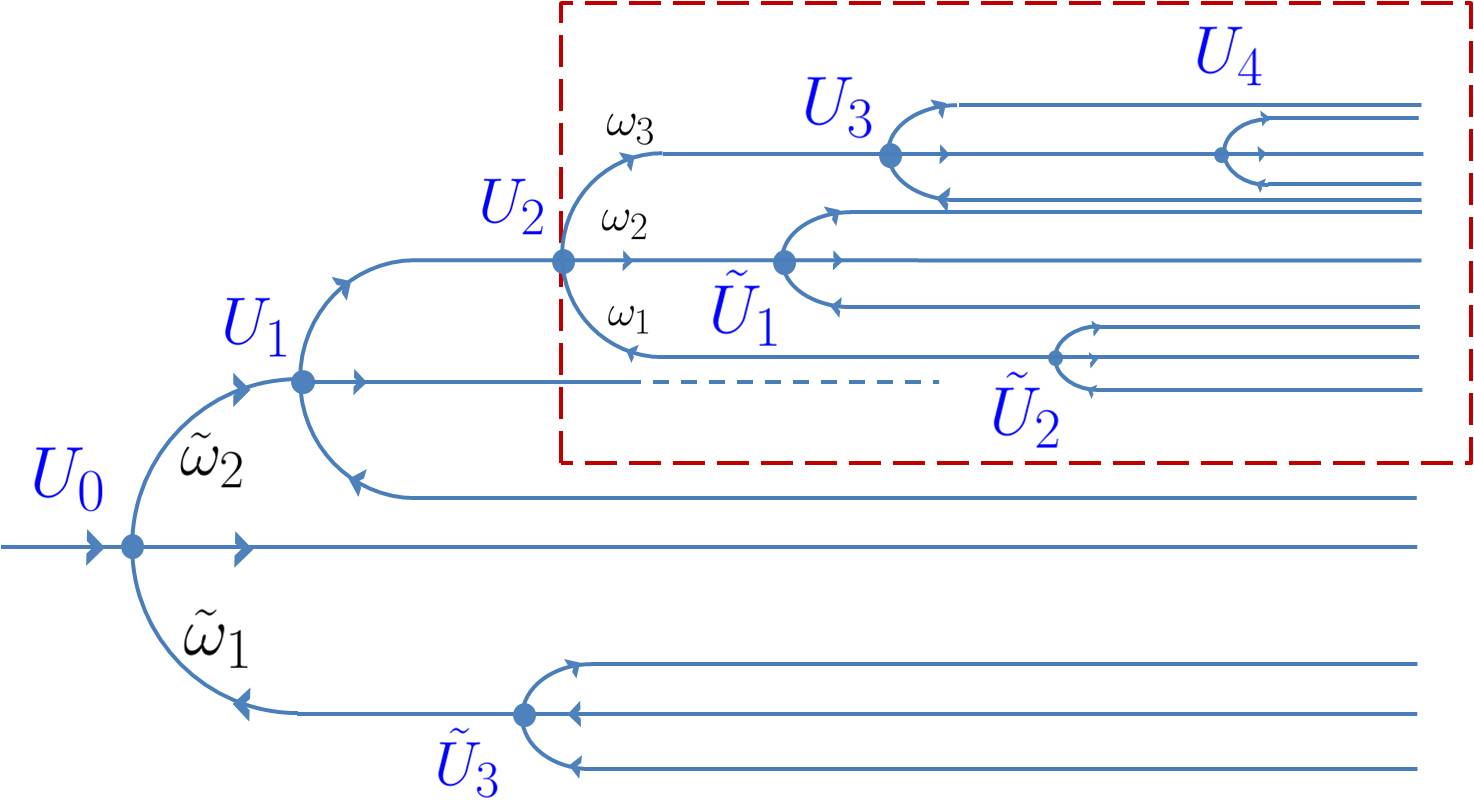}
             \end{center}
   \caption[]{ \footnotesize{A diagram with multiple branchings.}}\label{fig:ce}
   \end{figure}

As an additional example for the evaluation of diagrams as sums over effective paths, we give the explicit expression for  effective path weights associated to diagrams with the geometry of Fig.~\ref{fig:ce}.

For fixed indices on all segments, there are 105 different orderings of the interactions.\footnote{The interactions in the red dashed frame can be ordered in $15$ different ways, for each of which the interaction $\tilde{U}_3$ can be placed in $7$ different positions.} Their sum has the integral representation:
 \begin{equation}
 \begin{split}
  I_0\tonde{\grafe{U}}=\lim_{\epsilon \to 0}\int \frac{d \tilde{\omega}_1 d \tilde{\omega}_2  \hspace{.1 cm}\delta \tonde{\tilde{\omega}_1 +\tilde{\omega}_2 -\mathcal{E}_0}}{\tilde{\omega}_1^- (\tilde{\omega}_1^-+ \tilde{\mathcal{E}}_3)\tilde{\omega}_2^- (\tilde{\omega}_2^- + \mathcal{E}_1)} I_1\tonde{\omega= \tilde{\omega}_2 +  \mathcal{E}_1 + \mathcal{E}_2}, 
    \end{split}
    \end{equation}
      where $\omega^{-}_i \equiv \omega_i -i \epsilon$, and $I_1\tonde{\omega}$ is the integral representation of the sum of all the weights of the subdiagram in the dashed frame, with incoming energy $\omega$:
       \begin{equation}
I_1\tonde{\omega}=  \int \frac{d{\omega}_1 d{\omega}_2 d \omega_3  \hspace{.1 cm}\delta \tonde{{\omega}_1 +{\omega}_2+ \omega_3 -\omega}}{{\omega}_1^- ({\omega}_1^- + \tilde{ \mathcal{E}}_2){\omega}_2^- ({\omega}_2^- + \tilde{ \mathcal{E}}_1){\omega}_3^- ({\omega}_3^- + { \mathcal{E}}_3) ({\omega}_3^- + {\mathcal{E}}_3 +\mathcal{E}_4)}.
    \end{equation}

  By means of the residue theorem, $I_0$ can be  rewritten as the sum over only $8$ effective path weights:   
        \begin{equation}
        \label{appendix:effpath}
        \begin{split}
    I_0 \tonde{\grafe{U}}&=  \frac{1}{\tilde{\mathcal{E}}_3} \frac{1}{\mathcal{E}_0(\mathcal{E}_0+\mathcal{E}_1)}I_1(\mathcal{E}_0 +\mathcal{E}_1+\mathcal{E}_1)\\
    &-\frac{1}{\tilde{\mathcal{E}}_3} \frac{1}{(\tilde{\mathcal{E}}_3+\mathcal{E}_0)(\tilde{\mathcal{E}}_3+\mathcal{E}_0+\mathcal{E}_1)}I_1(\tilde{\mathcal{E}}_3+\mathcal{E}_0 +\mathcal{E}_1+\mathcal{E}_1)
    \end{split}
    \end{equation}      
   with
   \begin{equation}
    I_1(\omega)=  \frac{1}{\tilde{\mathcal{E}}_1\tilde{\mathcal{E}}_2 } \quadre{f(\omega)-f(\omega+ \tilde{\mathcal{E}}_1)-f(\omega+\tilde{\mathcal{E}}_2)+ f(\omega+\tilde{\mathcal{E}}_1+\tilde{\mathcal{E}}_2)}
    \end{equation}
and
\begin{equation}
    f(X)=\frac{1}{X (X +\mathcal{E}_3)(X+ \mathcal{E}_3+\mathcal{E}_4)}.
   \end{equation}
  
Note that as a function of the ${\cal E}_i$ and $\tilde{\cal E}_i$, $I_0$ has poles only due to denominators which involve the incoming energy ${\cal E}_0$, while $I_0$ remains regular as any of the $\tilde{\cal E}_{i}\to 0$, due to cancellations among different terms.

\begin{figure}[ht!]
     \begin{center}
 \captionsetup[subfigure]{labelformat=empty}
  \centering
  \   \includegraphics[width=.6\textwidth, angle=0]{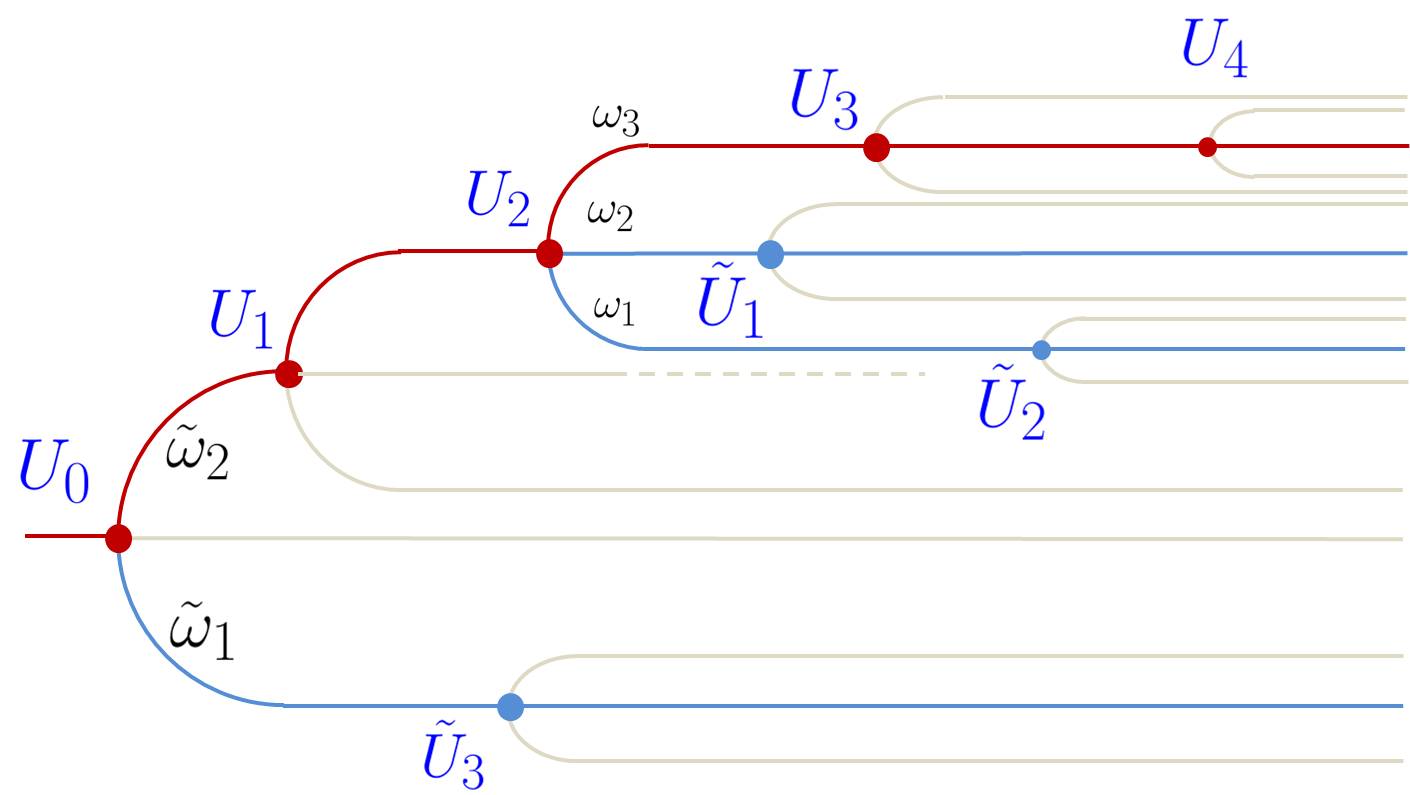}
             \end{center}
   \caption[]{\footnotesize{Diagram with colored branches. The branches with maximal and minimal (equal to zero) number of interactions along them are colored in red and gray, respectively.}}\label{fig:totef}
   \end{figure}
 
The minimal number of effective paths associated to a diagram equals to the product of the number of residua of any of the performed integrals. This number can be determined from the structure of the diagram using the following rules: First, one eliminates the final leaves which are not associated to auxiliary frequencies, since they do not contribute with poles in the integral representation (Fig.~\ref{fig:totef} represents the diagram of Fig.~\ref{fig:ce}, with these eliminated branches colored in gray). Then, one determines the directed path (branch) with the maximal number of interactions along it (red one in Fig.~\ref{fig:totef}). The auxiliary frequencies along this path are eliminated integrating the corresponding $\delta$-functions. All remaining branches contribute one more residua than interactions along the branch. In the example of Fig.~\ref{fig:totef}, the three branches that remain after eliminating the red one contribute $2$ residua each. The total number of effective paths is obtained by multiplying these numbers, which gives $2^3=8$ in the present case.

\begin{figure}[ht!]
     \centering
  \centering
  \subfloat[\hspace{2 cm}]{%
   \includegraphics[width=.3\textwidth, angle=0]{MaxBranched}
\label{MaxBranch2}
        } 
        \hspace{.8 cm}
            \subfloat[\hspace{0.5 cm}]{%
           \label{Colors}
           \includegraphics[width=0.32\textwidth]{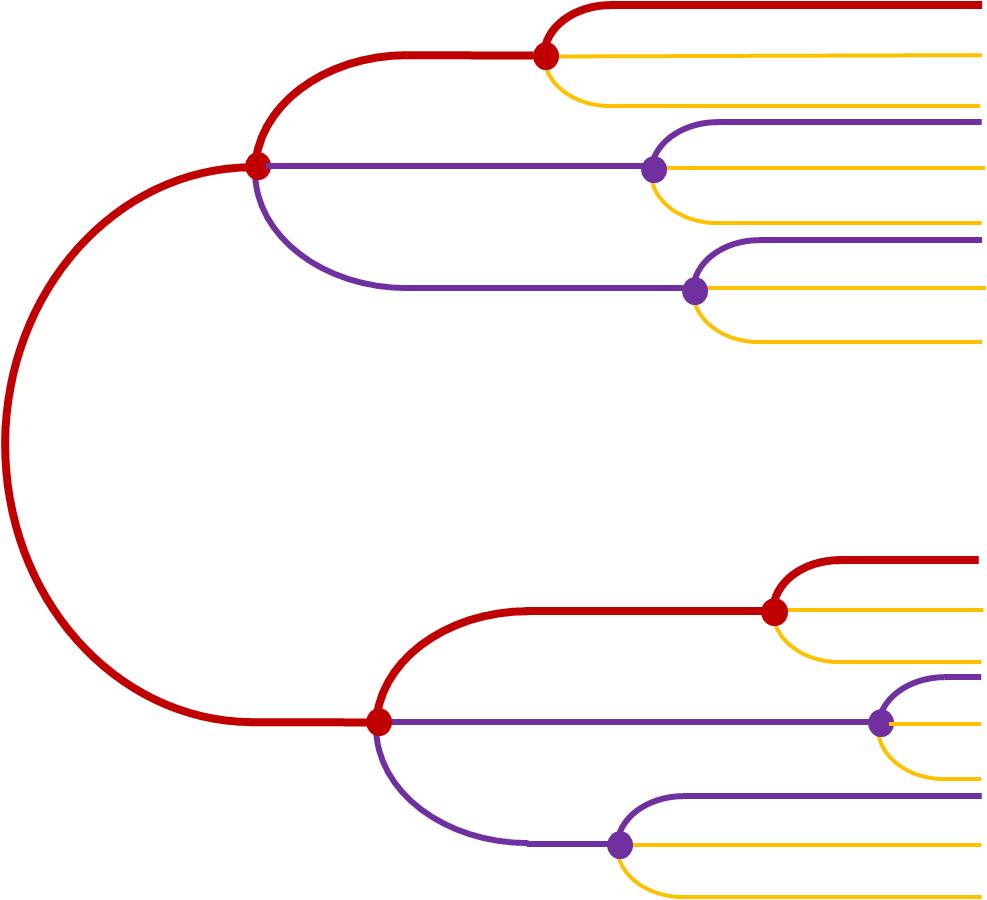}
        } %

  \caption{(a) Diagram with the maximal possible number of branchings. (b) Branches with the same number of interactions are drawn with the same color. }
     \label{fig:MaxBranch}
  \end{figure}

With the help of these  rules, we count the minimal number of effective paths associated to the maximally branched diagram with $N$ interactions, shown in Fig.~$\ref{MaxBranch2}$. We denote this number by $\modul{\mathcal{P}}$. 

The maximally branched diagram consists of two regular rooted trees with $L(N)\equiv {\log (N+1)}/{\log 3}$ generations. Since the weights of the two sub-diagrams factorize, we need to count only the effective paths associated to one of them, and square their number.  We therefore consider one sub-diagram, and organize its branches according to the number of interactions along it (in Fig.~\ref{Colors}, branches with the same number of interactions have the same color). The number $l$ of interactions along a branch ranges from $1$ to $L(N)-1$. There are $2 \cdot 3^{L(N)-1-l}$ branches with $l$ interactions; each of them contributes with $(l+1)$ residua, yielding a total number of
\begin{equation}\label{numeffweights}
\begin{split}
 \prod_{l=1}^{L(N)-1} 2  (l+1) 3^{L(N)-1-l}=L(N)! 3^{\frac{(L(N)-1)^2}{2}} \tonde{\frac{2}{\sqrt{3}}}^{L(N)-1}
 \end{split}
\end{equation}
terms. The total number of effective paths associated to the diagram of Fig.~\ref{MaxBranch2} is the square of this number, which grows as: 
\begin{equation}
 \modul{\mathcal{P}}= \text{exp} \quadre{{(\log N)^2}\log 3 + \ O \tonde{\log N \log (\log N)} }.
\end{equation}
As claimed in the main text, this number is sub-exponential in $N$.

\section{Probability of large deviations in products of correlated denominators}
\label{app:largedev}

Here we derive the probability of large deviations of effective path weights, i.e., the product of correlated denominators, as they occur in perturbation theory in the forward approximation. 

We denote by  $s_k= x_1 + \cdots +x_k$ the partial sums of i.i.d. random variables $x_i\equiv \mathcal{E}_i/\delta_\xi$. Let us assume the $x_i$ to be unit Gaussian variables with probability density
\begin{equation}
 \begin{split}
  f(x)= \frac{1}{\sqrt{2 \pi} } e^{-\frac{x^2}{2}}.
 \end{split}
\end{equation}
Consider  the distribution function $P_N(y)$ of the random variable 
\begin{equation}\label{vari}
Y_N \equiv \log\left( \prod_{i=1}^N \frac{1}{|s_i|}\right) = -\sum_{i=1}^N \log |s_i|,                                                                             
\end{equation}
and  its generating function $G_N(k)$
\begin{equation}
 G_N(k) \equiv \mathbb{E} \quadre{e^{-k Y_N}}.
\end{equation}

Let us compute $G_N$ for $N \gg 1$. We start by taking the expectation value over the joint distribution of $x_i=s_{i}-s_{i-1}$ ($s_0\equiv 0$):
\begin{equation}\label{genProd}
 G_N(k) = \int \prod_{i=1}^{N} f(s_i-s_{i-1}) e^{k \log |s_i|} ds_i= \int   {\mathcal{O}_k^{N-1} \quadre{f}}(s_N)|s_N|^k ds_{N},
\end{equation}
where the integral operator $\mathcal{O}_k \quadre{\cdot}$ acting on a function $g$ is given by
\begin{equation}
 \mathcal{O}_k\quadre{g}(s)= \int f(s-x) |x|^k g(x) dx.
\end{equation}
 
Consider now the basis of even functions:
\begin{equation}
g_n(x)= \frac{e^{-\frac{x^2}{2}} x^{2 n}}{\sqrt{2 \pi (2n)!}}, \quad \quad n=0, 1, \dots\,.
\end{equation}
In this basis the linear action of $\mathcal{O}_k$ is given by:
\begin{equation}\label{fir}
  \mathcal{O}_k\quadre{g_n}(x) = \frac{1}{2 \pi} \int e^{-\frac{1}{2}( x-y)^2} e^{-\frac{y^2}{2}} |y|^{k+2n}  dy = \sum_{m\geq 0} O_{mn}(k)g_{m}(x),
  \end{equation}
  
with the matrix
\bea\label{matele}
O_{mn}(k)  = \frac{1}{\sqrt{2\pi}} \frac{\Gamma \tonde{\frac{1+k}{2}+n+m}}{\sqrt{(2 m)! (2n)!}}.
\eea

From (\ref{genProd}) we thus readily obtain the following expression for $G_N(k)$:
 \begin{equation}\label{gentotalexp}
   \begin{split}
{G}_{N}(k)
&= \sum_{m=0}^{\infty} \left(O(k)^{N-1}\right)_{m0}  a_m,
   \end{split}
  \end{equation}
with $a_m(k) =\int_{-\infty}^\infty  g_m(s_N) |s_N|^k \, ds_N  = \frac{2^{\frac{k+1}{2}+m}}{\sqrt{2 \pi (2m)!}} \Gamma \tonde{\frac{k+1}{2}+m}$.

 The matrix $O_{mn}(k)$ can be interpreted as a $k$-dependent Hamiltonian describing a  particle hopping  on a semi-infinite open chain with sites labeled by integers $m=0,1,2, \dots\,$.

The large $N$ behavior of $\log G_N$ is dominated by the largest eigenvalue $\lambda_{\text{max}}(k)$ of  $\mathcal{O}$. Since for any $k>-1$, $O(k)$ is symmetric and positive definite, the Perron-Frobenius theorem ensures that $\lambda_{\text{max}}(k)$ is positive and unique, and
\begin{equation}\label{asimptotica}
\begin{split}
 G_N(k) \approx c(k)\quadre{{\lambda_{\text{max}}(k)}}^{N-1} ,
 \end{split}
\end{equation}
where 
$c(k) = \phi_{\text{max},0}\cdot \sum_{m\geq 0} a_m \phi_{\text{max},m} $, and 
$\phi_{\text{max}}$ is the normalized eigenvector corresponding to $\lambda_{\text{max}}$. 

\begin{figure}[h!]
    \centering
    \includegraphics[width=1.\textwidth, angle=0]{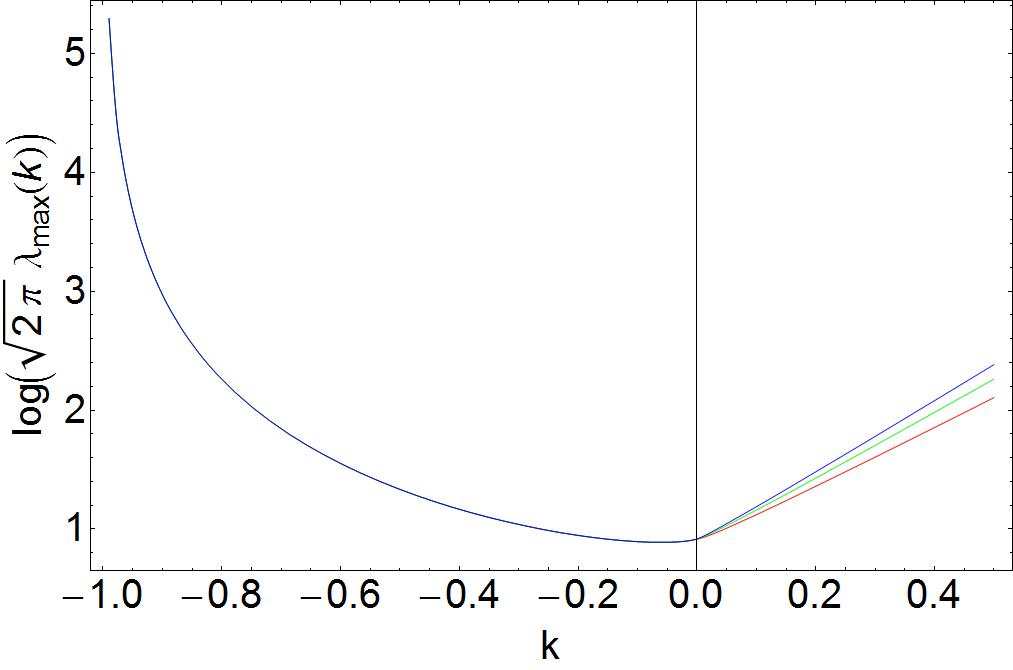}
\caption{{Maximal eigenvalue $\log \lambda_{\text{max}}(k)$ computed for truncated matrices $O(k)$ with basis sets of size $L=120$ (red), $L=200$ (green), $L=300$ (blue). Close to the singularity $k=-1$, $\lambda_{\text{max}}(k)$ converges rapidly with $L$.}}\label{Max1}
\end{figure}

Numerical results for the maximal eigenvalue are shown in Fig.~$\ref{Max1}$. They are obtained by truncating $O$ to an increasing set of basis states (or chain of sites) $m\leq L$. For $k$ close to the singularity $k=-1$ the results rapidly converge with increasing size $L$. 
In this region, we can extract information on the limiting curve $\lambda_{\text{max}}(k)$ from the truncated chain.  In particular, we see from the plot that both the function $\log \lambda_{\text{max}}(k)$ and its negative slope diverge at $k=-1$, which will also follow form the analysis below. Hence, $k\gtrsim -1$ is the relevant region for the saddle point approximation of Eq.~(\ref{cramer}), if very large deviations $\tilde{y} \gg 1$ are considered.

Due to the proximity to a logarithmic divergence at $k=-1$, to order $O(1+k)$ the eigenstate $\phi_{\text{max}}$ for $k \sim -1$ is localized 
 on the first site ($n=0$) of the corresponding hopping chain:
 \begin{equation}\label{order0vec}
 |\phi_{\text{max}} \rangle \simeq |0 \rangle,
 \end{equation}
with an eigenvalue 
\begin{equation}\label{order0val}
 \lambda_{\text{max}}(k) \simeq O(k)_{00} =\frac{1}{\sqrt{2 \pi}}\Gamma \tonde{\frac{1+k}{2}}.
\end{equation}

\begin{figure}[h!]
    \centering
    \includegraphics[width=1.\textwidth, angle=0]{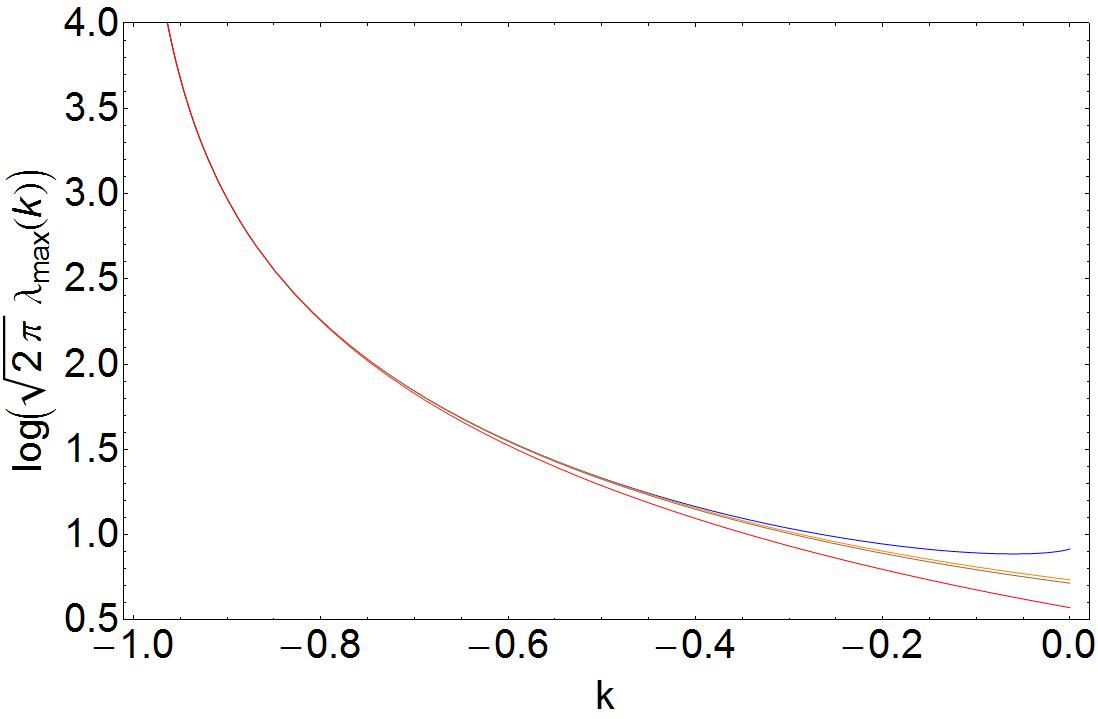}
\caption{{Comparison between $\log \lambda_{\text{max}}(k)$ obtained numerically for the truncated matrix (with $L=300$ basis functions) and the analytic expression $\log[\Gamma \tonde{\frac{k+1}{2}} +\sqrt{2 \pi}\delta \lambda(k)]$ with $\sqrt{2 \pi}\delta \lambda(k)$ expanded  at zeroth (red), first (brown) and second (orange) order in $(k+1)$. }}\label{comparison}
\end{figure}

Corrections to the maximal eigenvalue $(\ref{order0val})$ can be evaluated perturbatively in the matrix elements $O_{ik\neq 00}$ ($\ref{matele}$), which yields
\begin{equation}\label{fullLambda}
\begin{split}
 \lambda_{\text{max}}(k) &=\frac{1}{\sqrt{2 \pi}}\Gamma \tonde{\frac{1+k}{2}} + \lambda^{(2)}_{\text{max}}(k) + \lambda^{(3)}_{\text{max}}(k) + \dots \\
 &\equiv  \frac{1}{\sqrt{2 \pi}}\Gamma \tonde{\frac{1+k}{2}} +\delta \lambda(k).
 \end{split}
\end{equation}
One can show that $\delta\lambda(k)$ is analytic around $k=-1$ and satisfies $\delta\lambda(k \to -1) \to 0$. This is due to the fact that in $n$th order perturbation theory $\lambda_{\text{max}}^{(n)}$ is proportional to denominators of the form $1/O_{00}^{n-1} \sim (k+1)^{n-1}$.
The leading term in $\delta \lambda(k)$ results from:
\begin{equation}
 \lambda^{(2)}_{\text{max}}(k)= \sum_{m=1}^{\infty} \frac{ (2\pi)^{-\frac{1}{2}}\quadre{\Gamma \tonde{\frac{1+k}{2}+m}}^2}{\Gamma \tonde{\frac{1+k}{2}} \tonde{2m}! -  \Gamma \tonde{\frac{1+k}{2}+2m}}= \frac{1}{\sqrt{2 \pi}}\frac{\pi^2}{36} (k+1) + O (k+1)^2.
\end{equation}
A plot of the corrections to the maximal eigenvalue (\ref{order0val}) is given in Fig.~$\ref{comparison}$.

The inverse Laplace transform of the generating function is governed by
\bea\label{cramer2}
 \phi(\tilde{y},k) &:=& \tilde{y} k + \lim_{N\to \infty} \frac{G_N(k)}{N} \nn\\
 &=& \tilde{y} k + \log \quadre{\Gamma \tonde{\frac{k+1}{2}} + \sqrt{2 \pi} \delta\lambda(k)}- \frac{1}{2} \log 2 \pi.
\eea
It has a saddle point at $k=k^*(\tilde{y})$, determined by 
\bea\label{toinv2}
 \tilde{y}&=&-\frac{d}{d k} \grafe{\log \quadre{\Gamma \tonde{\frac{1+k}{2}} + \sqrt{2 \pi}\delta\lambda(k)}}_{k=k^*(\tilde{y})}\\
 &=& -\grafe{\frac{1}{2}\psi^{(0)} \tonde{\frac{1+k}{2}} \quadre{1+ \frac{\sqrt{2 \pi} \delta\lambda(k)}{\Gamma \tonde{\frac{1+k}{2}}}}^{-1} + \frac{\sqrt{2 \pi}\delta\lambda'(k)}{\Gamma \tonde{\frac{1+k}{2}}+\sqrt{2 \pi}\delta\lambda(k)}}_{k=k^*(\tilde{y})},\nn
\eea
where $\psi^{(0)}(x) \equiv d\log[\Gamma(x)]/dx$.

To isolate the singularity in $k=-1$ we use the Laurent expansion of $\psi^{(0)}(x)$ around $x=0$:
\begin{equation}
\psi^{(0)} \tonde{\frac{1+k}{2}}=-\frac{2}{k+1}  -\gamma+\frac{\pi^2}{12}(k+1) +\ O ( (k+1)^3), 
\end{equation}
where $\gamma$ is the Euler constant.
This allows us to recast ($\ref{toinv2}$) in the following form:
\begin{equation}
\tilde{y}= \frac{1}{k^*+1}+ Q(k^*+1).  
\end{equation}
Here, $Q(\cdot)$ is an analytic function with expansion:
\begin{equation}
 Q(x)= \frac{\gamma}{2}-\frac{\pi^2}{18}x+ \ O(x^2).
\end{equation}
 This yields the equation
\begin{equation}\label{itee}
 1+k^*= \frac{1}{\tilde{y}} \tonde{1-\frac{Q(k^*+1)}{\tilde{y}}}^{-1},
\end{equation}
which can be solved by iteration as an expansion in $1/\tilde y$:
\begin{equation}\label{serieZ}
 \begin{split}
 1+k^*(\tilde{y})&=\frac{1}{\tilde{y}} + \frac{\gamma}{2}\frac{1}{ \tilde{y}^2} +\tonde{\frac{\gamma^2}{4}-\frac{\pi^2}{18}} \frac{1}{\tilde{y}^3}+\ O \tonde{\frac{1}{\tilde{y}^4}}.   
 \end{split}
\end{equation}

Expanding ($\ref{cramer2}$) in powers of $k+1$ and substituting ($\ref{serieZ}$) we find: 
\begin{equation}
\begin{split}
 \phi(\tilde{y}, k^*(\tilde{y}))=& -\tilde{y} +\log \tilde{y}  -\frac{1}{2} \log \left(\frac{\pi}{2 e^2}\right) -\frac{\gamma}{2 \tilde{y}} +\frac{1}{8} \tonde{\frac{5 \pi^2}{18}-\gamma^2}\frac{1}{\tilde{y}^2} + \ O \tonde{\frac{1}{\tilde{y}^3}}. 
 \end{split}
\end{equation}
Finally, within the saddle point approximation to Eq.~(\ref{inverse}), for $\tilde{y} \gg 1$ we find the large deviation probability
\begin{equation}\label{zz22}
\begin{split}
P_N\left(-\log\left[\prod_{i=1}^N\frac{1}{|s_i|}\right]=N\tilde{y}\right)= C(\tilde{y},N) \tonde{{\frac{2e}{\sqrt{2 \pi}}}}^N \tilde{y}^N e^{-N\mathcal{F}(\tilde{y})} \quadre{1 + \frac{1}{N}},
 \end{split}
\end{equation}
where
\begin{equation}\label{zz3}
 \mathcal{F}(\tilde{y})=\tilde{y}  +\frac{\gamma}{2 \tilde{y}}-\frac{1}{8} \tonde{\frac{5 \pi^2}{18}-\gamma^2}\frac{1}{\tilde{y}^2} + \ O \tonde{\frac{1}{\tilde{y}^3}}.
\end{equation}
The prefactor 
\begin{equation}
 C(\tilde{y},N)= \tonde{\frac{1}{2 \pi N \phi^{''}_N(k^*(\tilde{y}))}}^{\frac{1}{2}} \frac{c(k^*(\tilde{y}))}{\lambda_{\text{max}}(k^*(\tilde{y}))}
\end{equation}
yields only logarithmic corrections to the exponent. 

As commented in the main text, when restricting to the linear term in (\ref{zz3}), the large deviation statistics for the correlated denominators coincides with that of {\em independent} identically distributed energy denominators. Indeed, from Eqs.~($\ref{asimptotica}$) and ($\ref{order0val}$) it follows that to leading order in $k+1$ the exponential growth of $G_N$ is almost equal to that of the generating function $g_N(k)=\quadre{2^{\frac{k+1}{2}}\Gamma \tonde{\frac{k+1}{2}}/\sqrt{2 \pi}}^N$ associated with products of $N$ independent Gaussian denominators with unit variance. For $\tilde{y} \gg 1$, the tail of the distribution is determined by the residue of the pole of the generation function at $k = -1$, which is identical in the two cases.  Repeating the above derivation of large deviations for independent denominators with generating function $g_N(k)$, one finds that it differs from (\ref{zz22}) at order $\ O \tonde{\frac{1}{\tilde{y}}}$: the tails for correlated denominators are suppressed by a factor $\text{exp}\tonde{-N \frac{\log 2}{2} \frac{1}{\tilde{y}}}$. The correction $\delta \lambda (k)$ in (\ref{fullLambda}) contributes to (\ref{zz3}) only at order  $\ O \tonde{1/\tilde{y}^2}$.

\section{Some useful combinatoric results for diagrams}
\label{app:combinatorics}
Let $\mathcal{T}_n$ be the number of tree-like diagrams with a root of connectivity $2$, and $n$ vertices with connectivity $4$. These trees are obtained by merging two trees of branching ratio $3$ at the root, and therefore
\begin{equation}\label{glue}
\mathcal{T}_n= \sum_{\begin{subarray}{1}
                      \hspace{.4 cm} n_1, n_2\geq 0\\
                        n_1+ n_2=n
                      \end{subarray}} T^{(n_1)}T^{(n_2)},
 \end{equation}
 where $T^{(m)}$ is the number of trees with $m$ vertices (including the root) and branching ratio $3$.
 This number satisfies the recursion equations
\begin{eqnarray}
T^{(0)}&=&1,\\
T^{(n)}&=&\sum_{n_1 +n_2+n_3=n-1} T^{(n_1)} T^{(n_2)} T^{(n_3)}.
\label{eq:recDn}
\end{eqnarray}
We can define the generating function
\begin{equation}
T(x)=\sum_{n\geq 0}x^n T^{(n)},
\end{equation}
which in virtue of (\ref{eq:recDn}) satisfies the polynomial equation
\begin{equation}
T(x)=1+xT(x)^3.
\end{equation}
Notice that the first singularity of $T(x)$ is a branch-cut at $x=4/27$, which implies the large-$n$ behavior $T^{(n)}\sim (27/4)^n$. However, we can find the $n$th order of the expansion for small $x$ using Lagrange's inversion theorem for the inverse function of
\begin{equation}
x(T)=\frac{T-1}{T^3},
\label{eq:xofD}
\end{equation}
expanding around $T=1$ ($x=0$).

This yields
\begin{equation}
T^{(n)}=\frac{1}{n}\lim_{T\to 1}\left[\frac{1}{(n-1)!}\frac{d^{n-1}}{dT^{n-1}}\left(\frac{T-1}{x(T)}\right) ^{n}\right]=\frac{1}{2n+1}\binom{3n}{n}.
\end{equation}
In general, for $k$-body interactions we have $T^{(n)}=\binom{(k-1)n}{n}/((k-2)n+1)$ diagrams. For $k=3$ these are the numbers of binary trees with $n$ vertices, or Catalan numbers.

There are two ways to solve Eq.~(\ref{glue}) and find $\mathcal{T}_n$. The first one is to notice that its generating function $\mathcal{T}(x)$ satisfies $\mathcal{T}(x)=T(x)^2$, write Eq.~(\ref{eq:xofD}) in terms of $\mathcal{T}$ and use Lagrange's inversion theorem again. Alternatively, one can use the explicit form of $T^{(n)}$ and apply a summation formula for the ratio of four $\Gamma$-functions   to obtain:
\begin{equation}
\mathcal{T}_n= \frac{3^{\frac{3}{2} + 3n}}{\pi} \frac{\Gamma \tonde{n + \frac{2}{3} }\Gamma \tonde{n + \frac{4}{3}}}{\Gamma \tonde{2n +3}}\sim \frac{3 }{4 }\sqrt{\frac{3}{\pi}} \frac{1}{n^{\frac{3}{2}}} \tonde{\frac{27}{4}}^n
\end{equation}
which is the result quoted in the text.

\bibliographystyle{elsarticle-num}
\bibliography{integralsbib-final}

 \end{document}